\documentclass[aip,jcp,reprint]{revtex4-1}
\usepackage{amsmath,amssymb,amsthm,dsfont,color,graphicx,listings,dcolumn,mathtools}
\usepackage[mathscr]{euscript}
\usepackage[dvipsnames]{xcolor}
\usepackage{soul}

\newcommand{\bigadd}[1]{{#1}}

\DeclareSymbolFont{matha}{OML}{txmi}{m}{it}% txfonts

\DeclareMathSymbol{\varv}{\mathord}{matha}{118}

\DeclarePairedDelimiterX\braket[2]{\langle}{\rangle}{#1 \delimsize\vert #2}
\renewcommand{\P}{\mathrm{\mathbf{P}}}
\newcommand{\1}{\mathds{1}}
\newcommand{\E}{\mathrm{\mathbf{E}}} %true expectation
\newcommand{\K}{\mathcal{K}} %Transition Operator
\renewcommand{\L}{\mathcal{L}} %Generic Operator
\newcommand{\eps}{\varepsilon}
\newcommand{\<}{\left \langle}
\renewcommand{\>}{\right \rangle}

\newtheorem{theorem}{Theorem}[section]
\newtheorem{assumption}[theorem]{Assumption}

\DeclareMathOperator*{\argmin}{arg\,min}

\begin{document}

\title{Galerkin Approximation of Dynamical Quantities using Trajectory Data}

\author{Erik H.\ Thiede}
\affiliation{Department of Chemistry and James Franck Institute, the University of Chicago, Chicago, IL 60637}
\email{thiede@uchicago.edu and dinner@uchicago.edu}
\author{Dimitrios Giannakis}
\affiliation{Courant Institute of Mathematical Sciences, New York University, New York, NY 10012, USA}
\email{dimitris@cims.nyu.edu and weare@nyu.edu}
\author{Aaron R.\ Dinner}
\affiliation{Department of Chemistry and James Franck Institute, the University of Chicago, Chicago, IL 60637}
%\email{dinner@uchicago.edu}
\author{Jonathan Weare}
\affiliation{Courant Institute of Mathematical Sciences, New York University, New York, NY 10012, USA}
%\email{weare@nyu.edu}

\begin{abstract}
    Understanding chemical mechanisms requires estimating dynamical statistics such as expected hitting times, reaction rates, and committors. 
    %In systems with well-defined metastable states and free energy barriers, these quantities can be estimated using enhanced sampling methods combined with classical rate theories.  
    %However, more general numerical methods are required when considering complex processes with rugged landscapes or multiple pathways.
	Here, we present a general framework for calculating these dynamical quantities by approximating boundary value problems using dynamical operators with a Galerkin expansion.
    %This allows the estimation of dynamical statistics even in the absence of long equilibrium trajectories realizing the dynamical process in question.
	A specific choice of basis set in the expansion corresponds to estimation of dynamical quantities using a Markov state model.
	\bigadd{More generally, the boundary conditions impose restrictions on the choice of basis sets.}
    \bigadd{We demonstrate how an alternative basis can be constructed using ideas from diffusion maps.}
    In our numerical experiments, this basis gives results of comparable or better accuracy to Markov state models.
    Additionally, we show that delay embedding can \bigadd{reduce the information lost when projecting the system's dynamics for model construction; this  improves estimates of dynamical statistics considerably over the standard practice of increasing the lag time.}
\end{abstract}

\maketitle

\section{Introduction}
\bigadd{
Molecular dynamics simulations allow chemical mechanisms to be studied with atomistic detail.  
By averaging over trajectories, one can estimate dynamical statistics such as mean first-passage times or committors.  
These quantities are integral to chemical rate theories.\cite{kramers1940brownian, hanggi1990reaction, vanden2006transition}  However, events of interest often occur on timescales several orders of magnitude longer than the timescales of microscopic fluctuations.  
In such cases, collecting chemical-kinetic statistics by integrating the system's equations of motion and directly computing averages (sample means) requires prohibitively large amounts of computational resources.
}

%Although computer simulations allow chemical dynamics to be studied with atomistic detail, predicting chemical kinetics from simulation remains difficult.
%While in principle most kinetic quantities can be calculated by simply propagating the dynamics forward in time, in practice this may require prohibitively large amounts of computational resources.
%For instance, atomistic molecular dynamics simulations require the force to be evaluated every few femtoseconds.
%However, many biomolecular processes such as protein folding or large-scale conformational change occur on timescales of milliseconds or longer. 
%Observing these events through direct simulation would require at least $10^{12}$ force evaluations.
%\bigadd{
%Simulations of these lengths are currently only possibly for smaller proteins.\cite{bernardi2015enhanced}
%}
%But since processes such as protein folding happen on timescales of milliseconds or longer, observing these events through direct simulation  requires at least $10^{12}$ force evaluations.
%Consequently, direct simulation is currently possible only for small proteins or rapid chemical processes

The traditional way to address this separation in timescales was through theories of activated processes.\cite{hanggi1990reaction, berezhkovskii2014multidimensional}  By assuming that the kinetics are dominated by passage through a single transition state, researchers were able to obtain approximate analytical forms for reaction rates and related quantities.  
These expressions can be connected with microscopic simulations by evaluating contributing statistics such as the potential of mean force and the diffusion tensor.\cite{ma2006dynamic,ovchinnikov2016simple,ghysels2017position}
 \bigadd{However, many processes involve multiple reaction pathways, such as the folding of larger proteins \cite{dinner1999thermodynamics, dinner2000understanding}.  In these cases it may not be possible in practice, or even in principle, to represent the system in a way that the assumptions underlying theories of activated processes are reasonable.}

\bigadd{
More recently, rates and other dynamical statistics have been estimated using transition path sampling algorithms.  
These focus sampling only on the pathways connecting metastable states.\cite{dellago1998transition, bolhuis2002transition}  
While this can be done efficiently if these pathways are short, if they are long and include multiple intermediates the sampling task becomes difficult.\cite{grunwald2008precision, gingrich2015preserving} 
Another approach is to use splitting schemes, which aim to efficiently direct sampling by intelligently splitting and reweighting short trajectory segments.\cite{huber1996weighted, van2003novel, faradjian2004computing, ffs_allen2006simulating, warmflash2007umbrella, vanden2009exact, dickson2009separating, guttenberg2012steered, bello2015exact, dinner2018trajectory}
% This approach underlies schemes such as weighted ensemble,\cite{huber1996weighted} transition interface sampling,\cite{van2003novel} forward flux sampling,\cite{ffs_allen2006simulating} nonequilibrium umbrella sampling,\cite{warmflash2007umbrella, dickson2009separating, dinner2018trajectory} and steered transition path sampling.\cite{guttenberg2012steered}.  [EHT: Is classifying all of these as splitting schemes acceptable?]
% In particular, nonequlibrium umbrella sampling has been recently extended into a general sampling scheme applicable to processes arbitrarily far from equilibrium or even a nonequilibrium steady state.\cite{dinner2018trajectory}
Some of these methods can yield results that are exact up to statistical precision, with minimal assumptions about the dynamics \cite{warmflash2007umbrella, vanden2009exact, dickson2009separating, guttenberg2012steered, bello2015exact, dinner2018trajectory}. However, the efficiency of these schemes is generally dependent on a reasonable choice of low-dimensional \textit{collective variable} (CV) space: a projection of the system's phase space.  Not only can this choice be nonobvious, it can be statistic specific.  Moreover, starting and stopping the molecular dynamics many times based on the values of the CVs may be impractical depending on the implementation of the molecular dynamics engine and the overhead associated with computational communication.
}

\bigadd{
A third approach is the construction of Markov state models (MSMs).\cite{schutte1999direct,swope2004describing,pande2010everything}  Here, the dynamics of the system are modeled as a discrete-state Markov chain with state-to-state transition probabilities estimated from previously sampled data.  
Projecting the dynamics onto a finite-dimensional model introduces a systematic bias, although this bias goes to zero in the correct limit of infinitely many states.\cite{sarich2010approximation}
While MSMs were initially developed as a technique for approximating the slowest eigenmodes of a system's dynamics,\cite{schutte1999direct} MSMs can also be used to calculate dynamical statistics for the study of kinetics.\cite{noe2008transition, noe2014introduction, keller2019markov}
Since MSM construction only requires time pairs separated by a single lag time, one has more freedom in how one generates the molecular dynamics data.
In particular, if the lag time is sufficiently short, MSMs can be used to estimate rates even in the absence of full reactive trajectories.
Constructing an efficient MSM requires projection onto CVs, and the systematic error in the resulting estimates can depend strongly on how they are defined.  However, the CV space can generally be higher dimensional since it is only used to define Markov states.
}

\bigadd{
It has been shown that calculating the system's eigenmodes with MSMs can be generalized to a basis expansion of the eigenmodes using an arbitrary basis set.\cite{weber2006meshless,sarich2010approximation,noe2013variational}
In this paper, we show that a similar generalization is possible for other dynamical statistics.
Rather than solving eigenproblems, these quantities solve linear boundary value problems.
This raises additional challenges: not only do the solutions obey specific boundary conditions, the resulting approximations are sensitive to the choice of lag time.  We provide numerical schemes to address these difficulties.
}

We organize our work as follows.
In Section~\ref{sec:background} we give background on the transition operator and review both MSMs and more general schemes for data-driven analysis of the spectrum of dynamical operators.
We then continue our review with the connection between operator equations and chemical kinetics in Section~\ref{sec:feynman_kac}.
In Section~\ref{sec:framework} we present our formalism.
\bigadd{
We discuss the choice of basis set in Section~\ref{sec:dmap_basis_definition} and introduce a new algorithm for constructing basis sets that obey the boundary conditions our formalism requires.
In Section~\ref{sec:delay_embedding} we show that delay embedding can recover information lost in projecting the system's dynamics onto a few degrees of freedom, negating the need for increasing the scheme's lag time to enforce Markovianity. 
}
We then demonstrate our algorithm on a collection of long trajectories of the Fip35 WW domain dataset in Section~\ref{sec:fip35}, and conclude in Section~\ref{sec:conclusion}.

\section{Background}\label{sec:background}

Many key quantities in chemical kinetics can be expressed through solutions to linear operator equations.  Key to this formalism is the \textit{transition operator}.
We begin by assuming that the system's dynamics are given by a Markov process $\xi^{(t)}$ that is time-homogeneous, i.e. that the dynamics are time-independent.  We do not put any restrictions on the nature of the system's state space.  For example, if $\xi$ is a diffusion process, the state space could be the space of real coordinates, $\mathbb{R}^n$.  Similarly, for a finite-state Markov chain, it would be a finite set of configurations.  \bigadd{We also do not assume that the dynamics are reversible or that the system is in a stationary state unless specifically noted.}

The transition operator at a lag time of $s$ is defined as
\begin{equation}\label{eq:defn_markov_operator}
\K_s f(x) = \E \left[f \left(\xi^{(s)}\right) | \xi^{(0)} = x\right],
\end{equation}
where $f$ is a function on the state space, and $\E$ denotes expectation.
Note that due to time-homogeneity, we could just as easily have defined the transition operator with the time pair $\left(\xi^{(t)}, \xi^{(t+s)}\right)$ in place of $\left(\xi^{(0)}, \xi^{(s)}\right)$. 
Depending on the context in question, $\K_s$ may also be referred to as the Markov or Koopman operator.\cite{eisner2015operator, klus2018data}
We use the term transition operator as it is well established in the mathematical literature and stresses the notion that $\K_s$ is the generalization of the transition matrix for finite-state Markov processes.
For instance, the requirement that the rows of a transition matrix sum to one generalizes to
\begin{equation}
\K_s 1 =  \E \left[1 | \xi^{(0)} = x\right] = 1.
\end{equation}
Studying the transition operator provides, in principle, a route to analyzing the system's dynamics.  
Unfortunately $\K_s$ is often either unknown or too complicated to be studied directly.
This has motivated research into data-driven approaches that instead treat $\K_s$ indirectly by analyzing sampled trajectories.

\subsection{Markov State Modeling}\label{ssec:msm}
One approach to studying chemical dynamics through the transition operator is the construction of Markov state models.\cite{schutte1999direct,swope2004describing,pande2010everything}
In this technique one constructs a Markov chain on a finite state space to model the true dynamics of the system.
The transition matrix of this Markov chain is then taken as a model for the true transition operator.

To construct an MSM from trajectory data, we partition the system's state space into $M$ nonoverlapping sets.  We refer to these sets as Markov states and denote them as $S_i$.
Now, let $\mu$ be an arbitrary probability measure.
If the system is initially distributed according to $\mu$, the probability of transitioning from a set $S_i$ to $S_j$ after a time $s$ is given by 
\begin{equation}\label{eq:msm_tmat_defn}
P_{ij} = \frac{\int \1_{S_i} (x) \K_s \1_{S_j}(x)  \mu(dx) }{\int \1_{S_i} (x) \mu(dx)},
\end{equation}
where $\1_{S_i}$ is the indicator function
\begin{equation}\label{eq:defn_indicator}
    \1_{S_i}(x) = \begin{cases}
    1\text{ for } x \text{ in } S_i \\
    0\text{ otherwise.}
\end{cases}
\end{equation}
Here $\int f(x) \mu(dx)$ is the expectation with respect to the probability measure $\mu$.\cite{billingsley2008probability}
When $\mu$ has a probability density function this integral is the same as the integral against the density, and in a finite state space it would be a weighted average over states.
This formalism lets us treat both continuous and discrete state spaces with one notation.

Because the sets $S_i$ partition the state space, a simple calculation shows that the elements in each row of $P_{ij}$ sum to one.  $P_{ij}$ therefore defines a transition matrix for a finite-state Markov process where state $i$ corresponds to the set $S_i$.
The dynamics of this process are a model for the true dynamics, and $P_{ij}$ is a model for the transition operator.

To build this model, we construct an estimate of $P_{ij}$ from sampled data.
A simple approach is to collect a dataset consisting of $N$ time pairs, $(X_n,Y_n)$.
Here the initial point $X_n$ is drawn from $\mu$, and  $Y_n$ is collected by starting at $X_n$ and propagating the dynamics for time $s$.
Note that since the choice of $\mu$ in \eqref{eq:msm_tmat_defn} is relatively arbitrary, it can be defined implicitly through the sampling procedure.
For instance, one can construct a dataset by extracting all pairs of points separated by the lag time $s$ from a collection of trajectories; since we have assumed the dynamics are time-homogeneous, the actual physical time at which $X_n$ was collected does not matter.
We then define $\mu$ to be the measure from which our initial points $X_n^{(0)}$ were sampled.
%
%With this dataset, $P_{ij}$ is now approximated as
%\begin{equation}\label{eq:msm_tmat_approx}
%\bar{P}_{ij} = \bar{C}_{ij} / \sum_j \bar{C}_{ij}\text{, where }
%\end{equation}
%where $\bar{C}_{ij}$ is the count matrix defined as 
%\begin{equation}
%\bar{C}_{ij} = \sum_{n=1}^N  \1_{S_j} \left(Y_n\right) \1_{S_i} \left(X_n\right).
%\end{equation}
%Like $P_{ij}$, \eqref{eq:msm_tmat_approx} defines a valid transition matrix.
%\bigadd{
%This is not the only approach for constructing estimates of $P_{ij}$.
%One commonly used approach modifies this procedure to ensure that $\bar{P}_{ij}$ obeys detailed balance. If $\mu$ is the stationary measure of the system, then this can be accomplished by replacing $\bar{C}_{ij}$ with $\bar{C}_{ij} + \bar{C}_{ji}$ in \eqref{eq:msm_tmat_approx}.\cite{noe2008probability}  For general choice of $\mu$, this introduces a systematic bias.  Consequently, recent approaches have proposed finding a maximum-likelihood estimate for $P_{ij}$ using a self-consistent iteration.\cite{bowman2009progress,prinz2011markov}
%}
With this dataset, $P_{ij}$ is now approximated as
\begin{equation}\label{eq:msm_tmat_approx}
\bar{P}_{ij} = \frac{\sum_{n=1}^N  \1_{S_j} \left(Y_n\right) \1_{S_i} \left(X_n\right)}{\sum_{n=1}^N \1_{S_i} \left(X_n\right)}
\end{equation}
Like $P_{ij}$, \eqref{eq:msm_tmat_approx} defines a valid transition matrix.
\bigadd{
This is not the only approach for constructing estimates of $P_{ij}$.
One commonly used approach modifies this procedure to ensure that $\bar{P}_{ij}$ gives reversible dynamics.  In this approach, one adds a self-consistent iteration that seeks to find the reversible transition matrix with the maximum likelihood given the data.\cite{bowman2009progress,prinz2011markov}
}

The MSM approach has many attractive features.
\bigadd{
Conceptually, an MSM gives a reduced model of the dynamics which helps the practitioner gain intuition.}
MSMs can also be used to calculate a wide class of dynamical quantities, including committors, reaction rates, and expected hitting times.\cite{noe2008transition, noe2014introduction, keller2019markov}
\bigadd{
Importantly, as constructing MSMs only requires datapoints separated by a short lag time, these long-time dynamical quantities can be evaluated using a collection of short trajectories.\cite{noe2009constructing}
%Moreover, since the MSM can be constructed using trajectories initialized from arbitrary probability distributions, sampling can, in theory, be focused in critical regions of the state space.  
In this paper, we focus exclusively on the latter application and consider MSMs as a technique for calculating the dynamical quantities required in rate theories.
}

The accuracy with which $P_{ij}$ approximates $\K_s$ depends strongly on the choice of the sets $S_i$, and choosing good sets is a nontrivial problem in high-dimensional state spaces.\cite{schwantes2013improvements,perez2013identification,schwantes2014perspective,schutte2015critical,shukla2015markov}
To address this issue,
states are generally constructed by projecting the system's state space onto a CV space. Sets are then defined by either gridding the CV space or clustering sampled configurations based on the values of their CVs.
Unfortunately, when gridding, the number of states grows exponentially with the dimension of the CV space.
\bigadd{
This is not necessarily the case for partitioning schemes based on data clustering and
recent work in this direction appears promising. \cite{prinz2011markov, berezovska2013consensus, sheong2014automatic, li2016effect, husic2017ward, husic2018minimum}  However, effectively clustering high-dimensional data is a nontrivial problem,\cite{steinbach2004challenges,kriegel2009clustering}
and constructing an MSM that accurately reflects the dynamics may still require knowledge of a good, relatively low-dimensional CV space.
\cite{schwantes2013improvements, perez2013identification, scherer2015pyemma}
}
% %As a consequence, constructing an MSM that accurately reflects the dynamics may still require knowledge of a good, relatively low-dimensional CV space.
% \bigadd{
% %Extracting such a subspace has been the focus of recent work,\cite{schwantes2013improvements, perez2013identification, scherer2015pyemma}
% %as have been efforts to find clustering algorithms that lead to better MSMs.\cite{berezovska2013consensus, sheong2014automatic, li2016effect, husic2017ward, husic2018minimum}
% Though recent work on developing improved clustering algorithms appears promising, \cite{prinz2011markov, berezovska2013consensus, sheong2014automatic, li2016effect, husic2017ward, husic2018minimum} constructing an MSM that accurately reflects the dynamics may still require knowledge of a good, relatively low-dimensional CV space.
% \cite{schwantes2013improvements, perez2013identification, scherer2015pyemma}

\subsection{Data-driven Solutions to Eigenfunctions of Dynamical Operators}\label{ssec:edmd}
A related approach to characterizing chemical systems is to estimate the eigenfunctions and eigenvalues of operators associated with the system's dynamics from sampled data.\cite{klus2018data}
These separate the dynamics by timescale: eigenfunctions with larger eigenvalues correlate with the system's slower degrees of freedom.
These eigenfunctions and eigenvalues can often be approximated from trajectory data, even when the transition operator is unknown.
\bigadd{
Multiple schemes that attempt this have been proposed, often independently, in different fields.\cite{molgedey1994separation, takano1995relaxation, hirao1997molecular,schutte1999direct, schutte2011markov, weber2006meshless, noe2013variational, giannakis2015spatiotemporal, giannakis2017data, williams2015data}
We will refer to the family of these techniques using the umbrella term \textit{Dynamical Operator Eigenfunction Analysis (DOEA)} for brevity and convenience.
Below, we summarize a simple DOEA scheme for the transition operator for the reader's convenience, largely following work in reference~\onlinecite{hirao1997molecular}.  Other schemes exist and we refer the reader to reference~\onlinecite{klus2018data} for further reading.
}

Here, we consider the solution to the eigenproblem
\begin{equation}\label{eq:koopman_eigenproblem}
\K_s \psi_l (x) = \lambda_l \psi_l(x).
\end{equation}
We approximate $\psi_l$ as a sum of basis functions $\phi_j$ with unknown coefficients $a_j$,
\begin{equation}\label{eq:efxn_basis_expansion}
\psi_l(x) = \sum_{j=1}^M a_j \phi_j(x).
\end{equation}
This is an example of Galerkin approximation of \eqref{eq:koopman_eigenproblem}, \cite{schutte1999direct} a formalism we cover more closely in Section~\ref{sec:framework}.
We now assume our data takes the form discussed in Section~\ref{ssec:msm}.
Substituting the basis expansion into \eqref{eq:koopman_eigenproblem}, multiplying by $\phi_i(x)$, and taking the expectation against $\mu$, we obtain the matrix equation
\begin{equation}\label{eq:esystem_coefficient_equation}
\sum_{j=1}^M  K_{ij}a_j =\lambda_l \sum_{j=1}^M S_{ij} a_j
\end{equation}
where $K_{ij}$ and $S_{ij}$ are defined as
\begin{align}
K_{ij} &= \int \phi_i(x) \K_s \phi_j(x) \mu(dx) \label{eq:koopman_matrix_defns} \\
S_{ij} &= \int \phi_i(x) \phi_j(x) \mu(dx)
\label{eq:stiffness_matrix_defns}
\end{align}
respectively.  The matrix elements can be approximated as
\begin{align}\label{eq:mc_koopman_approximation}
\bar{K}_{ij} =& \frac{1}{N}\sum_{n=1}^N \phi_i(X_n) \phi_j(Y_n) \\
\bar{S}_{ij} =&  \frac{1}{N}\sum_{n=1}^N \phi_i(X_n) \phi_j(X_n). \label{eq:mc_stiffness_approximation} 
\end{align}
We substitute these approximations into \eqref{eq:esystem_coefficient_equation} and solve for estimates of $a_i$ and $\lambda_l$. 
Equation~\eqref{eq:efxn_basis_expansion} can then be used to give an approximation for $\psi_l$.

\bigadd{
DOEA schemes are closely linked to MSMs. Using the indicator functions from Section~\ref{ssec:msm} is mathematically equivalent to solving for the eigenfunctions of $P_{ij}$.
Indeed, one of the first uses for MSMs was for approximating the eigenfunctions and eigenvalues of the transition operator.\cite{schutte1999direct, sarich2010approximation}
}

\bigadd{
The use of more general basis sets in DOEA allows information to be more easily extracted from high-dimensional CV spaces and gives added flexibility in algorithm design.\cite{molgedey1994separation, schwantes2013improvements, perez2013identification, boninsegna2015investigating, vitalini2015basis}
Time-lagged independent component analysis (TICA) corresponds to a basis of linear functions and is commonly applied as a preprocessing step to generate CVs for MSM construction.\cite{molgedey1994separation, schwantes2013improvements, perez2013identification}
%Recent work has focused on solving eigenproblems using a variational principle.\cite{noe2013variational, nüske2014variational,nuske2016variational, wu2017vamp, mardt2017vampnets} 
Variational principles can be exploited to obtain the eigenfunctions of $\K_s$ for reversible dynamics (variational approach of conformation dynamics, VAC) \cite{noe2013variational} and, more generally, for the singular value decomposition of $\K_s$ (the variational approach for Markov processes, VAMP).\cite{wu2017vamp, mardt2017vampnets}
These principles can allow the creation of cost functions that can be used to assess how well a basis recapitulates the spectral properties of $\K_s$.\cite{noe2013variational, nüske2014variational, wu2017vamp}
Furthermore, by directly minimizing these cost functions, one can construct nonlinear basis sets using complex machine learning approaches such as tensor-product algorithms or neural networks.\cite{nuske2016variational, mardt2017vampnets}
}

%However, whereas an MSM can be used to calculate a broad range of dynamical quantities, DOEA schemes only calculate the eigenfunctions and eigenvalues of the dynamical operator in question.
%\bigadd{
%DOEA schemes do not directly give the quantities required in traditional rate theories, such as committors and mean first-passage times.
%}
\bigadd{
While attempts have been made to define a theory of chemical dynamics purely in terms of the transition operator's eigenfunctions and eigenvalues,\cite{prinz2014spectral}  
most chemical theories require dynamical quantities such as committors and mean first-passage times.
In this work, we show that it is possible to construct estimates of these quantities using a general basis expansion. 
Just as DOEA schemes extend MSM estimates of spectral properties to general basis functions, our formalism generalizes MSM estimation of the quantities used in rate theory.
%Here, we take a different approach and directly estimate the dynamical quantities required.
%\bigadd{
%Our work can be considered a generalization of the estimates constructed from MSMs to a more general family of basis sets.  Alternatively, our work can be considered to be an extension of ideas in DOEA to a new class of operator equations.
%}
}

\section{The Generator and Chemical Kinetics}\label{sec:feynman_kac}
%Many key quantities in chemical kinetics have associated Feynman-Kac formulas.  
%In these equations, dynamical quantities are expressed as the solutions to operator equations acting on functions of the state space.
Many key quantities in chemical kinetics solve operator equations acting on functions of the state space.
%In our work, our aim is to solve these operator equations directly to get key quantities in chemical kinetics.
Below, we give a quick review of this formalism, detailing a few examples of chemically relevant quantities that can be expressed in this manner.
These include statistics such as the mean first-passage time, forwards and backwards committors, and autocorrelation times.
In particular, many of these operator equations are examples of Feynman-Kac formulas.
For an in-depth treatment of this formalism, we refer the reader to references~\onlinecite{delmoral2004feynman, karatzas2012brownian}.

In this work, we focus on analyzing data gathered from experiment or simulation. 
We expect the data to consist of a series of measurements collected at a fixed time interval.
Therefore, rather than considering the dynamics of $\xi^{(t)}$, we will consider the dynamics of a discrete-time process $\Xi^{(t)}$ constructed by recording $\xi$ every $\Delta t$ units of time.
If $\Delta t$ is sufficiently small, this should not appreciably change any kinetic quantities.

In the discussion that follows, we choose to work with the generator of $\Xi^{(t)}$, defined as 
\begin{equation}\label{eq:generator}
\mathcal{L} f(x) = \frac{\K_{\Delta t} f(x) - f(x)}{\Delta t},
\end{equation}
instead of the transition operator.  This makes no mathematical difference, but using $\mathcal{L}$ simplifies the presentation.  
We also stress that, with the exception of \eqref{eq:expr_com} below, the equations that follow hold only for a lag-time of $s=\Delta t$. 
For larger lag times, i.e. $s > \Delta t$, these equations only hold approximately.  This is discussed further in Section \ref{sec:delay_embedding}.

\subsection{Equations using the Generator}\label{ssec:fk_with_generator}
We begin by considering the mean first-passage time and forward committor, two central quantities in chemical kinetics.\cite{hanggi1990reaction, du1998transition, bolhuis2000reaction}
Let $A$ and $B$ be disjoint subsets of state space and let $\tau_A$ be the first time the system enters $A$:
\begin{equation}
\tau_A  = \min \left\{ t \geq 0 | \Xi^{(t)} \in A \right\}.
\end{equation}
The \textit{mean first-passage time} is the expectation of $\tau_A$, conditioned on the dynamics starting at $x$:
\begin{equation}
m_A (x)  = \E \left[\tau_A | \Xi^{(0)} = x\right].
\end{equation}
%In Arrhenius kinetics 
Note that $1 / m_A(x)$ is a commonly used definition of the rate.\cite{hanggi1990reaction}
The \textit{forward committor} is defined as the probability of entering $B$ before $A$, conditioned on starting at $x$:
\begin{equation}
q_+(x)  = \P \left[\tau_B < \tau_A | \Xi^{(0)} = x \right].
\end{equation}

Both of these quantities solve operator equations using the generator.
The mean first-passage obeys the operator equation
\begin{equation}
\begin{split}
\mathcal{L} m_A (x) &= - 1 \text{ for } x \text{ in } A^c \label{eq:gen_expr_htime_1} \\
m_A (x) &= 0 \text{ for } x \text{ in } A. 
\end{split}
\end{equation}
Here $A^c$ denotes the set of all state space configurations not in $A$.
Equation \eqref{eq:gen_expr_htime_1} can be derived by conditioning on the first step of the dynamics.
For all $x$ in $A^c$ we have
\begin{align*}
m_A(x) &=   \E \left[\tau_A | \Xi^{(0)} = x\right] \\
%&= \E \left[ \E \left[ \tau_A \big | \Xi^{(\Delta t)}\right] \bigg | \Xi^{(0)} =x \right]  \\
&=  \E \left[ m_A\left(\Xi^{(\Delta t)} \right) + \Delta t  \bigg | \Xi^{(0)} =x \right] \\
&=  \E \left[ m_A\left(\Xi^{(\Delta t)} \right) \bigg | \Xi^{(0)} =x \right] + \Delta t \\
&= \K_{\Delta t} m_A(x) + \Delta t
\end{align*}
where the second line follows from the time-homogeneity of $\Xi$.
Rearranging then gives \eqref{eq:gen_expr_htime_1}.

We can show that the forward committor obeys
\begin{align}\label{eq:gen_expr_fcommittor}
    \mathcal{L} q_+ (x) &= 0 \text{ for } x \text{ in } (A\cup B)^c \nonumber \\
    q_+ (x) &= 0\text{ for } x \text{ in } A \\
    q_+ (x) &= 1\text{ for } x \text{ in } B \nonumber 
\end{align}
by similar arguments.
We introduce the random variable
\begin{equation}
\mathbf{1}_{\tau_B < \tau_A} = \begin{cases}
1\text{ if } \tau_B < \tau_A \\
0 \text{ otherwise}.
\end{cases}
\end{equation}
For all $x$ outside $A$ and $B$, we can then write
\begin{align*}
q_+(x) &= \E \left[\mathbf{1}_{\tau_B < \tau_A}\big | \Xi^{(0)} = x\right] \\
%=& \E \left[ \E \left[ \mathbf{1}_{\tau_B < \tau_A} \big | \Xi^{(\Delta t)}\right] \bigg | \Xi^{(0)} =x \right]  \\
&=  \E \left[ q_+\left(\Xi^{(\Delta t)} \right) \bigg | \Xi^{(0)} =x \right] \\
&= \K_{\Delta t} q_+(x),
\end{align*}
which gives \eqref{eq:gen_expr_fcommittor} on rearranging.

\subsection{Expressions using Adjoints of the Generator}\label{ssec:fk_with_adjoints}
Additional quantities can be characterized using adjoints of the generator.
We reintroduce the sampling measure $\mu$ from Section~\ref{sec:background}, and define the inner product
\begin{equation}\label{eq:mu_inner_product}
    \<u,v\> = \int u(x) v(x) \mu(dx).
\end{equation}
Equipped with this inner product, the space of all functions that are square-integrable against $\mu$ forms a Hilbert space that that we denote as $L^2_\mu$.
The unweighted adjoint of $\mathcal{L}$ is the operator $\mathcal{L}^\dagger$ such that for all $u$ and $v$ in the Hilbert space,
\begin{equation}\label{eq:defn_adjoint}
\< \mathcal{L}^\dagger u, v \>  = \<u, \mathcal{L} v\>.
\end{equation}

We now assume that the system has a unique stationary measure.
The change of measure from $\mu$ to the stationary measure is defined as the function $\pi$ such that
\begin{equation}
\int  \E\left[ f\left(\Xi^{(t)}\right)| \Xi^{(0)}=x\right] \pi (x) \mu(dx) = \int f(x) \pi (x) \mu(dx)
\end{equation}
or equivalently,
\begin{equation}\label{eq:com_with_generator}
\int \mathcal{L} f(x)  \pi (x) \mu (dx) = 0
\end{equation}
holds for all functions $f$.
As an example, if the system's state space is Euclidean space and the dynamics are stationary at thermal equilibrium, we would have
\begin{equation*}
\pi(x) \mu(dx) \propto e^{\frac{H(x)}{k_B T}}dx
\end{equation*}
where $H(x)$ is the system's Hamiltonian, $T$ is the system's temperature, and $k_B$ is the Boltzmann constant.
However, this relation is not necessarily true for general state spaces or for nonequilibrium stationary states.

The change of measure to the stationary measure can be written as the solution to an expression with $\mathcal{L}^\dagger$.
Interpreting \eqref{eq:com_with_generator} as an inner product, the definition of the adjoint implies
\begin{equation*}
0 = \<\pi, \mathcal{L} f\> = \<\mathcal{L}^\dagger \pi, f\>
\end{equation*}
for all $f$, or equivalently
\begin{equation}\label{eq:expr_com}
    \mathcal{L}^\dagger \pi(x) = 0.
\end{equation}

Other equations may use weighted adjoints of $\mathcal{L}$.
Let $p$ be the change of measure from $\mu$ to another, currently unspecified measure.  The \textit{$p$-weighted adjoint} of $\mathcal{L}$ is the operator $\mathcal{L}_p^\dagger$ such that
\begin{equation}\label{eq:defn_weighted_adjoint}
\< u, p \mathcal{L} v \> = \< \mathcal{L}^\dagger_p u, p v \>.
\end{equation}
A few manipulations show that the weighted adjoint can be expressed as
\begin{equation}\label{eq:adjoint_expression}
\mathcal{L}^\dagger_p f(x) = \frac{1}{p(x)} \mathcal{L}^\dagger \left(f p\right)(x).
\end{equation}
This reduces to the unweighted adjoint when $p(x)=1$.

One example of a formula that uses a weighted adjoint is a relation for the \textit{backwards committor}.  The backwards committor is the probability that, if the system is observed at configuration $x$ and the system is in the stationary state, the system exited state $A$ more recently than state $B$.
It satisfies the equation
\begin{align}\label{eq:gen_expr_bcommittor}
    \mathcal{L}^\dagger_{\pi}  q_- (x) &= 0 \text{ for } x \text{ in } (A\cup B)^c \nonumber \\
    q_- (x) &= 1\text{ for } x \text{ in } A \\
    q_- (x) &= 0\text{ for } x \text{ in } B. \nonumber 
\end{align}

Finally, we note that some quantities in chemical dynamics require the solution to multiple operator equations.
For instance, in transition path theory\cite{vanden2006transition} the \textit{total reactive current} and \textit{reaction rate} between $A$ and $B$ require evaluating the backwards committor and the forward committor, followed by another application of the generator.
The total reactive current between the two sets is given by
\begin{align}\label{eq:defn_total_rxn_current}
    I_{AB} =& \int q_-(x) \1_C(x) \mathcal{L}  \left( \1_{C^c} q_+ \right)(x) \pi(x)\mu(dx)  \\
                  &-\int q_-(x) \1_{C^c}(x) \mathcal{L}  \left( \1_{C} q_+ \right)(x) \pi(x)\mu(dx).\nonumber
\end{align}
Here $C$ is a set that contains $B$ but not $A$.
The reaction rate constant is then given by  
\begin{equation}\label{eq:defn_tpt_rate}
    k_{AB} = \frac{I_{AB}}{\int q_-(x) \pi(x)\mu(dx)}.
\end{equation}
We derive these expressions in the Supplementary Information in Section~\ref{ssec:tpt_derivation} through arguments very similar to those presented in reference~\onlinecite{metzner2009transition}.

Evaluating the \textit{integrated autocorrelation time} (IAT) of a function requires estimating $\pi$, as well as solving an equation using the generator.
For a function with $\int f(x) \pi(x)\mu(dx)=0$, the IAT 
is the sum over the correlation function
\begin{equation}\label{eq:defn_iat}
   t_f =\left(2\sum_{i=0}^\infty  \frac{ \int f(x)\K_{i \Delta t}f(x)\pi(x)\mu(dx) }{\int \left(f(x ))\right)^2 \pi(x)\mu(dx)}  - 1\right)\Delta t
\end{equation}
and, using the Neumann series representation\cite{yosida1971functional} of the appropriate pseudo-inverse of $\mathcal{L}$, can be expressed as 
\begin{equation}
    t_f = 2\frac{\int f(x) \omega(x) \pi (x) \mu (dx)}{\int f(x)^2 \pi(x)\mu(dx) }- \Delta t, \label{eq:gen_expr_actime}
\end{equation}
where $\omega$ is the solution to the equation
\begin{align}\label{eq:defn_chi_actime}
\mathcal{L}  \omega(x) &= f(x)
\end{align}
constrained to have $\int \omega(x) \pi (x) \mu (dx) = 0$.

Note that although the quantities above give us information about the long-time behavior of the system, the formalism introduced here only requires information over short time intervals.
This suggests that solving these equations directly could lead to a numerical strategy for estimating these long-time statistics from short-time data.

\section{Dynamical Galerkin Approximation}\label{sec:framework}
\bigadd{
Inspired by the theory behind DOEA and MSMs, we seek to solve the equations in Section~\ref{sec:feynman_kac} in a data-driven manner.
}
We first note that the equations follow the general form 
\begin{equation}\label{eq:generic_inhomogeneous_eqn} 
\begin{split}
    \mathcal{L} g(x) =& h(x)  \text{ for } x \text{ in } D \\
    g(x) =& b(x) \text{ for } x \text{ in } D^c
\end{split}
\end{equation}
or
\begin{equation}\label{eq:generic_inhomogeneous_adjoint_eqn} 
\begin{split}
    \mathcal{L}_p^\dagger g(x) =& h(x)  \text{ for } x \text{ in } D \\
    g(x) =& b(x) \text{ for } x \text{ in } D^c.
\end{split}
\end{equation}
Here $D$ is a set in state space that constitutes the \textit{domain}, $g$ is the unknown solution, and $h$ and $b$ are known functions.
If the generator and its adjoints are known, these equations can in principle be solved numerically.\cite{lapelosa2013transition,lai2018point, khoo2018solving}
However, this is generally not the case, and even if the operators are known, the dimension of the full state space is often too high to allow numerical solution.
In our approach, we use approximations similar to  \eqref{eq:mc_koopman_approximation} and \eqref{eq:mc_stiffness_approximation} to estimate these quantities from trajectory data.
This procedure only requires collections of short trajectories of the system, and works when the dynamical operators are not known explicitly.

We first discuss operator equations using the generator; equations using an adjoint require only slight modification and are discussed at the end of the subsection.
We construct an approximation of the operator equation through the following steps.
\begin{enumerate}
\item \textit{Homogenize boundary conditions:}  If necessary, rewrite \eqref{eq:generic_inhomogeneous_eqn} as a problem with homogeneous boundary conditions using a guess for $g$.
\item \textit{Construct a Galerkin scheme:} Approximate the solution as a sum of basis functions and convert the result of step 1 into a matrix equation.
\item \textit{Approximate inner products with trajectory averages:}  Approximate the terms in the Galerkin scheme using trajectory averages and solve for an estimate of $g$.
\end{enumerate}
Since we use dynamical data to estimate the terms in a Galerkin approximation, we refer to our scheme as \textit{Dynamical Galerkin Approximation} (DGA).

\subsection{Homogenizing the Boundary Conditions}
First, we rewrite \eqref{eq:generic_inhomogeneous_eqn} as a problem with homogeneous boundary conditions.
\bigadd{
This allows us to enforce the boundary conditions in step 2 by working within a vector space where every function vanishes at the boundary of the domain.
}
If the boundary conditions are already homogeneous, either because $b$ is explicitly zero or because $D$ includes all of state space, this step can be skipped.
We introduce a guess function $r$ that is equal to $b$ on $D^c$.  We then rewrite \eqref{eq:generic_inhomogeneous_eqn} in terms of the difference between the guess and the true solution:
\begin{align}
\gamma(x) =& g(x) - r(x).
\end{align}
This converts \eqref{eq:generic_inhomogeneous_eqn} into  a problem with homogeneous boundary conditions:
\begin{align}
	\mathcal{L} \gamma (x) =& h(x) - \mathcal{L} r(x) \text{ for } x \text{ in } D \label{eq:generic_operator_equation} \\
    \gamma(x) =& 0 \text{ for } x \text{ in } D^c.\label{eq:generic_homogeneous_bc}
\end{align}
A naive guess can always be constructed as
\begin{equation}\label{eq:naive_guess}
r^{naive}(x) = \1_{D^c}(x) b(x),
\end{equation}
but if possible, one should attempt to choose $r$ so that $\gamma$ can be efficiently expressed using the basis functions introduced in step 2.

\subsection{Constructing the Galerkin Scheme}\label{ssec:gkn_step}
We now approximate the solution of \eqref{eq:generic_operator_equation} and \eqref{eq:generic_homogeneous_bc} via basis expansion using the formalism of Galerkin approximation.
Equation \eqref{eq:generic_operator_equation} implies that
\begin{equation}\label{eq:weak_formulation}
\<u \1_D, \mathcal{L} \gamma \> = \<u \1_D, h\> - \<u \1_D, \mathcal{L} r \> 
\end{equation}
holds for all $u$ in the Hilbert space $L^2_{\mu}$.
This is known as the \textit{weak formulation} of \eqref{eq:generic_operator_equation}.\cite{evans1998partial}

The space $L^2_{\mu}$ is typically infinite dimensional.  Consequently, we cannot expect to ensure that \eqref{eq:weak_formulation} holds for every function in $L^2_\mu$.
We therefore attempt to solve \eqref{eq:weak_formulation} only on a finite-dimensional subspace of $L^2_{\mu}$.
To do this, we introduce a set of $M$ linearly independent functions denoted $\left\{\phi_{1},...,\phi_M\right\}$ that obey the homogeneous boundary conditions; we refer to these as the \textit{basis functions}.
The space of all linear combinations of the basis functions forms a subspace in $L^2_{\mu}$
which we call the \textit{Galerkin subspace}, $G$.
By construction, every function in $G$ obeys the homogeneous boundary conditions.
We now project \eqref{eq:weak_formulation} onto this subspace, giving the approximate equation
\begin{equation}\label{eq:weak_formulation_subspace}
\<\tilde{u}, \L \tilde{\gamma} \> = \<\tilde{u}, h\> - \<\tilde{u}, \L r \>
\end{equation}
for all $\tilde{u}$ in $G$.
Here $\tilde{\gamma}$ is the projection of $\gamma$ onto $G$.
\bigadd{
Constructing $G$ using a linear combination of basis functions that obey the homogeneous boundary conditions ensures that $\tilde{\gamma}$ obeys the homogeneous boundary conditions as well.  
If we had constructed $G$ using arbitrary basis functions, this would not be true.	
}
As we increase the dimensionality of $G$, we expect the error between $\gamma$ and $\tilde{\gamma}$ to become arbitrarily small.
Since $\tilde{u}$ is in $G$, it can be written as a linear combination of basis functions.
Consequently, if 
\begin{equation*}
\<\phi_i, \L \tilde{\gamma} \> =  \<\phi_i, h \> - \<\phi_i, \L r \>
\end{equation*}
holds for all  $\phi_i$, then \eqref{eq:weak_formulation_subspace} holds for all $\tilde{u}$.
Moreover, the construction of $G$ implies that there exist unique coefficients $a_j$ such that 
\begin{equation}\label{eq:basis_expansion}
    \tilde{\gamma} (x) = \sum_{j=1}^M a_j \phi_j (x),
\end{equation}
enabling us to write
\begin{equation}\label{eq:exact_gkn_scheme}
\sum_{j=1}^M L_{ij} a_j = h_i - r_{i}
\end{equation}
where
\begin{align}
L_{ij} =&  \<\phi_i, \L \phi_j\> \label{eq:L_inner_product}\\
h_i =& \<\phi_i, h \> \label{eq:h_inner_product}\\
r_i =& \<\phi_i, \L r \>. \label{eq:Lr_inner_product}
\end{align}
If the terms in \eqref{eq:L_inner_product}-\eqref{eq:Lr_inner_product} are known, \eqref{eq:exact_gkn_scheme} can be solved for the coefficients $a_j$, and an estimate of $g$ can be constructed as
\begin{equation}\label{eq:expansion_of_soln}
\tilde{g}(x) = r(x) + \sum_{j=1}^M a_j \phi_j (x).
\end{equation}
Since $\tilde{\gamma}$ is zero on $D^c$ and $r$ obeys the inhomogeneous boundary conditions by construction,
\begin{equation}
\tilde{g} = r(x) = b(x)\text{ for } x \text{ in } D^c.
\end{equation}
Consequently, our estimate of $g$ obeys the boundary conditions.

A similar scheme can be constructed for equations with a weighted adjoint $\mathcal{L}^\dagger_p$ by adding one additional step to the procedure.
After homogenizing the boundary conditions, we multiply both sides of \eqref{eq:generic_operator_equation} by $p$.
We then proceed as before, and obtain \eqref{eq:exact_gkn_scheme} with terms defined as 
\begin{align}
L_{ij} =&  \<\phi_i, p \mathcal{L}^\dagger_p \phi_j\> = \<\mathcal{L} \phi_i, p  \phi_j\> \label{eq:L_inner_product_adjoint}\\
h_i =& \<\phi_i, p h \>  \label{eq:h_inner_product_adjoint}\\
r_i =& \<\phi_i, p \mathcal{L}^\dagger_p r \> = \<\mathcal{L} \phi_i, p r \> \label{eq:r_inner_product_adjoint} 
\end{align}
instead of \eqref{eq:L_inner_product}, \eqref{eq:h_inner_product}, and \eqref{eq:Lr_inner_product} respectively.

\subsection{Approximating Inner Products through Monte Carlo}\label{ssec:term_specification}
Solving for $a_j$ in~\eqref{eq:basis_expansion} requires estimates of the other terms in \eqref{eq:exact_gkn_scheme}.
In general, these terms cannot be evaluated directly, due to the complexity of the dynamical operators and the high dimensionality of these integrals.
However, we can estimate these terms using trajectory averages, in the style of the estimates in \eqref{eq:mc_koopman_approximation}.
Let $\rho_{\Delta t}$ be the joint probability measure of $\Xi^{(0)}$ and $\Xi^{(\Delta t)}$, such that for two sets $X$ and $Y$ in state space,
\begin{equation}
\int_{X,Y}\rho_{\Delta t}(dx,dy) =  \P[\Xi^{(0)} \in X,\Xi^{(\Delta t)} \in Y].
\end{equation}
We observe that 
\begin{align}\label{eq:twopoint_gen}
    \<u, \mathcal{L} v\> =& \int u(x)  \frac{\E \left[v \left(\Xi^{(\Delta t)}\right) | \Xi^{(0)} = x\right] - v(x) }{\Delta t}  \mu(dx) \nonumber \\
    =& \int u(x) \frac{v(y) - v(x)}{\Delta t} \rho_{\Delta t}(dx,dy).
\end{align}

We now assume that we have a dataset of the form described in Section~\ref{ssec:msm}, with a lag time of $\Delta t$.
Since each pair $\left(X_n, Y_n\right)$ is a draw from $\rho_{\Delta t} $, \eqref{eq:twopoint_gen} can be approximated using the Monte Carlo estimate 
\begin{equation}\label{eq:general_mc_koopman_approximation}
	\overline{\<u, \mathcal{L} v\>} = \frac{1}{N}\sum_{n=1}^N u(X_n) \frac{v(Y_n) - v(X_n) }{\Delta t}. 
\end{equation}
Similarly, inner products of the form $\<u, v \>$ can be estimated as
\begin{align}\label{eq:general_mc_ip_approximation}
\overline{\<u, v\>} = \frac{1}{N}\sum_{n=1}^N u(X_n) v(X_n).
\end{align}
If the Galerkin scheme arose from an equation with a weighted adjoint, evaluating the expectations in \eqref{eq:L_inner_product_adjoint} and \eqref{eq:r_inner_product_adjoint} may require knowing $p$ a priori.
However, if $p=\pi$, one can construct an estimate of $\pi$ by applying the DGA framework to equation \eqref{eq:expr_com}.

\subsection{Pseudocode}
The DGA procedure can thus be summarized as follows.
\begin{enumerate}
    \item Sample $N$ pairs of configurations $\left(X_n,Y_n \right)$, where $Y_n$ is the configuration resulting from propagating the system forward from $X_n$ for time $\Delta t$.
    \item Construct a set of $M$ basis functions $\phi_i$ obeying the homogeneous boundary conditions and, if needed, the guess function $r$.
    \item Estimate the terms in \eqref{eq:exact_gkn_scheme} using the expressions in \eqref{ssec:term_specification}.
    \item Solve the resulting matrix equations for the coefficients and substitute them into \eqref{eq:expansion_of_soln} to construct an estimate of the function of interest.
\end{enumerate}

Some DGA estimates may require additional manipulation to ensure physical meaning.
For instance, change of measures and expected hitting times are nonnegative, and committors are constrained to be between zero and one.  
These bounds are not guaranteed to hold for estimates constructed through DGA.
To correct this, we apply a simple postprocessing step, and round the DGA estimate to the nearest value in the range. 
Alternatively, constraints on the mean of the solution (e.g., that for $\omega$ below \eqref{eq:defn_chi_actime}) can be applied by subtracting a constant from the estimate.

Finally, many dynamical quantities require evaluation of additional inner products.
%These inner products can also be evaluated using the estimates in Section~\ref{ssec:term_specification}.  
For instance, to estimate the autocorrelation time, $t_f$, one must construct approximations to $\omega$ and $\pi$ and set $\omega$ to have zero mean against $\pi(x) \mu(dx)$.  
One would then evaluate the numerator and denominator of \eqref{eq:gen_expr_actime} using \eqref{eq:general_mc_ip_approximation}.

\bigadd{
To aid the reader in constructing estimates using this framework, we have written a Python package for creating DGA estimates.\cite{pyedgar}  This package also contains code for constructing the basis set we introduce in Section~\ref{sec:dmap_basis_definition}.  As part of the documentation, we have included Jupyter notebooks to aid the reader in reproducing the calculations in this work.
}

\subsection{Connection with Other Schemes}
As we have previously discussed, this formalism is closely related to DOEA.
Rather than considering the solution for a linear system, we could construct a Galerkin scheme for the eigenfunctions of $\mathcal{L}$.
Since $\mathcal{L}$ and $\K_s$ have the same eigenfunctions, in the limit of infinite sampling this would give equivalent results to the scheme in \ref{ssec:edmd}.
\bigadd{
DOEA techniques have also been extended to solve \eqref{eq:expr_com}.\cite{wu2017variationalkoopman}  A similar algorithm for addressing boundary conditions has also been suggested in the context of the data-driven study of partial differential equations and fluid flows.\cite{chen2012variants}
}

Our scheme is also closely related to Markov state modeling.
Let $\phi_i$ be a basis set of indicator functions on disjoint sets $S_i$ covering the state space.  
Under minor restrictions, applying DGA with this basis is equivalent to 
estimating the quantities in \ref{sec:feynman_kac} with an MSM.
We give a more thorough treatment in the Supplementary Information in Section~\ref{ssec:dga_msm_connection}; here we quickly motivate this connection by examining~\eqref{eq:L_inner_product} for this particulary choice of basis.
We note that we can divide both sides of \eqref{eq:exact_gkn_scheme} by $\int \phi(x) \mu\left(dx\right)$ without changing the solution.  For this choice of basis, we would then have
\begin{equation}
\frac{L_{ij}}{\int \phi(x) \mu (dx)} = \frac{1}{\Delta t}\left(P - I\right)_{ij}
\end{equation}
where $P$ is the MSM transition matrix defined in \eqref{eq:msm_tmat_defn} and $I$ is the identity matrix.
Because of this similarity, we refer to a basis set constructed in this manner as an ``MSM'' basis.

\section{Basis Construction using Diffusion Maps}\label{sec:dmap_basis_definition} 
One natural route to improving the accuracy of DGA schemes is to improve the set of basis functions $\phi$, thus reducing the error caused by projecting the operator equation onto the finite-dimensional subspace.  
Various approaches have been used to construct basis sets for describing dynamics in DOEA schemes.\cite{molgedey1994separation, schwantes2013improvements, perez2013identification,boninsegna2015investigating, vitalini2015basis}
\bigadd{However, if $D^c$ is nonempty these functions cannot be used in DGA.
In particular, the linear basis in TICA cannot be used.
% Constructing a basis that obeys the homogeneous boundary conditions for an arbitrary $D$ in high-dimensional CV space is a nontrivial problem.
%Constructing a basis set that obeys arbitrary homogeneous boundary conditions 
Here, we provide a simple method for constructing basis functions with homogeneous boundary conditions based on the technique of diffusion maps.\cite{coifman2006diffusion, berry2016variable}}
%[EHT: THIS IS NEW.]
\begin{figure*}
 \includegraphics[width=\textwidth]{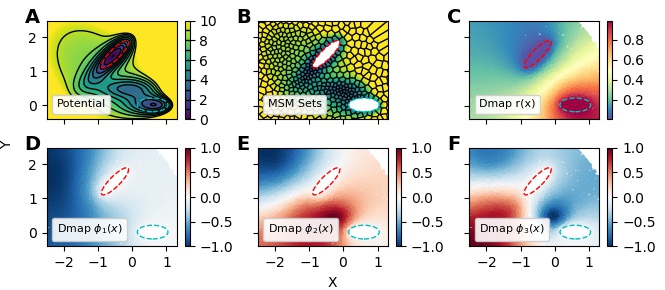}
  \caption{Example basis and guess functions constructed by the diffusion-map basis on the scaled M\"uller-Brown potential.  (A) The potential energy surface.  Black contour lines indicate the potential energy in units of $k_B T$, red and cyan dotted contours indicate the boundaries of states $A$ and $B$ respectively.
  (B) An MSM clustering with 500 sets on the domain; the color scale is the same as in (A).  Each MSM basis function is one inside a cell and zero otherwise.
  Sets inside states A and B are not shown to emphasize the boundary conditions.
  (C) Scatter plot of the guess function for the committor for hitting $B$ before $A$, constructed using~\eqref{eq:dmap_guess_generator}.  
  (D--F) Scatter plots of the first three basis functions constructed according to~\eqref{eq:defn_dmap_basis}. }
  \label{fig:demo_basis_functions}
\end{figure*}

Diffusion maps are a technique shown to have success in finding global descriptions of molecular systems from high-dimensional input data.\cite{ferguson2010systematic, rohrdanz2011determination, zheng2011delineation, ferguson2011nonlinear,long2014nonlinear,  kim2015systematic}
A simple implementation proceeds by constructing the transition matrix
\begin{equation}\label{eq:dmap_transition_mat}
P^{\rm DMAP}_{mn} = \frac{K_\eps(x_m, x_n)}{\sum_n K_\eps(x_m, x_n)},
\end{equation}
where $K_\eps$ is a kernel function that decays exponentially with the distance between datapoints $x_m$ and $x_n$ at a rate set by  $\eps$.
Multiple choices of $K_\eps$ exist; we give the algorithm used to construct the kernel in the Supplementary Information in Section~\ref{ssec:dmap_details}.
The eigenvectors of $P^{\rm DMAP}$ with $M$ highest positive eigenvalues were historically used to define a new coordinate system for dimensionality reduction.
They can also be used as a basis set for DOEA and similar analyses.\cite{boninsegna2015investigating, berry2015nonparametric, giannakis2017data}
\bigadd{
Here we extend this line of research, showing that diffusion maps can also be used to construct basis functions that obey homogeneous boundary conditions on arbitrary sets as required for use in DGA.
We note that the diffusion process represented by $P^{\rm DMAP}$ is not intended as an approximation of the dynamics, but rather as a tool for building the basis functions $\phi_i$.  In particular, while the $P^{\rm DMAP}$ matrix is typically reversible, this imposes no reversible constraint in the DGA scheme using the basis derived from $P^{\rm DMAP}$.
}

To construct a basis set that obeys nontrivial boundary conditions, we first take the submatrix of $P^{\rm DMAP}$ such that $x_m, x_n \in D$.
We then calculate the eigenvectors $\varphi_i$ of this submatrix that have the $M$ highest positive eigenvalues, and take as our basis
\begin{equation}\label{eq:defn_dmap_basis}
\phi_i(x) = \begin{cases}
    \varphi_i(x) \text{ for } x \text{ in } D, \\
0 \text{ otherwise}.
\end{cases}
\end{equation}
In addition to allowing us to define a basis set, $P^{\rm DMAP}$ gives a natural way of constructing guess functions that obey the boundary conditions.  Since \eqref{eq:dmap_transition_mat} is a transition matrix, it corresponds to a discrete Markov chain on the data.
Therefore, we can construct guesses by solving analogs to \eqref{eq:generic_inhomogeneous_eqn} using the dynamics specified by the diffusion map.
For equations using the generator, we solve the problem
\begin{align}\label{eq:dmap_guess_generator}
\sum_{n}(P^{\rm DMAP}-I)_{mn}  r_n  &= h(x_m) \text{ for } m\text{ in } D \\
r_m &= b(x_m) \text{ for } m \text{ in } D^c
\end{align}
where $I$ is the identity matrix.   Here the sum runs over all datapoints, not just those in $D$.
The resulting estimate obeys the boundary conditions for all datapoints sampled in $D^c$.

Equation \eqref{eq:dmap_guess_generator} can also be used to construct guesses for equations using weighted adjoints.
In principle, one could replace $P^{\rm DMAP}$ with its weighted adjoint against $p$ and solve the corresponding equation.  
However, $r_n$ still obeys the boundary conditions irrespective of the weighted adjoint used.
We therefore take the adjoint of $P^{\rm DMAP}$ with respect to its stationary measure.  Since the Markov chain associated with the diffusion map is reversible,\cite{coifman2006diffusion} $P^{\rm DMAP}$ is self-adjoint with respect to its stationary measure and we again solve \eqref{eq:dmap_guess_generator}.
We discuss how to perform out-of-sample extension on the basis and the guess functions in the Supplementary Information in Section~\ref{ssec:dmap_details}.
%\begin{figure*}
% \includegraphics[width=\textwidth]{sample_bfxns_r_0_N_10000_d_2}
%  \caption{Example basis and guess functions constructed by the diffusion-map basis on the scaled M\"uller-Brown potential.  (A) The potential energy surface.  Black contour lines indicate the potential energy in units of $k_B T$, red and cyan dotted contours indicate the boundaries of states $A$ and $B$ respectively.
%  (B) An MSM clustering with 500 sets on the domain; the color scale is the same as in (A).  Each MSM basis function is one inside a cell and zero otherwise.
%  Sets inside states A and B are not shown to emphasize the boundary conditions.
%  (C) Scatter plot of the guess function for the committor for hitting $B$ before $A$, constructed using~\eqref{eq:dmap_guess_generator}.  
%  (D--F) Scatter plots of the first three basis functions constructed according to~\eqref{eq:defn_dmap_basis}. }
%  \label{fig:demo_basis_functions}
%\end{figure*}

To help the reader visualize a diffusion-map basis, we analyze a collection of datapoints sampled from the M\"uller-Brown potential,\cite{muller1979location} scaled by 20 so that the barrier height is about 7 energy units; we set $k_B T=1$.
This potential is sampled using a Brownian particle with isotropic diffusion coefficient of $0.1$ using the BAOAB integrator for overdamped dynamics with a time step of 0.01 time units.\cite{leimkuhler2012rational}
Trajectories are initialized out of the stationary measure by uniformly picking 10000 starting locations on the interval $x\in\left(-2.5, 1.5\right), y\in\left(-2.5, 1.5\right)$.  Initial points with potential energies larger than $100$ are rejected and resampled to avoid numerical artifacts.  Each trajectory is then constructed by simulating the dynamics for 500 steps, saving the position every 100 steps.

We then define two states $A$ and $B$ (red and cyan dashed contours in Figure~\ref{fig:demo_basis_functions}, respectively) and
construct the basis and guess functions required for the committor.  The results, plotted in Figure~\ref{fig:demo_basis_functions}, demonstrate that the diffusion-map basis functions are smoothly varying with global support.
% \bigadd{
% One disadvantage of this basis construction procedure is that the same basis cannot be used for problems with different boundary conditions.  For instance, if one wishes to construct an estimate of both $m_A$ and $q_+$, one would have to construct two separate sets of basis functions from the diffusion map matrix.  This is in contrast to the MSM basis, where any of the problems in Section~\ref{sec:feynman_kac} can be solved as long as $D$ is the union of a collection of Markov states.  
% }

\subsection{Basis Set Performance in High-Dimensional CV spaces.}\label{ssec:basis_set_d_comparison}
\begin{figure*}
  \includegraphics[width=\textwidth]{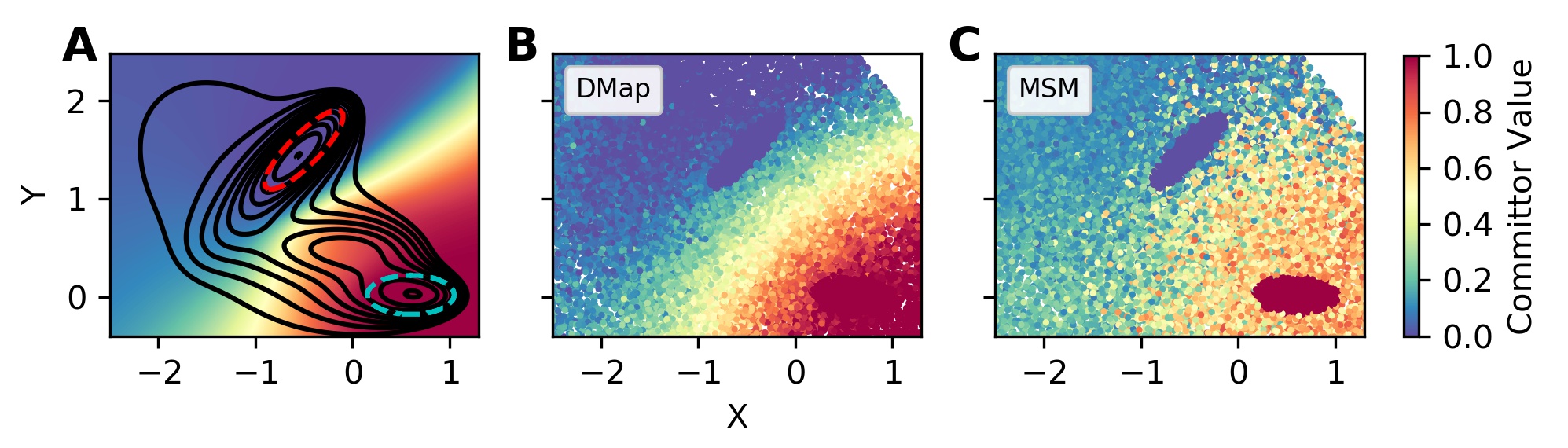}
  \caption{Example forward committors calculated using the diffusion-map and MSM bases on a high-dimensional toy problem.  The system is the same as in Figure~\ref{fig:demo_basis_functions}, with 18 additional nuisance dimensions.  (A) Forward committor function calculated using an accurate grid-based scheme.  The black lines indicate the contours of free energy in the $x$ and $y$ coordinates, and the red and cyan dashed contours indicate the two states.  Every subsequent dimension has a harmonic potential with force constant of 2.  (B--C) Estimated forward committor constructed using the diffusion map and MSM bases, respectively.
  }
  \label{fig:2D_MB_potential}
\end{figure*}

We now test the effect of dimension on the performance of the basis set by attempting to calculate the forward committor and total reactive flux for a series of toy systems based on the model above.
%To so, weWe first test the behavior of the diffusion-map basis set in high-dimensional CV spaces by attempting to calculate the forward committor function and the total reactive flux and comparing with a specific choice of MSM basis.
%Since the forward committor cannot be directly estimated for complex systems, we study the behavior on the toy potential discussed above.
To be able to vary the dimensionality of the system, we add up to $18$ harmonic ``nuisance'' degrees of freedom.  Specifically,
\begin{equation}
U\left(x,y,z_3,...,z_{d}\right)  = U_{\rm MB}(x,y) + \sum_{l=3}^{d} z_l^2,
\end{equation}
where $U_{\rm MB}$ is the scaled M\"uller-Brown potential discussed above.
We compare our results with references computed by a grid-based scheme described in the Supplementary Information.
Our reference for the committor is plotted in Figure~\ref{fig:2D_MB_potential}A.
We initialize the $x$ and $y$ dimensions as above; the initial values of the nuisance coordinates were drawn from their marginal distributions at equilibrium.
We then sampled the system using the same procedure as before.

\bigadd{
Throughout this section and all subsequent numerical comparisons, we compare the diffusion-map basis with a basis of indicator functions.  
Since, with minor restrictions, using a basis of indicator functions is equivalent to calculating the same dynamical quantities using a MSM, we estimate committors, mean first-passage times, and stationary distributions by constructing a MSM in PyEMMA and using established formulas.\cite{metzner2009transition, noe2014introduction, keller2019markov}
In general, it is not our intention to compare an optimal diffusion-map basis to an optimal MSM basis.  Multiple diffusion-map and clustering schemes exist and performing an exhaustive comparison would require comparison over multiple methods and hyperparameters.  We leave such a comparison for future work, and only seek to present reasonable examples of both schemes.
}

\bigadd{
%\subsubsection{MSM Construction Procedure}
MSM clusters are constructed using $k$-means, as implemented in PyEMMA.\cite{scherer2015pyemma}
While MSMs are generally constructed by clustering points globally, this does not guarantee that a given clustering satisfies a specific set of boundary conditions.
Consequently, we modify the set definition procedure slightly.
}

\bigadd{
We first construct $M$ clusters on the domain $D$, and then cluster $D^c$ separately. 
The number of states inside $D^c$ is chosen so that states inside $D^c$ have approximately the same number of samples on average as states in the domain. 
For the current calculation, this corresponded to approximately one state inside set $A$ or $B$ for every five states outside the domain; we round to a ratio of $1/5$ for numerical simplicity.
We note that clustering on the interior of $D^c$ does not affect calculated committors or mean first-passage times.}
We use 500 basis functions for both the MSM and diffusion-map basis sets.
\bigadd{ 
We give plots supporting this choice in Section~\ref{ssec:mb_basis_size} of the Supplementary Information.
}

\bigadd{
In modern Markov state modeling, one commonly constructs the transition matrix only over a well-connected subset of states named the active set.\cite{prinz2011markov, beauchamp2011msmbuilder2}
We have follow this practice, and exclude points outside the active set from any error analyses of the resulting MSMs.
We believe this gives the MSM basis an advantage over the diffusion-map basis in our comparisons, as we are explicitly ignoring points where it fails to provide an answer and would presumably give poor results.}

\bigadd{
It is also common to ensure that the resulting matrix obeys detailed balance through a maximum likelihood procedure.\cite{bowman2009progress, prinz2011markov}
We choose not to do this because we do not wish to assume reversibility in our formalism.
%, and we wish to study the effect of basis choice purely without additional assumptions.
Moreover, our calculations have also shown that enforcing reversibility can introduce a statistical bias that dominates the error in any estimates.  We give numerical examples of this phenomenon in Section~\ref{ssec:detailed_balance_results} of the Supplementary Information.  
}

\begin{figure}
  \includegraphics{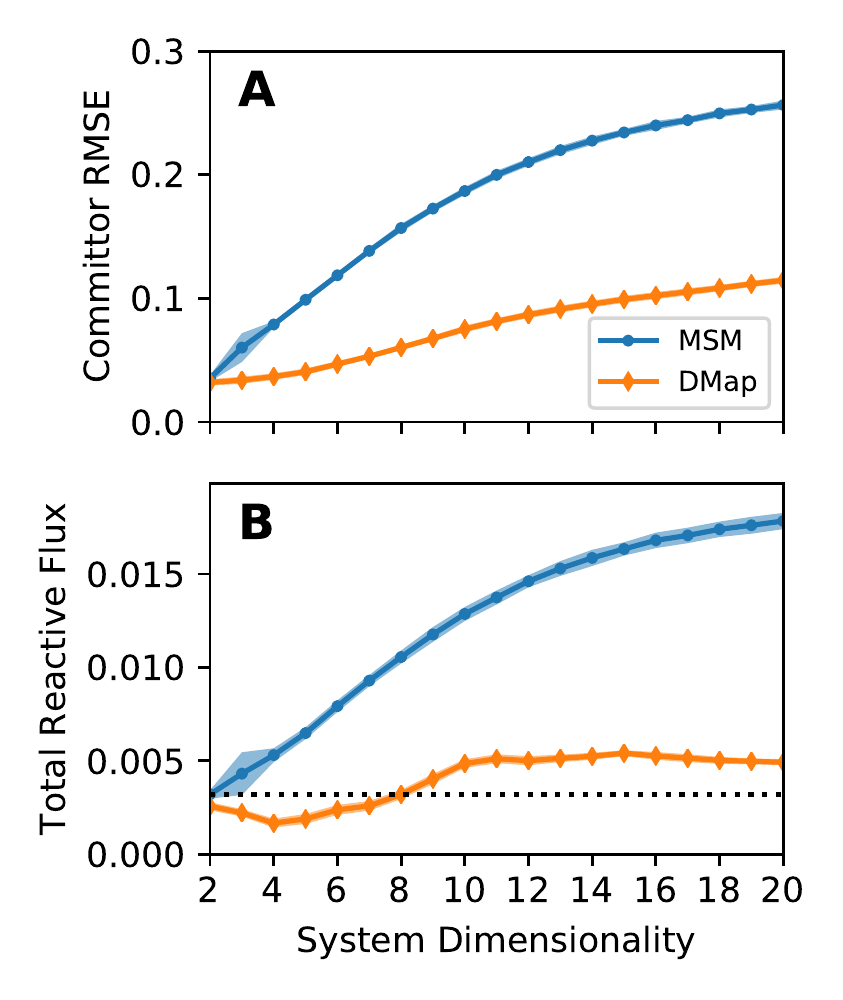}
  \caption{Comparison of basis performance the dimensionality of the toy system increases.  (A) The average error in the forward committor between states $B$ and $A$ in Figure \ref{fig:2D_MB_potential} for both the MSM and the diffusion-map basis functions, as a function of the number of nuisance degrees of freedom.  (B) Estimated reactive using both MSM and the diffusion-map basis functions as function of the same.  In both plots shading indicates the standard deviation over 30 datasets.  The dotted line in (B) is the reactive flux as calculated by an accurate reference scheme.}
  \label{fig:committor_error_w_dimension}
\end{figure}

In Figure~\ref{fig:committor_error_w_dimension}A, we plot root-mean-square error (RMSE) between the estimated and reference forward committors as a function of the number of nuisance degrees of freedom.
While for low-dimensional systems the MSM and the diffusion-map basis give comparable results, as we increase the dimensionality, the MSM gives increasingly worse answers.
To aid in understanding these results, we plot example forward committor estimates for the 20-dimensional system in Figure~\ref{fig:2D_MB_potential}.  We see that the diffusion-map basis manages to capture the general trends in the reference in Figure \ref{fig:2D_MB_potential}A.  In contrast, the MSM basis gives considerably noisier results.

We also estimate the total reactive flux across the same dataset, setting $C$ and $C^c$ in \eqref{eq:total_rxn_current_traj_level} to be the sets on either side of the calculated isocommittor (Figure~\ref{fig:committor_error_w_dimension}B).
The large errors that we observe in the reactive flux occur due to the nature of the dataset.
If data were collected from a long equilibrium trajectory, it would not be necessary to estimate $\pi(x)$ separately, and we could set $\pi(x)=1$.  In that case, provided the number of MSM states was sufficient, the MSM reactive flux reverts to direct estimation of the number of reactive trajectories per unit time.
This would give an accurate reactive flux regardless of the quality of the estimated forwards or backwards committors.

%\begin{figure*}
%  \includegraphics[width=\textwidth]{sample_committor_comparison}
%  \caption{Example forward committors calculated using the diffusion-map and MSM bases on a high-dimensional toy problem.  The system is the same as in Figure~\ref{fig:demo_basis_functions}, with 18 additional nuisance dimensions.  (A) Forward committor function calculated using an accurate grid-based scheme.  The black lines indicate the contours of free energy in the $x$ and $y$ coordinates, and the red and cyan dashed contours indicate the two states.  Every subsequent dimension has a harmonic potential with force constant of 2.  (B--C) Estimated forward committor constructed using the diffusion map and MSM bases, respectively.
%  }
%  \label{fig:2D_MB_potential}
%\end{figure*}

\section{Addressing Projection Error Through Delay Embedding}\label{sec:delay_embedding}

Our results suggest that improving basis set choice can yield DGA schemes with better accuracy in higher-dimensional CV spaces.  However, even large CV spaces are considerably lower-dimensional than the system's full state space.
Consequently, they may still omit key degrees of freedom needed to describe the long-time dynamics.
In both MSMs and DOEA, this projection error is often addressed by increasing the lag time of the transition operator.\cite{prinz2011markov, suarez2016accurate, nuske2016variational, mardt2017vampnets}
This is based on the assumption that degrees of freedom omitted from the CV space equilibrate quickly and contribute little to the long-time dynamics.
However, there is no guarantee that this assumption holds for any specific choice of CVs.
\bigadd{This is reflected by existing bounds on the approximation error for DOEA.  For instance, the relative error in the estimate of the dominant eigenvalue is bounded above by the projection error of the basis onto the dominant eigenfunction; this bound does not decrease with the lag time.\cite{djurdjevac2012estimating} }

Moreover, \bigadd{whereas changing the lag time does not affect the eigenfunctions in~\eqref{eq:koopman_eigenproblem},}
the equations in Section~\ref{sec:feynman_kac} hold only for a lag time of $\Delta t$.
Using a longer time is effectively making the approximation 
\begin{equation}\label{eq:approx_generator}
\mathcal{L} f(x) \approx \frac{\K_s f(x) - f(x)}{s}.
\end{equation}
%This approximation leaves the eigenfunctions unaffected, as the eigenfunctions for the transition operator are independent of $s$.
This causes a systematic bias in the estimates of the dynamical quantities discussed in Section~\ref{sec:feynman_kac}.
While for small lag times this bias is likely negligible, it may become large as the lag time increases.
For instance, estimates of the mean first-passage time grow linearly with $s$ as the lag time goes to infinity.\cite{suarez2016accurate}

Here, we propose an alternative strategy for dealing with projection error.  Rather than looking at larger time lags, we use past configurations in CV space to account for contributions from the removed degrees of freedom.
This idea is central to the Mori-Zwanzig formalism.\cite{zwanzig2001nonequilibrium}
Here, we use \textit{delay embedding} to include history information. 
Let $\zeta^{(t)}$ be the projection of $\Xi^{(t)}$ at time $t$.  We define the delay-embedded process with $d$ delays as
\begin{equation}\label{eq:defn_delay_embedding}
\theta^{(t)} = \left(\zeta^{(t)}, \zeta^{(t-\Delta t)}, \zeta^{(t-2\Delta t)},\ldots, \zeta^{(t-d\Delta t)}\right).
\end{equation}
Delay embedding has a long history in the study of deterministic, finite-dimensional systems, where it has been shown that delay embedding can recapture attractor manifolds up to diffeomorphism.\cite{takens1981detecting, aeyels1981generic} 
\bigadd{Weaker mathematical results have been extended to stochastic systems,\cite{muldoon1998delay, stark1997takens} although these are not sufficient to guarantee its effectiveness in all cases.}

\bigadd{Delay embedding has been used previously with dimensionality reduction on both experimental\cite{berry2013time} and simulated chemical systems,\cite{wang2016nonlinear, wang2018recovery}} and has also been used in applications of DOEA in geophysics.\cite{giannakis2017data} In references~\onlinecite{giannakis2017data, fung2016dynamics} it was argued that delay embedding can improve statistical accuracy for noise-corrupted and time-uncertain data.
Other methods of augmenting the dynamical process with history information have been used in the construction of MSMs. 
\bigadd{
	In reference~\onlinecite{suarez2014simultaneous}, each trajectory was augmented with a labeling variable indicating its origin state.
	%This approach can be combined with delay-embedding, although we leave that to future work.
	In reference~\onlinecite{suarez2016accurate}, it was suggested to write transition probabilities as a function of both the current and the preceding MSM state.
	This corresponds to a specific choice of basis on a delay embedded process.
}

\begin{figure*}
  \includegraphics[width=\textwidth]{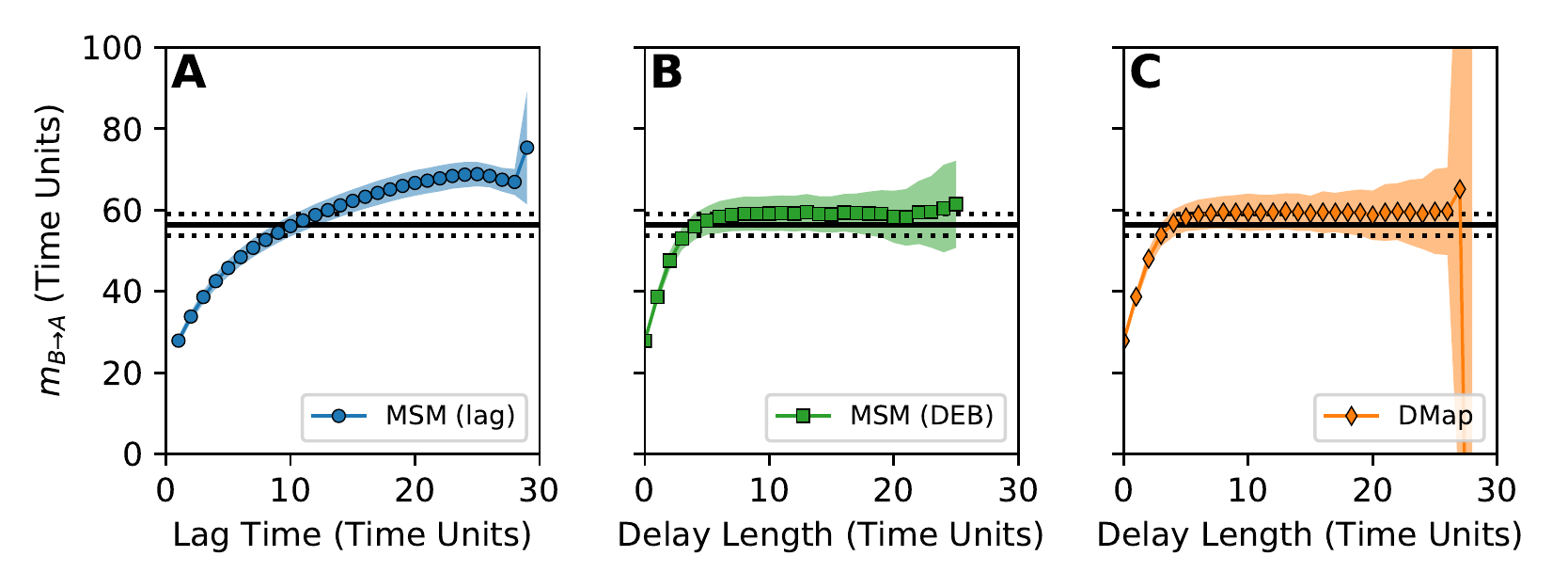}
  \caption{
		Comparison of methods for dealing with the projection error in an incomplete CV space.
		In all subplots we estimate the mean first-passage time from state $B=\left\{y < 0.15\right\}$ to state $A=\left\{y > 1.15\right\}$ using a DGA scheme on only the $y$ coordinate of the M\"uller-Brown potential.
		(A) Estimate constructed using an MSM basis with increasing lag time in \eqref{eq:approx_generator}, as a function of the lag time.
		(B) Estimate constructed using an MSM basis, but applying delay embedding rather than increasing the lag time, as a function of the delay length.
		(C) Estimate constructed using the diffusion-map basis with delay embedding, as a function of the delay length.
		In each plot, the symbols show the mean over 30 identically constructed trajectories, and the shading indicates the standard deviation across trajectories.
		The black solid line is an estimate of the mean first-passage time calculated using the reference scheme in the Supplementary Information, and the dashed error bars represent the standard deviation of the mean first-passage time over state $B$.
	    }
  \label{fig:delay_vs_lag_1d}
\end{figure*}
Here we show that delay embedding can be used to improve dynamical estimates in DGA.
To apply DGA to the delay-embedded process, we must extend the functions $h$ and $b$ in \eqref{eq:generic_inhomogeneous_eqn} and \eqref{eq:generic_inhomogeneous_eqn} to the delay-embedded space.
We do this by using the value of the function on the central timepoint,
\begin{equation}
f\left(\theta^{(t)}\right) = f\left(\zeta^{(t-\lfloor d/2 \rfloor \Delta t}\right)
\end{equation}
where $\lfloor \ldots \rfloor$ denotes rounding down to the nearest integer.
The states $D$ and $D^c$ in the delay-embedded space are extended similarly.
One can easily show that this preserves dynamical quantities such as mean first-passage times and committors.
The basis set is then constructed directly on $\theta$, and the DGA formalism is applied as before.

We test the effect of delay embedding in the presence of projection error by constructing DGA schemes on the same system as in Section~\ref{sec:dmap_basis_definition} and taking as our CV space only the $y$-coordinate.
For this study we revise our dataset to include 2000 trajectories, each sampled for 3000 time steps.  
While using longer trajectories changes the density such that it is closer to equilibrium, it allows us to test longer lag times and delay lengths.
To ensure that our states are well-defined in this new CV space, we redefine state $A$ to be the set $\{y > 1.15\}$, and state $B$ to be the set $\{y < 0.15\}$.
We then estimate the mean first-passage time into state $A$, conditioned on starting in state $B$ at equilibrium.
\begin{equation*}
m_{B \to A} = \frac{\int \1_{B}(x) m_{A} (x) \pi(x) \mu (dx)}{\int \1_{B}(x) \pi(x) \mu(dx)}.
\end{equation*}
We construct estimates using an MSM basis with varying lag time, an MSM basis with delay embedding, and a diffusion-map basis with delay embedding.

In Figure~\ref{fig:delay_vs_lag_1d}, we plot the average mean first-passage time as a function of the lag time and the trajectory length used in the delay embedding.  
We compare the resulting estimates with an estimate of the mean first-passage time constructed using our grid-based scheme.  In addition, an implied timescale analysis for the two MSM schemes is given in the Supplementary Information in Section~\ref{ssec:deb_supp}.

The mean first-passage time estimated from the MSM basis with the lag time steadily increases as the lag time becomes longer (Figure \ref{fig:delay_vs_lag_1d}A), as predicted in reference~\onlinecite{suarez2016accurate}.
In contrast, the estimates obtained from delay embedding both converge as the delay length increases, albeit to a value slightly larger than the reference.
We believe this small error is because we treat the dynamics as having a discrete time step, while the reference curve approximates the mean first-passage time for a continuous-time Brownian dynamics.  \bigadd{In particular, The latter includes events in which the system enters and exits the target state within the duration of a discrete-time time step, but such events are missing from the discrete-time dynamics.}

In all three schemes, we see anomalous behavior as the length of the lag time or delay length increases.
This is due to an increase in statistical error when the delay length or lag time becomes close to the length of the trajectory.
If each trajectory has $N$ datapoints, performing a delay-embedding with $d$ delays means that each trajectory only gives $N-d$ samples. 
When $N$ and $d$ are of the same order of magnitude, this leads to increased statistical error in the estimates in \ref{ssec:term_specification}, to the point of making the resulting linear algebra problem ill-posed.
The diffusion-map basis fluctuates to unreasonable values at long delay lengths, and the MSM basis fails completely, truncating the curve in Figures \ref{fig:delay_vs_lag_1d}B and C.
Similarly, the lagged MSM has an anomalous downturn in the average mean first-passage time near 26 time units.
\bigadd{We give additional plots supporting this theory in the Supplementary Information in Section~\ref{ssec:deb_supp}.
}

\bigadd{
Finally, we observe that the delay length required for the estimate to converge is substantially smaller than the mean first-passage time.  This suggests that delay embedding can be effectively used on short trajectories to get estimates of long-time quantities.
}

\section{Application to the Fip35 WW Domain}\label{sec:fip35}
To further assess our methods, we now apply them to molecular dynamics data and seek to evaluate committors and mean first-passage times.
In contrast to the simulations above, we do not have accurate reference values and cannot directly calculate the error in our estimates.
Instead, we observe that both the mean first-passage time and forward committor are conditional expectations, and obey the following relations \cite{durrett2010probability}
\begin{align*}
m_A(x) =&  \argmin_{f(x)} \E\left[\left(\tau_A - f(x)\right)^2\right] \\
q_+(x) =& \argmin_{f(x)} \E\left[\left(\mathbf{1}_{\tau_B < \tau_A} - f(x)\right)^2\right].
\end{align*}
\bigadd{
% We can therefore estimate these expectations using trajectory averages to give a measure of the error in our estimate.
This suggests a scheme for assessing the quality of our estimates.  
If we have access to long trajectories, each point in the trajectory has an associated sample of $\tau_A$ and $\mathbf{1}_{\tau_B < \tau_A}$.
We define the two empirical cost functions
\begin{align}
\text{COST}_{m_A} =&  \frac{1}{N}\sum_{n=1}^{N}(\bar{m}_A(x_n) -\tau_{{A},n})^2 \label{eq:cost_mfpt} \\
\text{COST}_{q_+} =&  \frac{1}{N}\sum_{n=1}^{N}(\bar{q}_+(x_n) - \mathbf{1}_{{\tau_B < \tau_A},n})^2. \label{eq:cost_comm}
\end{align}
Here $x_n$ is a collection of samples from a long trajectory, $\tau_{{A},n}$ is the time from $x_n$ to $A$, and $\mathbf{1}_{{\tau_B
 < \tau_{{A},n}}}$ is one if the sampled trajectory next reaches $B$ and zero if it next reaches $A$.  
The numerical estimates of the mean first-passage time and committor are written as $\bar{m}_A$ and $\bar{q}$, respectively.
In the limit of $N\to\infty$, the true mean first-passage time and committor would minimize \eqref{eq:cost_mfpt} and \eqref{eq:cost_comm}.
We consequently expect lower values of our cost functions to indicate improved estimates. 
For a perfect estimate, however, these cost functions would not go to zero.  Rather, in the limit of infinite sampling, \eqref{eq:cost_mfpt} and \eqref{eq:cost_comm} would converge to the variances of $\tau_A$ and $\mathbf{1}_{\tau_B < \tau_A}$.
For the procedure to be valid, it is important that the cost estimates are not constructed using the same dataset used to build the dynamical estimates.  This avoids spurious correlations between the dynamical estimate and the estimated cost.
}

We applied our methods to the Fip35 WW domain trajectories described by D.E. Shaw Research in references~\onlinecite{shaw2010atomic, piana2011computational}.
The dataset consists of six trajectories, each of length 100000 ns with frames output every 0.2 ns.
\bigadd{
Each trajectory has multiple folding and unfolding events, allowing us to evaluate the empirical cost functions.
}
To avoid correlations between the DGA estimate and the calculated cost, we perform a test/train split and divide the data into two halves.  We choose three trajectories to construct our estimate, and use the other three to approximate the expectations in \eqref{eq:cost_mfpt} and \eqref{eq:cost_comm}.  
Repeating this for each possible choice of trajectories creates a total of 20 unique test/train splits.

To reduce the memory requirements in constructing the diffusion map kernel matrix, we subsampled the trajectories, keeping every 100th frame.  This allowed us to test the scheme over a broad range of hyperparameters.  We expect that in practical applications a finer time resolution would be used, and any additional computational expense could be offset by using landmark diffusion maps.\cite{long2017landmark}

To define the folded and unfolded states, we follow reference \onlinecite{berezovska2013consensus} and calculate $r_{\beta 1}$ and $r_{\beta 1}$, the minimum root-mean-square-displacement for each of the two $\beta$ hairpins, defined as amino acids 7-23 and 18-29, respectively.\cite{berezovska2013consensus}
We define the folded configuration as having both $r_{\beta 1}<0.2$ nm and $r_{\beta 2}<0.13$ nm and the unfolded configuration as having $0.4$ nm $< r_{\beta 1}<1.0$ nm and $0.3$ nm $< r_{\beta 2}<0.75$ nm.
For convenience, we refer to these states as $A$ and $B$ throughout this section.
We then attempt to estimate the forward committor between the two states and the mean first-passage time into $A$ using the same methods as in Section \ref{sec:delay_embedding}.

\bigadd{
We take as our CVs the pairwise distances between every other $\alpha$-carbon, leading to a 153-dimensional space.  
In previous studies, dimensionality-reduction schemes such as TICA have been applied prior to MSM construction.  We choose not to do this, as we are interested in the performance of the schemes in large CV spaces.  This also helps control the number of hyperparameters and algorithm design choices, although we think the interaction between dimensionality-reduction schemes and families of basis sets merits future investigation.
}

\begin{figure*}
  \includegraphics[width=\textwidth]{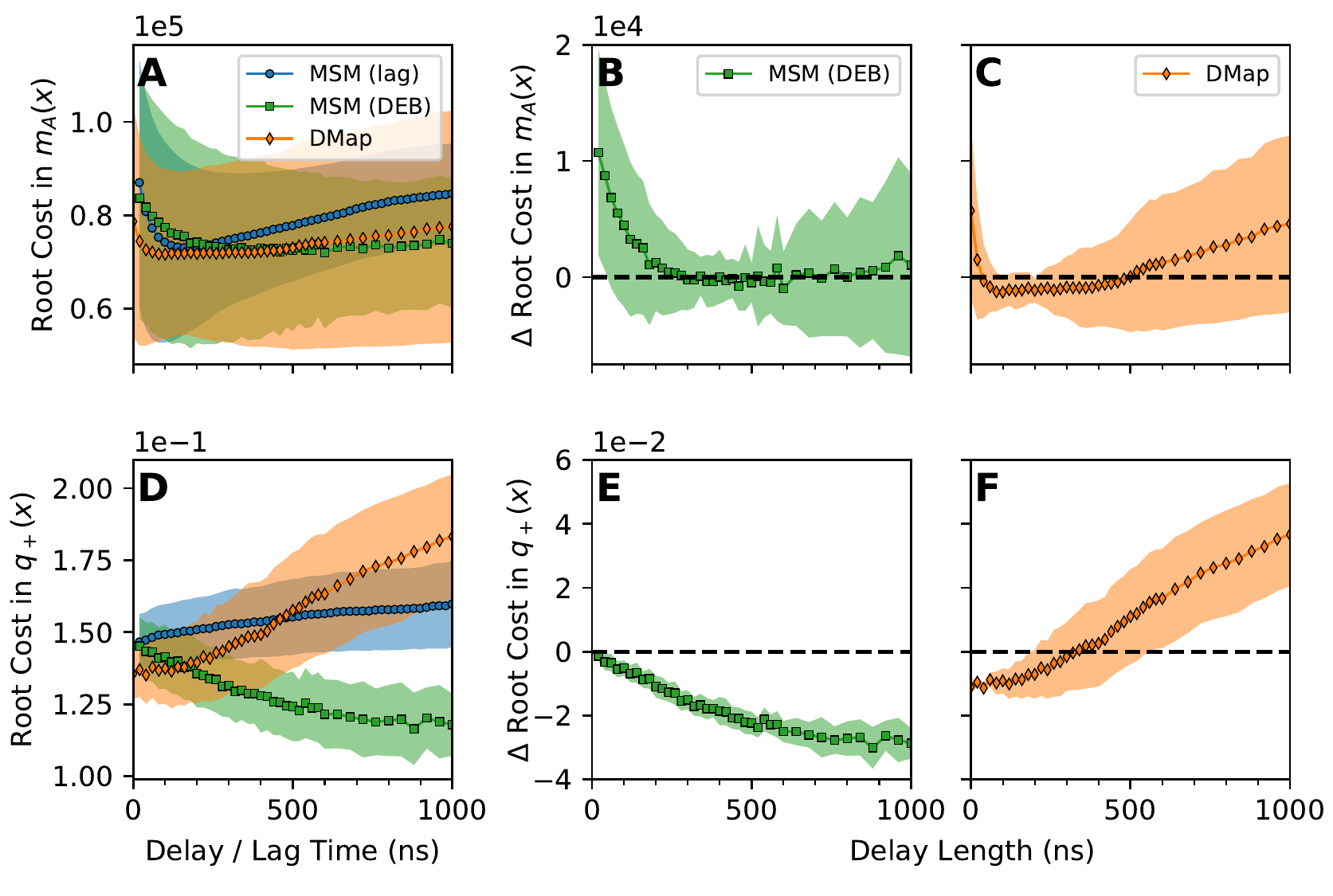}
  \caption{Results from a DGA calculation on a dataset of six long folding and unfolding trajectories of the Fip35 WW domain.  (A,D) The root cost in the mean first-passage time and forward committor respectively, calculated using an MSM basis with increasing lag time, an MSM basis with delay embedding, and diffusion map basis with delay embedding, averaged over all test/train splits.  
  (B,C,E,F) Difference in root cost relative to the best parameter choice for the estimate constructed using the MSM basis with increasing lag time. Negative values are better. (B) Difference in cost for the mean first-passage time estimated with an MSM basis with delay embedding.  (C) The same as in (B) but with the diffusion map basis instead.  (E) Difference in cost for the committor estimated with an MSM basis with delay embedding. (F) The same as in (E) but with the diffusion map basis instead.
%   The left column gives results for the MSM basis with delay embedding, and the right for the diffusion-map basis.  The top two figures gives the change in cost for the mean first-passage time, and the bottom two for the forwards committor.
  In all plots the symbols are the average over test/train splits, and the shading indicates the standard deviation across test/train splits.}
  \label{fig:fip35_results}
\end{figure*}
Our results are given in Figure \ref{fig:fip35_results}.
In panels A and B, we give the mean value of the cost for the mean first-passage time and forward committor over all test/train splits, as calculated using 200 basis functions for each algorithm.
\bigadd{
The number of basis functions was chosen to give the best result for the MSM scheme with increasing lag over any lag time, although we see only very minor differences in behavior for larger basis sets.
}
The large standard deviations primarily reflect variation in the cost across different test/train splits, rather than any difference between the methods.
This suggests the presence of large numerical noise in our results.   

To get a more accurate comparison, we instead look at the expected improvement in cost between schemes for a given test/train split.
To quantify whether an improvement occurs, we first determine the best parameter choice for the MSM basis with increasing lag.
We estimate the cost for the MSM basis with delay embedding and for the diffusion-map basis, and calculate the difference in cost versus the lagged MSM scheme for each test/train split.
We then average and calculate the standard deviation over pairs, and plot the results in figures \ref{fig:fip35_results}C through \ref{fig:fip35_results}F.
As the difference is calculated against the best parameter choice for the lagged MSM scheme, they are intrinsically conservative: in practice, one should not expect to have the optimal lagged MSM parameters.

In our numerical experiments, we see that the diffusion map seems to give the best results for relatively short delay lengths.  
However, the diffusion-map basis performs progressively worse as the delay length increases.  
The mechanism causing this loss in accuracy requires further analysis.
This tentatively suggests the use of the diffusion-map basis for datasets consisting of very short trajectories, where using long delays may be infeasible.
In contrast, our results with the delay-embedded MSM basis are more ambiguous.  For the mean first-passage time, we do not see significant improvement over the results from the lagged MSM results.
We do see noticeable improvement in the estimated forward committor probability as the delay length increases.
However, we observe that the delay lengths required to improve upon the diffusion map result are comparable in magnitude to the average time required for the trajectory to reach either the $A$ or $B$ states.
Indeed, we only see an improvement over the diffusion-map result at a delay length of 180 ns, and we observe that the longest the trajectory spends outside of both state $A$ or state $B$ is 223 ns.  
This negates any advantage of using datasets of short trajectories.

\bigadd{
	Caution is warranted in interpreting these results.  We see large variances between different test/train splits, suggesting that despite having 300 $\mu s$ of data in each training dataset, we are still in a relatively data-poor regime.  Similarly, we cannot make an authoritative recommendation for any particular scheme for calculating dynamical quantities without further research.
	Such a study would not only require more simulation data, but also a comparison of multiple clustering and diffusion map schemes across several hyperparameters and their interaction with various dimensionality-reduction schemes.  We leave this task for future work.  
	However, our initial results are promising, suggesting that further development of DGA schemes and basis sets is warranted.
}

\section{Conclusions}\label{sec:conclusion}
In this paper, we introduce a new framework for estimating dynamical statistics from trajectory data.
We express the quantity of interest as the solution to an operator equation using the generator or one of its adjoints.  We then apply a Galerkin approximation, projecting the unknown function onto a finite-dimensional basis set.
This allows us to approximate the problem as a system of linear equations, whose matrix elements we approximate using Monte Carlo integration on dynamical data. 
We refer to this framework as \textit{Dynamical Galerkin Approximation} (DGA).
These estimates can be constructed using collections of short trajectories initialized from relatively arbitrary distributions.
Using a basis set of indicator functions on nonoverlapping sets recovers MSM estimates of dynamical quantities.
Our work is closely related to existing work on estimating the eigenfunctions of dynamical operators in a data-driven manner.

To demonstrate the utility of alternative basis sets, we introduce a new method for constructing basis functions based on diffusion maps.
Results on a toy system shows that this basis has the potential to give improved results in high-dimensional CV spaces.
We also combine our formalism with delay-embedding, a technique for recovering degrees of freedom omitted in constructing a CV space.
Applying it to an incomplete, one-dimensional projection of our test system, we see that delay embedding can improve on the current practice of increasing the lag time of the dynamical operator.
%We apply our technique to a one-dimensional projection of our test system, where we construct our schemes on a collective variable known to be incomplete.  
%In contrast to the common practice of increasing the lag time in the approximation of the dynamical operator, delay embedding causes our estimate to converge to the correct result.

%To study our methods in a 
We then applied the method to long folding trajectories of the Fip35 WW domain to study the performance of the schemes in a large CV space on a nontrivial biomolecule.
Our results suggest that the diffusion-map basis gives the best performance for short delay times, giving results that  are as good or better than the best time-lagged MSM parameter choice.
Moreover, our results suggest that combining the MSM basis with delay embedding gives promising results, particularly for long delay lengths.
However, long delay lengths are required to see an improvement over the diffusion-map basis, potentially negating any computational advantage in using short trajectories to estimate committors and mean first-passage times.

\bigadd{
We believe our work raises new theoretical and algorithmic questions.  Most immediately, we hope our preliminary numerical results motivate the need for new approaches to building basis sets and guess functions obeying the necessary boundary conditions. Further theoretical work is also required to assess the validity of using delay embedding in our schemes.
Finally, we believe it is worth searching for connections between our work and VAC and VAMP theory.\cite{noe2013variational, nüske2014variational,nuske2016variational, wu2017vamp, mardt2017vampnets} 
%For instance, one might ask how variational scores can be best extended to evaluate the quality of DGA basis sets.
%A more complex question is whether a variational extension of our formalism exists.  The variational formulation of DOEA schemes allowed treatment by more advanced basis sets, and we expect a variational approach to the equations in Section~\ref{sec:feynman_kac} to give comparable advances.
In particular, a variational reformulation of the DGA scheme would allow substantially more flexible representation of solutions.
With these further developments, we believe DGA schemes have the potential to give further improved estimates of dynamical quantities for difficult molecular problems.
}

%\bigadd{
%[FIND POSITION FOR THIS]
%Similarly, we caution against naive application of variational methods for assessing basis quality, as many of these methods are derived for function spaces without nontrivial boundary conditions.
%For this reason, rather than comparing our basis sets using variational scores, we will instead attempt to directly assess the error in the estimated dynamical quantity.
%We leave variational analysis of basis sets in this framework to future work, and discuss potential directions in the conclusion
%}

\section{Acknowledgements}
This work was supported by National Institutes of Health Award R01 GM109455 and by the Molecular Software Sciences Institute (MolSSI) Software Fellows program.  Computing resources where provided by the University of Chicago Research Computing Center (RCC).  The Fip35 WW domain data was provided by D.E. Shaw Research.  Most of this work was completed while JW was a member of the Statistics Department and the James Franck Institute at the University of Chicago.  We thank Charles Matthews, Justin Finkel, and Benoit Roux for helpful discussions, \bigadd{as well as Fabian Paul, Frank No{\'e}, and the anonymous reviewers for their constructive feedback on earlier versions of the manuscript}.

\bibliography{gk_dmap_paper}

\providecommand{\noopsort}[1]{}\providecommand{\singleletter}[1]{#1}%
\begin{thebibliography}{109}%
\makeatletter
\providecommand \@ifxundefined [1]{%
 \@ifx{#1\undefined}
}%
\providecommand \@ifnum [1]{%
 \ifnum #1\expandafter \@firstoftwo
 \else \expandafter \@secondoftwo
 \fi
}%
\providecommand \@ifx [1]{%
 \ifx #1\expandafter \@firstoftwo
 \else \expandafter \@secondoftwo
 \fi
}%
\providecommand \natexlab [1]{#1}%
\providecommand \enquote  [1]{``#1''}%
\providecommand \bibnamefont  [1]{#1}%
\providecommand \bibfnamefont [1]{#1}%
\providecommand \citenamefont [1]{#1}%
\providecommand \href@noop [0]{\@secondoftwo}%
\providecommand \href [0]{\begingroup \@sanitize@url \@href}%
\providecommand \@href[1]{\@@startlink{#1}\@@href}%
\providecommand \@@href[1]{\endgroup#1\@@endlink}%
\providecommand \@sanitize@url [0]{\catcode `\\12\catcode `\$12\catcode
  `\&12\catcode `\#12\catcode `\^12\catcode `\_12\catcode `\%12\relax}%
\providecommand \@@startlink[1]{}%
\providecommand \@@endlink[0]{}%
\providecommand \url  [0]{\begingroup\@sanitize@url \@url }%
\providecommand \@url [1]{\endgroup\@href {#1}{\urlprefix }}%
\providecommand \urlprefix  [0]{URL }%
\providecommand \Eprint [0]{\href }%
\providecommand \doibase [0]{http://dx.doi.org/}%
\providecommand \selectlanguage [0]{\@gobble}%
\providecommand \bibinfo  [0]{\@secondoftwo}%
\providecommand \bibfield  [0]{\@secondoftwo}%
\providecommand \translation [1]{[#1]}%
\providecommand \BibitemOpen [0]{}%
\providecommand \bibitemStop [0]{}%
\providecommand \bibitemNoStop [0]{.\EOS\space}%
\providecommand \EOS [0]{\spacefactor3000\relax}%
\providecommand \BibitemShut  [1]{\csname bibitem#1\endcsname}%
\let\auto@bib@innerbib\@empty
%</preamble>
\bibitem [{\citenamefont {Kramers}(1940)}]{kramers1940brownian}%
  \BibitemOpen
  \bibfield  {author} {\bibinfo {author} {\bibfnamefont {H.~A.}\ \bibnamefont
  {Kramers}},\ }\bibfield  {title} {\enquote {\bibinfo {title} {Brownian motion
  in a field of force and the diffusion model of chemical reactions},}\
  }\href@noop {} {\bibfield  {journal} {\bibinfo  {journal} {Physica}\ }\textbf
  {\bibinfo {volume} {7}},\ \bibinfo {pages} {284--304} (\bibinfo {year}
  {1940})}\BibitemShut {NoStop}%
\bibitem [{\citenamefont {H{\"a}nggi}, \citenamefont {Talkner},\ and\
  \citenamefont {Borkovec}(1990)}]{hanggi1990reaction}%
  \BibitemOpen
  \bibfield  {author} {\bibinfo {author} {\bibfnamefont {P.}~\bibnamefont
  {H{\"a}nggi}}, \bibinfo {author} {\bibfnamefont {P.}~\bibnamefont {Talkner}},
  \ and\ \bibinfo {author} {\bibfnamefont {M.}~\bibnamefont {Borkovec}},\
  }\bibfield  {title} {\enquote {\bibinfo {title} {Reaction-rate theory: fifty
  years after {K}ramers},}\ }\href@noop {} {\bibfield  {journal} {\bibinfo
  {journal} {Reviews of Modern Physics}\ }\textbf {\bibinfo {volume} {62}},\
  \bibinfo {pages} {251} (\bibinfo {year} {1990})}\BibitemShut {NoStop}%
\bibitem [{\citenamefont {Vanden-Eijnden}(2006)}]{vanden2006transition}%
  \BibitemOpen
  \bibfield  {author} {\bibinfo {author} {\bibfnamefont {E.}~\bibnamefont
  {Vanden-Eijnden}},\ }\bibfield  {title} {\enquote {\bibinfo {title}
  {Transition path theory},}\ }in\ \href@noop {} {\emph {\bibinfo {booktitle}
  {Computer Simulations in Condensed Matter Systems: From Materials to Chemical
  Biology Volume 1}}}\ (\bibinfo  {publisher} {Springer},\ \bibinfo {year}
  {2006})\ pp.\ \bibinfo {pages} {453--493}\BibitemShut {NoStop}%
\bibitem [{\citenamefont {Berezhkovskii}\ \emph {et~al.}(2014)\citenamefont
  {Berezhkovskii}, \citenamefont {Szabo}, \citenamefont {Greives},\ and\
  \citenamefont {Zhou}}]{berezhkovskii2014multidimensional}%
  \BibitemOpen
  \bibfield  {author} {\bibinfo {author} {\bibfnamefont {A.~M.}\ \bibnamefont
  {Berezhkovskii}}, \bibinfo {author} {\bibfnamefont {A.}~\bibnamefont
  {Szabo}}, \bibinfo {author} {\bibfnamefont {N.}~\bibnamefont {Greives}}, \
  and\ \bibinfo {author} {\bibfnamefont {H.-X.}\ \bibnamefont {Zhou}},\
  }\bibfield  {title} {\enquote {\bibinfo {title} {Multidimensional reaction
  rate theory with anisotropic diffusion},}\ }\href@noop {} {\bibfield
  {journal} {\bibinfo  {journal} {The Journal of Chemical Physics}\ }\textbf
  {\bibinfo {volume} {141}},\ \bibinfo {pages} {11B616\_1} (\bibinfo {year}
  {2014})}\BibitemShut {NoStop}%
\bibitem [{\citenamefont {Ma}, \citenamefont {Nag},\ and\ \citenamefont
  {Dinner}(2006)}]{ma2006dynamic}%
  \BibitemOpen
  \bibfield  {author} {\bibinfo {author} {\bibfnamefont {A.}~\bibnamefont
  {Ma}}, \bibinfo {author} {\bibfnamefont {A.}~\bibnamefont {Nag}}, \ and\
  \bibinfo {author} {\bibfnamefont {A.~R.}\ \bibnamefont {Dinner}},\ }\bibfield
   {title} {\enquote {\bibinfo {title} {Dynamic coupling between coordinates in
  a model for biomolecular isomerization},}\ }\href@noop {} {\bibfield
  {journal} {\bibinfo  {journal} {The Journal of chemical physics}\ }\textbf
  {\bibinfo {volume} {124}},\ \bibinfo {pages} {144911} (\bibinfo {year}
  {2006})}\BibitemShut {NoStop}%
\bibitem [{\citenamefont {Ovchinnikov}, \citenamefont {Nam},\ and\
  \citenamefont {Karplus}(2016)}]{ovchinnikov2016simple}%
  \BibitemOpen
  \bibfield  {author} {\bibinfo {author} {\bibfnamefont {V.}~\bibnamefont
  {Ovchinnikov}}, \bibinfo {author} {\bibfnamefont {K.}~\bibnamefont {Nam}}, \
  and\ \bibinfo {author} {\bibfnamefont {M.}~\bibnamefont {Karplus}},\
  }\bibfield  {title} {\enquote {\bibinfo {title} {A simple and accurate method
  to calculate free energy profiles and reaction rates from restrained
  molecular simulations of diffusive processes},}\ }\href@noop {} {\bibfield
  {journal} {\bibinfo  {journal} {The Journal of Physical Chemistry B}\
  }\textbf {\bibinfo {volume} {120}},\ \bibinfo {pages} {8457--8472} (\bibinfo
  {year} {2016})}\BibitemShut {NoStop}%
\bibitem [{\citenamefont {Ghysels}\ \emph {et~al.}(2017)\citenamefont
  {Ghysels}, \citenamefont {Venable}, \citenamefont {Pastor},\ and\
  \citenamefont {Hummer}}]{ghysels2017position}%
  \BibitemOpen
  \bibfield  {author} {\bibinfo {author} {\bibfnamefont {A.}~\bibnamefont
  {Ghysels}}, \bibinfo {author} {\bibfnamefont {R.~M.}\ \bibnamefont
  {Venable}}, \bibinfo {author} {\bibfnamefont {R.~W.}\ \bibnamefont {Pastor}},
  \ and\ \bibinfo {author} {\bibfnamefont {G.}~\bibnamefont {Hummer}},\
  }\bibfield  {title} {\enquote {\bibinfo {title} {Position-dependent diffusion
  tensors in anisotropic media from simulation: oxygen transport in and through
  membranes},}\ }\href@noop {} {\bibfield  {journal} {\bibinfo  {journal}
  {Journal of chemical theory and computation}\ }\textbf {\bibinfo {volume}
  {13}},\ \bibinfo {pages} {2962--2976} (\bibinfo {year} {2017})}\BibitemShut
  {NoStop}%
\bibitem [{\citenamefont {Dinner}\ and\ \citenamefont
  {Karplus}(1999)}]{dinner1999thermodynamics}%
  \BibitemOpen
  \bibfield  {author} {\bibinfo {author} {\bibfnamefont {A.~R.}\ \bibnamefont
  {Dinner}}\ and\ \bibinfo {author} {\bibfnamefont {M.}~\bibnamefont
  {Karplus}},\ }\bibfield  {title} {\enquote {\bibinfo {title} {The
  thermodynamics and kinetics of protein folding: a lattice model analysis of
  multiple pathways with intermediates},}\ }\href@noop {} {\bibfield  {journal}
  {\bibinfo  {journal} {The Journal of Physical Chemistry B}\ }\textbf
  {\bibinfo {volume} {103}},\ \bibinfo {pages} {7976--7994} (\bibinfo {year}
  {1999})}\BibitemShut {NoStop}%
\bibitem [{\citenamefont {Dinner}\ \emph {et~al.}(2000)\citenamefont {Dinner},
  \citenamefont {{\v{S}}ali}, \citenamefont {Smith}, \citenamefont {Dobson},\
  and\ \citenamefont {Karplus}}]{dinner2000understanding}%
  \BibitemOpen
  \bibfield  {author} {\bibinfo {author} {\bibfnamefont {A.~R.}\ \bibnamefont
  {Dinner}}, \bibinfo {author} {\bibfnamefont {A.}~\bibnamefont {{\v{S}}ali}},
  \bibinfo {author} {\bibfnamefont {L.~J.}\ \bibnamefont {Smith}}, \bibinfo
  {author} {\bibfnamefont {C.~M.}\ \bibnamefont {Dobson}}, \ and\ \bibinfo
  {author} {\bibfnamefont {M.}~\bibnamefont {Karplus}},\ }\bibfield  {title}
  {\enquote {\bibinfo {title} {Understanding protein folding via free-energy
  surfaces from theory and experiment},}\ }\href@noop {} {\bibfield  {journal}
  {\bibinfo  {journal} {Trends in Biochemical Sciences}\ }\textbf {\bibinfo
  {volume} {25}},\ \bibinfo {pages} {331--339} (\bibinfo {year}
  {2000})}\BibitemShut {NoStop}%
\bibitem [{\citenamefont {Dellago}\ \emph {et~al.}(1998)\citenamefont
  {Dellago}, \citenamefont {Bolhuis}, \citenamefont {Csajka},\ and\
  \citenamefont {Chandler}}]{dellago1998transition}%
  \BibitemOpen
  \bibfield  {author} {\bibinfo {author} {\bibfnamefont {C.}~\bibnamefont
  {Dellago}}, \bibinfo {author} {\bibfnamefont {P.~G.}\ \bibnamefont
  {Bolhuis}}, \bibinfo {author} {\bibfnamefont {F.~S.}\ \bibnamefont {Csajka}},
  \ and\ \bibinfo {author} {\bibfnamefont {D.}~\bibnamefont {Chandler}},\
  }\bibfield  {title} {\enquote {\bibinfo {title} {Transition path sampling and
  the calculation of rate constants},}\ }\href@noop {} {\bibfield  {journal}
  {\bibinfo  {journal} {The Journal of Chemical Physics}\ }\textbf {\bibinfo
  {volume} {108}},\ \bibinfo {pages} {1964--1977} (\bibinfo {year}
  {1998})}\BibitemShut {NoStop}%
\bibitem [{\citenamefont {Bolhuis}\ \emph {et~al.}(2002)\citenamefont
  {Bolhuis}, \citenamefont {Chandler}, \citenamefont {Dellago},\ and\
  \citenamefont {Geissler}}]{bolhuis2002transition}%
  \BibitemOpen
  \bibfield  {author} {\bibinfo {author} {\bibfnamefont {P.~G.}\ \bibnamefont
  {Bolhuis}}, \bibinfo {author} {\bibfnamefont {D.}~\bibnamefont {Chandler}},
  \bibinfo {author} {\bibfnamefont {C.}~\bibnamefont {Dellago}}, \ and\
  \bibinfo {author} {\bibfnamefont {P.~L.}\ \bibnamefont {Geissler}},\
  }\bibfield  {title} {\enquote {\bibinfo {title} {Transition path sampling:
  {T}hrowing ropes over rough mountain passes, in the dark},}\ }\href@noop {}
  {\bibfield  {journal} {\bibinfo  {journal} {Annual Review of Physical
  Chemistry}\ }\textbf {\bibinfo {volume} {53}},\ \bibinfo {pages} {291--318}
  (\bibinfo {year} {2002})}\BibitemShut {NoStop}%
\bibitem [{\citenamefont {Gr{\"u}nwald}, \citenamefont {Dellago},\ and\
  \citenamefont {Geissler}(2008)}]{grunwald2008precision}%
  \BibitemOpen
  \bibfield  {author} {\bibinfo {author} {\bibfnamefont {M.}~\bibnamefont
  {Gr{\"u}nwald}}, \bibinfo {author} {\bibfnamefont {C.}~\bibnamefont
  {Dellago}}, \ and\ \bibinfo {author} {\bibfnamefont {P.~L.}\ \bibnamefont
  {Geissler}},\ }\bibfield  {title} {\enquote {\bibinfo {title} {Precision
  shooting: {S}ampling long transition pathways},}\ }\href@noop {} {\bibfield
  {journal} {\bibinfo  {journal} {The Journal of Chemical Physics}\ }\textbf
  {\bibinfo {volume} {129}},\ \bibinfo {pages} {194101} (\bibinfo {year}
  {2008})}\BibitemShut {NoStop}%
\bibitem [{\citenamefont {Gingrich}\ and\ \citenamefont
  {Geissler}(2015)}]{gingrich2015preserving}%
  \BibitemOpen
  \bibfield  {author} {\bibinfo {author} {\bibfnamefont {T.~R.}\ \bibnamefont
  {Gingrich}}\ and\ \bibinfo {author} {\bibfnamefont {P.~L.}\ \bibnamefont
  {Geissler}},\ }\bibfield  {title} {\enquote {\bibinfo {title} {Preserving
  correlations between trajectories for efficient path sampling},}\ }\href@noop
  {} {\bibfield  {journal} {\bibinfo  {journal} {The Journal of Chemical
  Physics}\ }\textbf {\bibinfo {volume} {142}},\ \bibinfo {pages} {06B614\_1}
  (\bibinfo {year} {2015})}\BibitemShut {NoStop}%
\bibitem [{\citenamefont {Huber}\ and\ \citenamefont
  {Kim}(1996)}]{huber1996weighted}%
  \BibitemOpen
  \bibfield  {author} {\bibinfo {author} {\bibfnamefont {G.~A.}\ \bibnamefont
  {Huber}}\ and\ \bibinfo {author} {\bibfnamefont {S.}~\bibnamefont {Kim}},\
  }\bibfield  {title} {\enquote {\bibinfo {title} {Weighted-ensemble {B}rownian
  dynamics simulations for protein association reactions},}\ }\href@noop {}
  {\bibfield  {journal} {\bibinfo  {journal} {Biophysical Journal}\ }\textbf
  {\bibinfo {volume} {70}},\ \bibinfo {pages} {97} (\bibinfo {year}
  {1996})}\BibitemShut {NoStop}%
\bibitem [{\citenamefont {van Erp}, \citenamefont {Moroni},\ and\ \citenamefont
  {Bolhuis}(2003)}]{van2003novel}%
  \BibitemOpen
  \bibfield  {author} {\bibinfo {author} {\bibfnamefont {T.~S.}\ \bibnamefont
  {van Erp}}, \bibinfo {author} {\bibfnamefont {D.}~\bibnamefont {Moroni}}, \
  and\ \bibinfo {author} {\bibfnamefont {P.~G.}\ \bibnamefont {Bolhuis}},\
  }\bibfield  {title} {\enquote {\bibinfo {title} {A novel path sampling method
  for the calculation of rate constants},}\ }\href@noop {} {\bibfield
  {journal} {\bibinfo  {journal} {The Journal of Chemical Physics}\ }\textbf
  {\bibinfo {volume} {118}},\ \bibinfo {pages} {7762--7774} (\bibinfo {year}
  {2003})}\BibitemShut {NoStop}%
\bibitem [{\citenamefont {Faradjian}\ and\ \citenamefont
  {Elber}(2004)}]{faradjian2004computing}%
  \BibitemOpen
  \bibfield  {author} {\bibinfo {author} {\bibfnamefont {A.~K.}\ \bibnamefont
  {Faradjian}}\ and\ \bibinfo {author} {\bibfnamefont {R.}~\bibnamefont
  {Elber}},\ }\bibfield  {title} {\enquote {\bibinfo {title} {Computing time
  scales from reaction coordinates by milestoning},}\ }\href@noop {} {\bibfield
   {journal} {\bibinfo  {journal} {The Journal of Chemical Physics}\ }\textbf
  {\bibinfo {volume} {120}},\ \bibinfo {pages} {10880--10889} (\bibinfo {year}
  {2004})}\BibitemShut {NoStop}%
\bibitem [{\citenamefont {Allen}, \citenamefont {Frenkel},\ and\ \citenamefont
  {ten Wolde}(2006)}]{ffs_allen2006simulating}%
  \BibitemOpen
  \bibfield  {author} {\bibinfo {author} {\bibfnamefont {R.~J.}\ \bibnamefont
  {Allen}}, \bibinfo {author} {\bibfnamefont {D.}~\bibnamefont {Frenkel}}, \
  and\ \bibinfo {author} {\bibfnamefont {P.~R.}\ \bibnamefont {ten Wolde}},\
  }\bibfield  {title} {\enquote {\bibinfo {title} {Simulating rare events in
  equilibrium or nonequilibrium stochastic systems},}\ }\href@noop {}
  {\bibfield  {journal} {\bibinfo  {journal} {The Journal of chemical physics}\
  }\textbf {\bibinfo {volume} {124}},\ \bibinfo {pages} {024102} (\bibinfo
  {year} {2006})}\BibitemShut {NoStop}%
\bibitem [{\citenamefont {Warmflash}, \citenamefont {Bhimalapuram},\ and\
  \citenamefont {Dinner}(2007)}]{warmflash2007umbrella}%
  \BibitemOpen
  \bibfield  {author} {\bibinfo {author} {\bibfnamefont {A.}~\bibnamefont
  {Warmflash}}, \bibinfo {author} {\bibfnamefont {P.}~\bibnamefont
  {Bhimalapuram}}, \ and\ \bibinfo {author} {\bibfnamefont {A.~R.}\
  \bibnamefont {Dinner}},\ }\bibfield  {title} {\enquote {\bibinfo {title}
  {Umbrella sampling for nonequilibrium processes},}\ }\href@noop {} {\bibfield
   {journal} {\bibinfo  {journal} {The Journal of Chemical Physics}\ }\textbf
  {\bibinfo {volume} {127}},\ \bibinfo {pages} {114109} (\bibinfo {year}
  {2007})}\BibitemShut {NoStop}%
\bibitem [{\citenamefont {Vanden-Eijnden}\ and\ \citenamefont
  {Venturoli}(2009)}]{vanden2009exact}%
  \BibitemOpen
  \bibfield  {author} {\bibinfo {author} {\bibfnamefont {E.}~\bibnamefont
  {Vanden-Eijnden}}\ and\ \bibinfo {author} {\bibfnamefont {M.}~\bibnamefont
  {Venturoli}},\ }\bibfield  {title} {\enquote {\bibinfo {title} {Exact rate
  calculations by trajectory parallelization and tilting},}\ }\href@noop {}
  {\bibfield  {journal} {\bibinfo  {journal} {The Journal of chemical physics}\
  }\textbf {\bibinfo {volume} {131}},\ \bibinfo {pages} {044120} (\bibinfo
  {year} {2009})}\BibitemShut {NoStop}%
\bibitem [{\citenamefont {Dickson}, \citenamefont {Warmflash},\ and\
  \citenamefont {Dinner}(2009)}]{dickson2009separating}%
  \BibitemOpen
  \bibfield  {author} {\bibinfo {author} {\bibfnamefont {A.}~\bibnamefont
  {Dickson}}, \bibinfo {author} {\bibfnamefont {A.}~\bibnamefont {Warmflash}},
  \ and\ \bibinfo {author} {\bibfnamefont {A.~R.}\ \bibnamefont {Dinner}},\
  }\bibfield  {title} {\enquote {\bibinfo {title} {Separating forward and
  backward pathways in nonequilibrium umbrella sampling},}\ }\href@noop {}
  {\bibfield  {journal} {\bibinfo  {journal} {The Journal of Chemical Physics}\
  }\textbf {\bibinfo {volume} {131}},\ \bibinfo {pages} {154104} (\bibinfo
  {year} {2009})}\BibitemShut {NoStop}%
\bibitem [{\citenamefont {Guttenberg}, \citenamefont {Dinner},\ and\
  \citenamefont {Weare}(2012)}]{guttenberg2012steered}%
  \BibitemOpen
  \bibfield  {author} {\bibinfo {author} {\bibfnamefont {N.}~\bibnamefont
  {Guttenberg}}, \bibinfo {author} {\bibfnamefont {A.~R.}\ \bibnamefont
  {Dinner}}, \ and\ \bibinfo {author} {\bibfnamefont {J.}~\bibnamefont
  {Weare}},\ }\bibfield  {title} {\enquote {\bibinfo {title} {Steered
  transition path sampling},}\ }\href@noop {} {\bibfield  {journal} {\bibinfo
  {journal} {The Journal of Chemical Physics}\ }\textbf {\bibinfo {volume}
  {136}},\ \bibinfo {pages} {06B609} (\bibinfo {year} {2012})}\BibitemShut
  {NoStop}%
\bibitem [{\citenamefont {Bello-Rivas}\ and\ \citenamefont
  {Elber}(2015)}]{bello2015exact}%
  \BibitemOpen
  \bibfield  {author} {\bibinfo {author} {\bibfnamefont {J.~M.}\ \bibnamefont
  {Bello-Rivas}}\ and\ \bibinfo {author} {\bibfnamefont {R.}~\bibnamefont
  {Elber}},\ }\bibfield  {title} {\enquote {\bibinfo {title} {Exact
  milestoning},}\ }\href@noop {} {\bibfield  {journal} {\bibinfo  {journal}
  {The Journal of Chemical Physics}\ }\textbf {\bibinfo {volume} {142}},\
  \bibinfo {pages} {03B602\_1} (\bibinfo {year} {2015})}\BibitemShut {NoStop}%
\bibitem [{\citenamefont {Dinner}\ \emph {et~al.}(2018)\citenamefont {Dinner},
  \citenamefont {Mattingly}, \citenamefont {Tempkin}, \citenamefont {Koten},\
  and\ \citenamefont {Weare}}]{dinner2018trajectory}%
  \BibitemOpen
  \bibfield  {author} {\bibinfo {author} {\bibfnamefont {A.~R.}\ \bibnamefont
  {Dinner}}, \bibinfo {author} {\bibfnamefont {J.~C.}\ \bibnamefont
  {Mattingly}}, \bibinfo {author} {\bibfnamefont {J.~O.}\ \bibnamefont
  {Tempkin}}, \bibinfo {author} {\bibfnamefont {B.~V.}\ \bibnamefont {Koten}},
  \ and\ \bibinfo {author} {\bibfnamefont {J.}~\bibnamefont {Weare}},\
  }\bibfield  {title} {\enquote {\bibinfo {title} {Trajectory stratification of
  stochastic dynamics},}\ }\href@noop {} {\bibfield  {journal} {\bibinfo
  {journal} {SIAM Review}\ }\textbf {\bibinfo {volume} {60}},\ \bibinfo {pages}
  {909--938} (\bibinfo {year} {2018})}\BibitemShut {NoStop}%
\bibitem [{\citenamefont {Sch{\"u}tte}\ \emph {et~al.}(1999)\citenamefont
  {Sch{\"u}tte}, \citenamefont {Fischer}, \citenamefont {Huisinga},\ and\
  \citenamefont {Deuflhard}}]{schutte1999direct}%
  \BibitemOpen
  \bibfield  {author} {\bibinfo {author} {\bibfnamefont {C.}~\bibnamefont
  {Sch{\"u}tte}}, \bibinfo {author} {\bibfnamefont {A.}~\bibnamefont
  {Fischer}}, \bibinfo {author} {\bibfnamefont {W.}~\bibnamefont {Huisinga}}, \
  and\ \bibinfo {author} {\bibfnamefont {P.}~\bibnamefont {Deuflhard}},\
  }\bibfield  {title} {\enquote {\bibinfo {title} {A direct approach to
  conformational dynamics based on hybrid {M}onte {C}arlo},}\ }\href@noop {}
  {\bibfield  {journal} {\bibinfo  {journal} {Journal of Computational
  Physics}\ }\textbf {\bibinfo {volume} {151}},\ \bibinfo {pages} {146--168}
  (\bibinfo {year} {1999})}\BibitemShut {NoStop}%
\bibitem [{\citenamefont {Swope}, \citenamefont {Pitera},\ and\ \citenamefont
  {Suits}(2004)}]{swope2004describing}%
  \BibitemOpen
  \bibfield  {author} {\bibinfo {author} {\bibfnamefont {W.~C.}\ \bibnamefont
  {Swope}}, \bibinfo {author} {\bibfnamefont {J.~W.}\ \bibnamefont {Pitera}}, \
  and\ \bibinfo {author} {\bibfnamefont {F.}~\bibnamefont {Suits}},\ }\bibfield
   {title} {\enquote {\bibinfo {title} {Describing protein folding kinetics by
  molecular dynamics simulations. 1. {T}heory},}\ }\href@noop {} {\bibfield
  {journal} {\bibinfo  {journal} {The Journal of Physical Chemistry B}\
  }\textbf {\bibinfo {volume} {108}},\ \bibinfo {pages} {6571--6581} (\bibinfo
  {year} {2004})}\BibitemShut {NoStop}%
\bibitem [{\citenamefont {Pande}, \citenamefont {Beauchamp},\ and\
  \citenamefont {Bowman}(2010)}]{pande2010everything}%
  \BibitemOpen
  \bibfield  {author} {\bibinfo {author} {\bibfnamefont {V.~S.}\ \bibnamefont
  {Pande}}, \bibinfo {author} {\bibfnamefont {K.}~\bibnamefont {Beauchamp}}, \
  and\ \bibinfo {author} {\bibfnamefont {G.~R.}\ \bibnamefont {Bowman}},\
  }\bibfield  {title} {\enquote {\bibinfo {title} {Everything you wanted to
  know about {M}arkov state models but were afraid to ask},}\ }\href@noop {}
  {\bibfield  {journal} {\bibinfo  {journal} {Methods}\ }\textbf {\bibinfo
  {volume} {52}},\ \bibinfo {pages} {99--105} (\bibinfo {year}
  {2010})}\BibitemShut {NoStop}%
\bibitem [{\citenamefont {Sarich}, \citenamefont {No{\'e}},\ and\ \citenamefont
  {Sch{\"u}tte}(2010)}]{sarich2010approximation}%
  \BibitemOpen
  \bibfield  {author} {\bibinfo {author} {\bibfnamefont {M.}~\bibnamefont
  {Sarich}}, \bibinfo {author} {\bibfnamefont {F.}~\bibnamefont {No{\'e}}}, \
  and\ \bibinfo {author} {\bibfnamefont {C.}~\bibnamefont {Sch{\"u}tte}},\
  }\bibfield  {title} {\enquote {\bibinfo {title} {On the approximation quality
  of {M}arkov state models},}\ }\href@noop {} {\bibfield  {journal} {\bibinfo
  {journal} {Multiscale Modeling \& Simulation}\ }\textbf {\bibinfo {volume}
  {8}},\ \bibinfo {pages} {1154--1177} (\bibinfo {year} {2010})}\BibitemShut
  {NoStop}%
\bibitem [{\citenamefont {No{\'e}}\ and\ \citenamefont
  {Fischer}(2008)}]{noe2008transition}%
  \BibitemOpen
  \bibfield  {author} {\bibinfo {author} {\bibfnamefont {F.}~\bibnamefont
  {No{\'e}}}\ and\ \bibinfo {author} {\bibfnamefont {S.}~\bibnamefont
  {Fischer}},\ }\bibfield  {title} {\enquote {\bibinfo {title} {Transition
  networks for modeling the kinetics of conformational change in
  macromolecules},}\ }\href@noop {} {\bibfield  {journal} {\bibinfo  {journal}
  {Current Opinion in Structural Biology}\ }\textbf {\bibinfo {volume} {18}},\
  \bibinfo {pages} {154--162} (\bibinfo {year} {2008})}\BibitemShut {NoStop}%
\bibitem [{\citenamefont {No{\'e}}\ and\ \citenamefont
  {Prinz}(2014)}]{noe2014introduction}%
  \BibitemOpen
  \bibfield  {author} {\bibinfo {author} {\bibfnamefont {F.}~\bibnamefont
  {No{\'e}}}\ and\ \bibinfo {author} {\bibfnamefont {J.-H.}\ \bibnamefont
  {Prinz}},\ }\bibfield  {title} {\enquote {\bibinfo {title} {Analysis of
  {M}arkov models},}\ }in\ \href@noop {} {\emph {\bibinfo {booktitle} {An
  Introduction to Markov State Models and Their Application to Long Timescale
  Molecular Simulation, vol. 797 of Advances in Experimental Medicine and
  Biology}}},\ \bibinfo {editor} {edited by\ \bibinfo {editor} {\bibfnamefont
  {G.~R.}\ \bibnamefont {Bowman}}, \bibinfo {editor} {\bibfnamefont {V.~S.}\
  \bibnamefont {Pande}}, \ and\ \bibinfo {editor} {\bibfnamefont
  {F.}~\bibnamefont {No{\'e}}}}\ (\bibinfo  {publisher} {Springer},\ \bibinfo
  {year} {2014})\ Chap.~\bibinfo {chapter} {6}\BibitemShut {NoStop}%
\bibitem [{\citenamefont {Keller}, \citenamefont {Aleksic},\ and\ \citenamefont
  {Donati}(2019)}]{keller2019markov}%
  \BibitemOpen
  \bibfield  {author} {\bibinfo {author} {\bibfnamefont {B.~G.}\ \bibnamefont
  {Keller}}, \bibinfo {author} {\bibfnamefont {S.}~\bibnamefont {Aleksic}}, \
  and\ \bibinfo {author} {\bibfnamefont {L.}~\bibnamefont {Donati}},\
  }\bibfield  {title} {\enquote {\bibinfo {title} {{M}arkov state models in
  drug design},}\ }in\ \href@noop {} {\emph {\bibinfo {booktitle} {Biomolecular
  Simulations in Drug Discovery}}},\ \bibinfo {editor} {edited by\ \bibinfo
  {editor} {\bibfnamefont {F.~L.}\ \bibnamefont {Gervasio}}\ and\ \bibinfo
  {editor} {\bibfnamefont {V.}~\bibnamefont {Spiwok}}}\ (\bibinfo  {publisher}
  {Wiley-VCH},\ \bibinfo {year} {2019})\ Chap.~\bibinfo {chapter}
  {4}\BibitemShut {NoStop}%
\bibitem [{\citenamefont {Weber}(2006)}]{weber2006meshless}%
  \BibitemOpen
  \bibfield  {author} {\bibinfo {author} {\bibfnamefont {M.}~\bibnamefont
  {Weber}},\ }\emph {\bibinfo {title} {Meshless Methods in Conformation
  Dynamics}},\ \href@noop {} {Ph.D. thesis},\ \bibinfo  {school} {Freie
  Universit{\"a}t Berlin} (\bibinfo {year} {2006})\BibitemShut {NoStop}%
\bibitem [{\citenamefont {No{\'e}}\ and\ \citenamefont
  {N{\"u}ske}(2013)}]{noe2013variational}%
  \BibitemOpen
  \bibfield  {author} {\bibinfo {author} {\bibfnamefont {F.}~\bibnamefont
  {No{\'e}}}\ and\ \bibinfo {author} {\bibfnamefont {F.}~\bibnamefont
  {N{\"u}ske}},\ }\bibfield  {title} {\enquote {\bibinfo {title} {A variational
  approach to modeling slow processes in stochastic dynamical systems},}\
  }\href@noop {} {\bibfield  {journal} {\bibinfo  {journal} {Multiscale
  Modeling \& Simulation}\ }\textbf {\bibinfo {volume} {11}},\ \bibinfo {pages}
  {635--655} (\bibinfo {year} {2013})}\BibitemShut {NoStop}%
\bibitem [{\citenamefont {Eisner}\ \emph {et~al.}(2015)\citenamefont {Eisner},
  \citenamefont {Farkas}, \citenamefont {Haase},\ and\ \citenamefont
  {Nagel}}]{eisner2015operator}%
  \BibitemOpen
  \bibfield  {author} {\bibinfo {author} {\bibfnamefont {T.}~\bibnamefont
  {Eisner}}, \bibinfo {author} {\bibfnamefont {B.}~\bibnamefont {Farkas}},
  \bibinfo {author} {\bibfnamefont {M.}~\bibnamefont {Haase}}, \ and\ \bibinfo
  {author} {\bibfnamefont {R.}~\bibnamefont {Nagel}},\ }\href@noop {} {\emph
  {\bibinfo {title} {Operator Theoretic Aspects of Ergodic Theory}}},\ Vol.\
  \bibinfo {volume} {272}\ (\bibinfo  {publisher} {Springer},\ \bibinfo {year}
  {2015})\BibitemShut {NoStop}%
\bibitem [{\citenamefont {Klus}\ \emph {et~al.}(2018)\citenamefont {Klus},
  \citenamefont {N{\"u}ske}, \citenamefont {Koltai}, \citenamefont {Wu},
  \citenamefont {Kevrekidis}, \citenamefont {Sch{\"u}tte},\ and\ \citenamefont
  {No{\'e}}}]{klus2018data}%
  \BibitemOpen
  \bibfield  {author} {\bibinfo {author} {\bibfnamefont {S.}~\bibnamefont
  {Klus}}, \bibinfo {author} {\bibfnamefont {F.}~\bibnamefont {N{\"u}ske}},
  \bibinfo {author} {\bibfnamefont {P.}~\bibnamefont {Koltai}}, \bibinfo
  {author} {\bibfnamefont {H.}~\bibnamefont {Wu}}, \bibinfo {author}
  {\bibfnamefont {I.}~\bibnamefont {Kevrekidis}}, \bibinfo {author}
  {\bibfnamefont {C.}~\bibnamefont {Sch{\"u}tte}}, \ and\ \bibinfo {author}
  {\bibfnamefont {F.}~\bibnamefont {No{\'e}}},\ }\bibfield  {title} {\enquote
  {\bibinfo {title} {Data-driven model reduction and transfer operator
  approximation},}\ }\href@noop {} {\bibfield  {journal} {\bibinfo  {journal}
  {Journal of Nonlinear Science}\ }\textbf {\bibinfo {volume} {28}},\ \bibinfo
  {pages} {985--1010} (\bibinfo {year} {2018})}\BibitemShut {NoStop}%
\bibitem [{\citenamefont {Billingsley}(2008)}]{billingsley2008probability}%
  \BibitemOpen
  \bibfield  {author} {\bibinfo {author} {\bibfnamefont {P.}~\bibnamefont
  {Billingsley}},\ }\href@noop {} {\emph {\bibinfo {title} {Probability and
  Measure}}}\ (\bibinfo  {publisher} {John Wiley \& Sons},\ \bibinfo {year}
  {2008})\BibitemShut {NoStop}%
\bibitem [{\citenamefont {Bowman}\ \emph {et~al.}(2009)\citenamefont {Bowman},
  \citenamefont {Beauchamp}, \citenamefont {Boxer},\ and\ \citenamefont
  {Pande}}]{bowman2009progress}%
  \BibitemOpen
  \bibfield  {author} {\bibinfo {author} {\bibfnamefont {G.~R.}\ \bibnamefont
  {Bowman}}, \bibinfo {author} {\bibfnamefont {K.~A.}\ \bibnamefont
  {Beauchamp}}, \bibinfo {author} {\bibfnamefont {G.}~\bibnamefont {Boxer}}, \
  and\ \bibinfo {author} {\bibfnamefont {V.~S.}\ \bibnamefont {Pande}},\
  }\bibfield  {title} {\enquote {\bibinfo {title} {Progress and challenges in
  the automated construction of {M}arkov state models for full protein
  systems},}\ }\href@noop {} {\bibfield  {journal} {\bibinfo  {journal} {The
  Journal of Chemical Physics}\ }\textbf {\bibinfo {volume} {131}},\ \bibinfo
  {pages} {124101} (\bibinfo {year} {2009})}\BibitemShut {NoStop}%
\bibitem [{\citenamefont {Prinz}\ \emph {et~al.}(2011)\citenamefont {Prinz},
  \citenamefont {Wu}, \citenamefont {Sarich}, \citenamefont {Keller},
  \citenamefont {Senne}, \citenamefont {Held}, \citenamefont {Chodera},
  \citenamefont {Sch{\"u}tte},\ and\ \citenamefont
  {No{\'e}}}]{prinz2011markov}%
  \BibitemOpen
  \bibfield  {author} {\bibinfo {author} {\bibfnamefont {J.-H.}\ \bibnamefont
  {Prinz}}, \bibinfo {author} {\bibfnamefont {H.}~\bibnamefont {Wu}}, \bibinfo
  {author} {\bibfnamefont {M.}~\bibnamefont {Sarich}}, \bibinfo {author}
  {\bibfnamefont {B.}~\bibnamefont {Keller}}, \bibinfo {author} {\bibfnamefont
  {M.}~\bibnamefont {Senne}}, \bibinfo {author} {\bibfnamefont
  {M.}~\bibnamefont {Held}}, \bibinfo {author} {\bibfnamefont {J.~D.}\
  \bibnamefont {Chodera}}, \bibinfo {author} {\bibfnamefont {C.}~\bibnamefont
  {Sch{\"u}tte}}, \ and\ \bibinfo {author} {\bibfnamefont {F.}~\bibnamefont
  {No{\'e}}},\ }\bibfield  {title} {\enquote {\bibinfo {title} {Markov models
  of molecular kinetics: Generation and validation},}\ }\href@noop {}
  {\bibfield  {journal} {\bibinfo  {journal} {The Journal of Chemical Physics}\
  }\textbf {\bibinfo {volume} {134}},\ \bibinfo {pages} {174105} (\bibinfo
  {year} {2011})}\BibitemShut {NoStop}%
\bibitem [{\citenamefont {No{\'e}}\ \emph {et~al.}(2009)\citenamefont
  {No{\'e}}, \citenamefont {Sch{\"u}tte}, \citenamefont {Vanden-Eijnden},
  \citenamefont {Reich},\ and\ \citenamefont {Weikl}}]{noe2009constructing}%
  \BibitemOpen
  \bibfield  {author} {\bibinfo {author} {\bibfnamefont {F.}~\bibnamefont
  {No{\'e}}}, \bibinfo {author} {\bibfnamefont {C.}~\bibnamefont
  {Sch{\"u}tte}}, \bibinfo {author} {\bibfnamefont {E.}~\bibnamefont
  {Vanden-Eijnden}}, \bibinfo {author} {\bibfnamefont {L.}~\bibnamefont
  {Reich}}, \ and\ \bibinfo {author} {\bibfnamefont {T.~R.}\ \bibnamefont
  {Weikl}},\ }\bibfield  {title} {\enquote {\bibinfo {title} {Constructing the
  equilibrium ensemble of folding pathways from short off-equilibrium
  simulations},}\ }\href@noop {} {\bibfield  {journal} {\bibinfo  {journal}
  {Proceedings of the National Academy of Sciences}\ }\textbf {\bibinfo
  {volume} {106}},\ \bibinfo {pages} {19011--19016} (\bibinfo {year}
  {2009})}\BibitemShut {NoStop}%
\bibitem [{\citenamefont {Schwantes}\ and\ \citenamefont
  {Pande}(2013)}]{schwantes2013improvements}%
  \BibitemOpen
  \bibfield  {author} {\bibinfo {author} {\bibfnamefont {C.~R.}\ \bibnamefont
  {Schwantes}}\ and\ \bibinfo {author} {\bibfnamefont {V.~S.}\ \bibnamefont
  {Pande}},\ }\bibfield  {title} {\enquote {\bibinfo {title} {Improvements in
  {M}arkov state model construction reveal many non-native interactions in the
  folding of {NTL9}},}\ }\href@noop {} {\bibfield  {journal} {\bibinfo
  {journal} {Journal of Chemical Theory and Computation}\ }\textbf {\bibinfo
  {volume} {9}},\ \bibinfo {pages} {2000--2009} (\bibinfo {year}
  {2013})}\BibitemShut {NoStop}%
\bibitem [{\citenamefont {P{\'e}rez-Hern{\'a}ndez}\ \emph
  {et~al.}(2013)\citenamefont {P{\'e}rez-Hern{\'a}ndez}, \citenamefont {Paul},
  \citenamefont {Giorgino}, \citenamefont {De~Fabritiis},\ and\ \citenamefont
  {No{\'e}}}]{perez2013identification}%
  \BibitemOpen
  \bibfield  {author} {\bibinfo {author} {\bibfnamefont {G.}~\bibnamefont
  {P{\'e}rez-Hern{\'a}ndez}}, \bibinfo {author} {\bibfnamefont
  {F.}~\bibnamefont {Paul}}, \bibinfo {author} {\bibfnamefont {T.}~\bibnamefont
  {Giorgino}}, \bibinfo {author} {\bibfnamefont {G.}~\bibnamefont
  {De~Fabritiis}}, \ and\ \bibinfo {author} {\bibfnamefont {F.}~\bibnamefont
  {No{\'e}}},\ }\bibfield  {title} {\enquote {\bibinfo {title} {Identification
  of slow molecular order parameters for {M}arkov model construction},}\
  }\href@noop {} {\bibfield  {journal} {\bibinfo  {journal} {The Journal of
  Chemical Physics}\ }\textbf {\bibinfo {volume} {139}},\ \bibinfo {pages}
  {07B604\_1} (\bibinfo {year} {2013})}\BibitemShut {NoStop}%
\bibitem [{\citenamefont {Schwantes}, \citenamefont {McGibbon},\ and\
  \citenamefont {Pande}(2014)}]{schwantes2014perspective}%
  \BibitemOpen
  \bibfield  {author} {\bibinfo {author} {\bibfnamefont {C.~R.}\ \bibnamefont
  {Schwantes}}, \bibinfo {author} {\bibfnamefont {R.~T.}\ \bibnamefont
  {McGibbon}}, \ and\ \bibinfo {author} {\bibfnamefont {V.~S.}\ \bibnamefont
  {Pande}},\ }\bibfield  {title} {\enquote {\bibinfo {title} {Perspective:
  {M}arkov models for long-timescale biomolecular dynamics},}\ }\href@noop {}
  {\bibfield  {journal} {\bibinfo  {journal} {The Journal of Chemical Physics}\
  }\textbf {\bibinfo {volume} {141}},\ \bibinfo {pages} {09B201} (\bibinfo
  {year} {2014})}\BibitemShut {NoStop}%
\bibitem [{\citenamefont {Sch{\"u}tte}\ and\ \citenamefont
  {Sarich}(2015)}]{schutte2015critical}%
  \BibitemOpen
  \bibfield  {author} {\bibinfo {author} {\bibfnamefont {C.}~\bibnamefont
  {Sch{\"u}tte}}\ and\ \bibinfo {author} {\bibfnamefont {M.}~\bibnamefont
  {Sarich}},\ }\bibfield  {title} {\enquote {\bibinfo {title} {A critical
  appraisal of {M}arkov state models},}\ }\href@noop {} {\bibfield  {journal}
  {\bibinfo  {journal} {The European Physical Journal Special Topics}\ }\textbf
  {\bibinfo {volume} {224}},\ \bibinfo {pages} {2445--2462} (\bibinfo {year}
  {2015})}\BibitemShut {NoStop}%
\bibitem [{\citenamefont {Shukla}\ \emph {et~al.}(2015)\citenamefont {Shukla},
  \citenamefont {Hernández}, \citenamefont {Weber},\ and\ \citenamefont
  {Pande}}]{shukla2015markov}%
  \BibitemOpen
  \bibfield  {author} {\bibinfo {author} {\bibfnamefont {D.}~\bibnamefont
  {Shukla}}, \bibinfo {author} {\bibfnamefont {C.~X.}\ \bibnamefont
  {Hernández}}, \bibinfo {author} {\bibfnamefont {J.~K.}\ \bibnamefont
  {Weber}}, \ and\ \bibinfo {author} {\bibfnamefont {V.~S.}\ \bibnamefont
  {Pande}},\ }\bibfield  {title} {\enquote {\bibinfo {title} {Markov state
  models provide insights into dynamic modulation of protein function},}\
  }\href@noop {} {\bibfield  {journal} {\bibinfo  {journal} {Accounts of
  Chemical Research}\ }\textbf {\bibinfo {volume} {48}},\ \bibinfo {pages}
  {414--422} (\bibinfo {year} {2015})}\BibitemShut {NoStop}%
\bibitem [{\citenamefont {Berezovska}, \citenamefont {Prada-Gracia},\ and\
  \citenamefont {Rao}(2013)}]{berezovska2013consensus}%
  \BibitemOpen
  \bibfield  {author} {\bibinfo {author} {\bibfnamefont {G.}~\bibnamefont
  {Berezovska}}, \bibinfo {author} {\bibfnamefont {D.}~\bibnamefont
  {Prada-Gracia}}, \ and\ \bibinfo {author} {\bibfnamefont {F.}~\bibnamefont
  {Rao}},\ }\bibfield  {title} {\enquote {\bibinfo {title} {Consensus for the
  {F}ip35 folding mechanism?}}\ }\href@noop {} {\bibfield  {journal} {\bibinfo
  {journal} {The Journal of Chemical Physics}\ }\textbf {\bibinfo {volume}
  {139}},\ \bibinfo {pages} {07B608\_1} (\bibinfo {year} {2013})}\BibitemShut
  {NoStop}%
\bibitem [{\citenamefont {Sheong}\ \emph {et~al.}(2014)\citenamefont {Sheong},
  \citenamefont {Silva}, \citenamefont {Meng}, \citenamefont {Zhao},\ and\
  \citenamefont {Huang}}]{sheong2014automatic}%
  \BibitemOpen
  \bibfield  {author} {\bibinfo {author} {\bibfnamefont {F.~K.}\ \bibnamefont
  {Sheong}}, \bibinfo {author} {\bibfnamefont {D.-A.}\ \bibnamefont {Silva}},
  \bibinfo {author} {\bibfnamefont {L.}~\bibnamefont {Meng}}, \bibinfo {author}
  {\bibfnamefont {Y.}~\bibnamefont {Zhao}}, \ and\ \bibinfo {author}
  {\bibfnamefont {X.}~\bibnamefont {Huang}},\ }\bibfield  {title} {\enquote
  {\bibinfo {title} {Automatic state partitioning for multibody systems
  ({A}{P}{M}): an efficient algorithm for constructing {M}arkov state models to
  elucidate conformational dynamics of multibody systems},}\ }\href@noop {}
  {\bibfield  {journal} {\bibinfo  {journal} {Journal of Chemical Theory and
  Computation}\ }\textbf {\bibinfo {volume} {11}},\ \bibinfo {pages} {17--27}
  (\bibinfo {year} {2014})}\BibitemShut {NoStop}%
\bibitem [{\citenamefont {Li}\ and\ \citenamefont {Dong}(2016)}]{li2016effect}%
  \BibitemOpen
  \bibfield  {author} {\bibinfo {author} {\bibfnamefont {Y.}~\bibnamefont
  {Li}}\ and\ \bibinfo {author} {\bibfnamefont {Z.}~\bibnamefont {Dong}},\
  }\bibfield  {title} {\enquote {\bibinfo {title} {Effect of clustering
  algorithm on establishing {M}arkov state model for molecular dynamics
  simulations},}\ }\href@noop {} {\bibfield  {journal} {\bibinfo  {journal}
  {Journal of Chemical Information and Modeling}\ }\textbf {\bibinfo {volume}
  {56}},\ \bibinfo {pages} {1205--1215} (\bibinfo {year} {2016})}\BibitemShut
  {NoStop}%
\bibitem [{\citenamefont {Husic}\ and\ \citenamefont
  {Pande}(2017)}]{husic2017ward}%
  \BibitemOpen
  \bibfield  {author} {\bibinfo {author} {\bibfnamefont {B.~E.}\ \bibnamefont
  {Husic}}\ and\ \bibinfo {author} {\bibfnamefont {V.~S.}\ \bibnamefont
  {Pande}},\ }\bibfield  {title} {\enquote {\bibinfo {title} {Ward clustering
  improves cross-validated {M}arkov state models of protein folding},}\
  }\href@noop {} {\bibfield  {journal} {\bibinfo  {journal} {Journal of
  Chemical Theory and Computation}\ }\textbf {\bibinfo {volume} {13}},\
  \bibinfo {pages} {963--967} (\bibinfo {year} {2017})}\BibitemShut {NoStop}%
\bibitem [{\citenamefont {Husic}\ \emph {et~al.}(2018)\citenamefont {Husic},
  \citenamefont {McKiernan}, \citenamefont {Wayment-Steele}, \citenamefont
  {Sultan},\ and\ \citenamefont {Pande}}]{husic2018minimum}%
  \BibitemOpen
  \bibfield  {author} {\bibinfo {author} {\bibfnamefont {B.~E.}\ \bibnamefont
  {Husic}}, \bibinfo {author} {\bibfnamefont {K.~A.}\ \bibnamefont
  {McKiernan}}, \bibinfo {author} {\bibfnamefont {H.~K.}\ \bibnamefont
  {Wayment-Steele}}, \bibinfo {author} {\bibfnamefont {M.~M.}\ \bibnamefont
  {Sultan}}, \ and\ \bibinfo {author} {\bibfnamefont {V.~S.}\ \bibnamefont
  {Pande}},\ }\bibfield  {title} {\enquote {\bibinfo {title} {A minimum
  variance clustering approach produces robust and interpretable coarse-grained
  models},}\ }\href@noop {} {\bibfield  {journal} {\bibinfo  {journal} {Journal
  of Chemical Theory and Computation}\ }\textbf {\bibinfo {volume} {14}},\
  \bibinfo {pages} {1071--1082} (\bibinfo {year} {2018})}\BibitemShut {NoStop}%
\bibitem [{\citenamefont {Steinbach}, \citenamefont {Ert{\"o}z},\ and\
  \citenamefont {Kumar}(2004)}]{steinbach2004challenges}%
  \BibitemOpen
  \bibfield  {author} {\bibinfo {author} {\bibfnamefont {M.}~\bibnamefont
  {Steinbach}}, \bibinfo {author} {\bibfnamefont {L.}~\bibnamefont
  {Ert{\"o}z}}, \ and\ \bibinfo {author} {\bibfnamefont {V.}~\bibnamefont
  {Kumar}},\ }\bibfield  {title} {\enquote {\bibinfo {title} {The challenges of
  clustering high dimensional data},}\ }in\ \href@noop {} {\emph {\bibinfo
  {booktitle} {New Directions in Statistical Physics}}}\ (\bibinfo  {publisher}
  {Springer},\ \bibinfo {year} {2004})\ pp.\ \bibinfo {pages}
  {273--309}\BibitemShut {NoStop}%
\bibitem [{\citenamefont {Kriegel}, \citenamefont {Kr{\"o}ger},\ and\
  \citenamefont {Zimek}(2009)}]{kriegel2009clustering}%
  \BibitemOpen
  \bibfield  {author} {\bibinfo {author} {\bibfnamefont {H.-P.}\ \bibnamefont
  {Kriegel}}, \bibinfo {author} {\bibfnamefont {P.}~\bibnamefont {Kr{\"o}ger}},
  \ and\ \bibinfo {author} {\bibfnamefont {A.}~\bibnamefont {Zimek}},\
  }\bibfield  {title} {\enquote {\bibinfo {title} {Clustering high-dimensional
  data: A survey on subspace clustering, pattern-based clustering, and
  correlation clustering},}\ }\href@noop {} {\bibfield  {journal} {\bibinfo
  {journal} {ACM Transactions on Knowledge Discovery from Data (TKDD)}\
  }\textbf {\bibinfo {volume} {3}},\ \bibinfo {pages} {1} (\bibinfo {year}
  {2009})}\BibitemShut {NoStop}%
\bibitem [{\citenamefont {Scherer}\ \emph {et~al.}(2015)\citenamefont
  {Scherer}, \citenamefont {Trendelkamp-Schroer}, \citenamefont {Paul},
  \citenamefont {Pérez-Hernández}, \citenamefont {Hoffmann}, \citenamefont
  {Plattner}, \citenamefont {Wehmeyer}, \citenamefont {Prinz},\ and\
  \citenamefont {Noé}}]{scherer2015pyemma}%
  \BibitemOpen
  \bibfield  {author} {\bibinfo {author} {\bibfnamefont {M.~K.}\ \bibnamefont
  {Scherer}}, \bibinfo {author} {\bibfnamefont {B.}~\bibnamefont
  {Trendelkamp-Schroer}}, \bibinfo {author} {\bibfnamefont {F.}~\bibnamefont
  {Paul}}, \bibinfo {author} {\bibfnamefont {G.}~\bibnamefont
  {Pérez-Hernández}}, \bibinfo {author} {\bibfnamefont {M.}~\bibnamefont
  {Hoffmann}}, \bibinfo {author} {\bibfnamefont {N.}~\bibnamefont {Plattner}},
  \bibinfo {author} {\bibfnamefont {C.}~\bibnamefont {Wehmeyer}}, \bibinfo
  {author} {\bibfnamefont {J.-H.}\ \bibnamefont {Prinz}}, \ and\ \bibinfo
  {author} {\bibfnamefont {F.}~\bibnamefont {Noé}},\ }\bibfield  {title}
  {\enquote {\bibinfo {title} {Py{E}{M}{M}{A} 2: a software package for
  estimation, validation, and analysis of {M}arkov models},}\ }\href@noop {}
  {\bibfield  {journal} {\bibinfo  {journal} {Journal of Chemical Theory and
  Computation}\ }\textbf {\bibinfo {volume} {11}},\ \bibinfo {pages}
  {5525--5542} (\bibinfo {year} {2015})}\BibitemShut {NoStop}%
\bibitem [{\citenamefont {Molgedey}\ and\ \citenamefont
  {Schuster}(1994)}]{molgedey1994separation}%
  \BibitemOpen
  \bibfield  {author} {\bibinfo {author} {\bibfnamefont {L.}~\bibnamefont
  {Molgedey}}\ and\ \bibinfo {author} {\bibfnamefont {H.~G.}\ \bibnamefont
  {Schuster}},\ }\bibfield  {title} {\enquote {\bibinfo {title} {Separation of
  a mixture of independent signals using time delayed correlations},}\
  }\href@noop {} {\bibfield  {journal} {\bibinfo  {journal} {Physical Review
  Letters}\ }\textbf {\bibinfo {volume} {72}},\ \bibinfo {pages} {3634}
  (\bibinfo {year} {1994})}\BibitemShut {NoStop}%
\bibitem [{\citenamefont {Takano}\ and\ \citenamefont
  {Miyashita}(1995)}]{takano1995relaxation}%
  \BibitemOpen
  \bibfield  {author} {\bibinfo {author} {\bibfnamefont {H.}~\bibnamefont
  {Takano}}\ and\ \bibinfo {author} {\bibfnamefont {S.}~\bibnamefont
  {Miyashita}},\ }\bibfield  {title} {\enquote {\bibinfo {title} {Relaxation
  modes in random spin systems},}\ }\href@noop {} {\bibfield  {journal}
  {\bibinfo  {journal} {Journal of the Physical Society of Japan}\ }\textbf
  {\bibinfo {volume} {64}},\ \bibinfo {pages} {3688--3698} (\bibinfo {year}
  {1995})}\BibitemShut {NoStop}%
\bibitem [{\citenamefont {Hirao}, \citenamefont {Koseki},\ and\ \citenamefont
  {Takano}(1997)}]{hirao1997molecular}%
  \BibitemOpen
  \bibfield  {author} {\bibinfo {author} {\bibfnamefont {H.}~\bibnamefont
  {Hirao}}, \bibinfo {author} {\bibfnamefont {S.}~\bibnamefont {Koseki}}, \
  and\ \bibinfo {author} {\bibfnamefont {H.}~\bibnamefont {Takano}},\
  }\bibfield  {title} {\enquote {\bibinfo {title} {Molecular dynamics study of
  relaxation modes of a single polymer chain},}\ }\href@noop {} {\bibfield
  {journal} {\bibinfo  {journal} {Journal of the Physical Society of Japan}\
  }\textbf {\bibinfo {volume} {66}},\ \bibinfo {pages} {3399--3405} (\bibinfo
  {year} {1997})}\BibitemShut {NoStop}%
\bibitem [{\citenamefont {Sch{\"u}tte}\ \emph {et~al.}(2011)\citenamefont
  {Sch{\"u}tte}, \citenamefont {No{\'e}}, \citenamefont {Lu}, \citenamefont
  {Sarich},\ and\ \citenamefont {Vanden-Eijnden}}]{schutte2011markov}%
  \BibitemOpen
  \bibfield  {author} {\bibinfo {author} {\bibfnamefont {C.}~\bibnamefont
  {Sch{\"u}tte}}, \bibinfo {author} {\bibfnamefont {F.}~\bibnamefont
  {No{\'e}}}, \bibinfo {author} {\bibfnamefont {J.}~\bibnamefont {Lu}},
  \bibinfo {author} {\bibfnamefont {M.}~\bibnamefont {Sarich}}, \ and\ \bibinfo
  {author} {\bibfnamefont {E.}~\bibnamefont {Vanden-Eijnden}},\ }\bibfield
  {title} {\enquote {\bibinfo {title} {Markov state models based on
  milestoning},}\ }\href@noop {} {\bibfield  {journal} {\bibinfo  {journal}
  {The Journal of Chemical Physics}\ }\textbf {\bibinfo {volume} {134}},\
  \bibinfo {pages} {05B609} (\bibinfo {year} {2011})}\BibitemShut {NoStop}%
\bibitem [{\citenamefont {Giannakis}, \citenamefont {Slawinska},\ and\
  \citenamefont {Zhao}(2015)}]{giannakis2015spatiotemporal}%
  \BibitemOpen
  \bibfield  {author} {\bibinfo {author} {\bibfnamefont {D.}~\bibnamefont
  {Giannakis}}, \bibinfo {author} {\bibfnamefont {J.}~\bibnamefont
  {Slawinska}}, \ and\ \bibinfo {author} {\bibfnamefont {Z.}~\bibnamefont
  {Zhao}},\ }\bibfield  {title} {\enquote {\bibinfo {title} {Spatiotemporal
  feature extraction with data-driven {K}oopman operators},}\ }in\ \href@noop
  {} {\emph {\bibinfo {booktitle} {Feature Extraction: Modern Questions and
  Challenges}}}\ (\bibinfo {year} {2015})\ pp.\ \bibinfo {pages}
  {103--115}\BibitemShut {NoStop}%
\bibitem [{\citenamefont {Giannakis}(2017)}]{giannakis2017data}%
  \BibitemOpen
  \bibfield  {author} {\bibinfo {author} {\bibfnamefont {D.}~\bibnamefont
  {Giannakis}},\ }\bibfield  {title} {\enquote {\bibinfo {title} {Data-driven
  spectral decomposition and forecasting of ergodic dynamical systems},}\
  }\href@noop {} {\bibfield  {journal} {\bibinfo  {journal} {Applied and
  Computational Harmonic Analysis}\ } (\bibinfo {year} {2017})}\BibitemShut
  {NoStop}%
\bibitem [{\citenamefont {Williams}, \citenamefont {Kevrekidis},\ and\
  \citenamefont {Rowley}(2015)}]{williams2015data}%
  \BibitemOpen
  \bibfield  {author} {\bibinfo {author} {\bibfnamefont {M.~O.}\ \bibnamefont
  {Williams}}, \bibinfo {author} {\bibfnamefont {I.~G.}\ \bibnamefont
  {Kevrekidis}}, \ and\ \bibinfo {author} {\bibfnamefont {C.~W.}\ \bibnamefont
  {Rowley}},\ }\bibfield  {title} {\enquote {\bibinfo {title} {A data--driven
  approximation of the {K}oopman operator: Extending dynamic mode
  decomposition},}\ }\href@noop {} {\bibfield  {journal} {\bibinfo  {journal}
  {Journal of Nonlinear Science}\ }\textbf {\bibinfo {volume} {25}},\ \bibinfo
  {pages} {1307--1346} (\bibinfo {year} {2015})}\BibitemShut {NoStop}%
\bibitem [{\citenamefont {Boninsegna}\ \emph {et~al.}(2015)\citenamefont
  {Boninsegna}, \citenamefont {Gobbo}, \citenamefont {No{\'e}},\ and\
  \citenamefont {Clementi}}]{boninsegna2015investigating}%
  \BibitemOpen
  \bibfield  {author} {\bibinfo {author} {\bibfnamefont {L.}~\bibnamefont
  {Boninsegna}}, \bibinfo {author} {\bibfnamefont {G.}~\bibnamefont {Gobbo}},
  \bibinfo {author} {\bibfnamefont {F.}~\bibnamefont {No{\'e}}}, \ and\
  \bibinfo {author} {\bibfnamefont {C.}~\bibnamefont {Clementi}},\ }\bibfield
  {title} {\enquote {\bibinfo {title} {Investigating molecular kinetics by
  variationally optimized diffusion maps},}\ }\href@noop {} {\bibfield
  {journal} {\bibinfo  {journal} {Journal of Chemical Theory and Computation}\
  }\textbf {\bibinfo {volume} {11}},\ \bibinfo {pages} {5947--5960} (\bibinfo
  {year} {2015})}\BibitemShut {NoStop}%
\bibitem [{\citenamefont {Vitalini}, \citenamefont {No{\'e}},\ and\
  \citenamefont {Keller}(2015)}]{vitalini2015basis}%
  \BibitemOpen
  \bibfield  {author} {\bibinfo {author} {\bibfnamefont {F.}~\bibnamefont
  {Vitalini}}, \bibinfo {author} {\bibfnamefont {F.}~\bibnamefont {No{\'e}}}, \
  and\ \bibinfo {author} {\bibfnamefont {B.}~\bibnamefont {Keller}},\
  }\bibfield  {title} {\enquote {\bibinfo {title} {A basis set for peptides for
  the variational approach to conformational kinetics},}\ }\href@noop {}
  {\bibfield  {journal} {\bibinfo  {journal} {Journal of Chemical Theory and
  Computation}\ }\textbf {\bibinfo {volume} {11}},\ \bibinfo {pages}
  {3992--4004} (\bibinfo {year} {2015})}\BibitemShut {NoStop}%
\bibitem [{\citenamefont {Wu}\ and\ \citenamefont
  {No{\'e}}(2017)}]{wu2017vamp}%
  \BibitemOpen
  \bibfield  {author} {\bibinfo {author} {\bibfnamefont {H.}~\bibnamefont
  {Wu}}\ and\ \bibinfo {author} {\bibfnamefont {F.}~\bibnamefont {No{\'e}}},\
  }\bibfield  {title} {\enquote {\bibinfo {title} {Variational approach for
  learning {M}arkov processes from time series data},}\ }\href@noop {}
  {\bibfield  {journal} {\bibinfo  {journal} {arXiv preprint arXiv:1707.04659}\
  } (\bibinfo {year} {2017})}\BibitemShut {NoStop}%
\bibitem [{\citenamefont {Mardt}\ \emph {et~al.}(2018)\citenamefont {Mardt},
  \citenamefont {Pasquali}, \citenamefont {Wu},\ and\ \citenamefont
  {No{\'e}}}]{mardt2017vampnets}%
  \BibitemOpen
  \bibfield  {author} {\bibinfo {author} {\bibfnamefont {A.}~\bibnamefont
  {Mardt}}, \bibinfo {author} {\bibfnamefont {L.}~\bibnamefont {Pasquali}},
  \bibinfo {author} {\bibfnamefont {H.}~\bibnamefont {Wu}}, \ and\ \bibinfo
  {author} {\bibfnamefont {F.}~\bibnamefont {No{\'e}}},\ }\bibfield  {title}
  {\enquote {\bibinfo {title} {{V}{A}{M}{P}nets for deep learning of molecular
  kinetics},}\ }\href@noop {} {\bibfield  {journal} {\bibinfo  {journal}
  {Nature Communications}\ }\textbf {\bibinfo {volume} {9}},\ \bibinfo {pages}
  {5} (\bibinfo {year} {2018})}\BibitemShut {NoStop}%
\bibitem [{\citenamefont {N{\"u}ske}\ \emph {et~al.}(2014)\citenamefont
  {N{\"u}ske}, \citenamefont {Keller}, \citenamefont {P{\'e}rez-Hern{\'a}ndez},
  \citenamefont {Mey},\ and\ \citenamefont {No{\'e}}}]{nüske2014variational}%
  \BibitemOpen
  \bibfield  {author} {\bibinfo {author} {\bibfnamefont {F.}~\bibnamefont
  {N{\"u}ske}}, \bibinfo {author} {\bibfnamefont {B.~G.}\ \bibnamefont
  {Keller}}, \bibinfo {author} {\bibfnamefont {G.}~\bibnamefont
  {P{\'e}rez-Hern{\'a}ndez}}, \bibinfo {author} {\bibfnamefont {A.~S.}\
  \bibnamefont {Mey}}, \ and\ \bibinfo {author} {\bibfnamefont
  {F.}~\bibnamefont {No{\'e}}},\ }\bibfield  {title} {\enquote {\bibinfo
  {title} {Variational approach to molecular kinetics},}\ }\href@noop {}
  {\bibfield  {journal} {\bibinfo  {journal} {Journal of Chemical Theory and
  Computation}\ }\textbf {\bibinfo {volume} {10}},\ \bibinfo {pages}
  {1739--1752} (\bibinfo {year} {2014})}\BibitemShut {NoStop}%
\bibitem [{\citenamefont {N{\"u}ske}\ \emph {et~al.}(2016)\citenamefont
  {N{\"u}ske}, \citenamefont {Schneider}, \citenamefont {Vitalini},\ and\
  \citenamefont {No{\'e}}}]{nuske2016variational}%
  \BibitemOpen
  \bibfield  {author} {\bibinfo {author} {\bibfnamefont {F.}~\bibnamefont
  {N{\"u}ske}}, \bibinfo {author} {\bibfnamefont {R.}~\bibnamefont
  {Schneider}}, \bibinfo {author} {\bibfnamefont {F.}~\bibnamefont {Vitalini}},
  \ and\ \bibinfo {author} {\bibfnamefont {F.}~\bibnamefont {No{\'e}}},\
  }\bibfield  {title} {\enquote {\bibinfo {title} {Variational tensor approach
  for approximating the rare-event kinetics of macromolecular systems},}\
  }\href@noop {} {\bibfield  {journal} {\bibinfo  {journal} {The Journal of
  Chemical Physics}\ }\textbf {\bibinfo {volume} {144}},\ \bibinfo {pages}
  {054105} (\bibinfo {year} {2016})}\BibitemShut {NoStop}%
\bibitem [{\citenamefont {Prinz}, \citenamefont {Chodera},\ and\ \citenamefont
  {No{\'e}}(2014)}]{prinz2014spectral}%
  \BibitemOpen
  \bibfield  {author} {\bibinfo {author} {\bibfnamefont {J.-H.}\ \bibnamefont
  {Prinz}}, \bibinfo {author} {\bibfnamefont {J.~D.}\ \bibnamefont {Chodera}},
  \ and\ \bibinfo {author} {\bibfnamefont {F.}~\bibnamefont {No{\'e}}},\
  }\bibfield  {title} {\enquote {\bibinfo {title} {Spectral rate theory for
  two-state kinetics},}\ }\href@noop {} {\bibfield  {journal} {\bibinfo
  {journal} {Physical Review X}\ }\textbf {\bibinfo {volume} {4}},\ \bibinfo
  {pages} {011020} (\bibinfo {year} {2014})}\BibitemShut {NoStop}%
\bibitem [{\citenamefont {Del~Moral}(2004)}]{delmoral2004feynman}%
  \BibitemOpen
  \bibfield  {author} {\bibinfo {author} {\bibfnamefont {P.}~\bibnamefont
  {Del~Moral}},\ }\href@noop {} {\emph {\bibinfo {title} {Feynman-Kac
  Formulae}}}\ (\bibinfo  {publisher} {Springer},\ \bibinfo {year}
  {2004})\BibitemShut {NoStop}%
\bibitem [{\citenamefont {Karatzas}\ and\ \citenamefont
  {Shreve}(2012)}]{karatzas2012brownian}%
  \BibitemOpen
  \bibfield  {author} {\bibinfo {author} {\bibfnamefont {I.}~\bibnamefont
  {Karatzas}}\ and\ \bibinfo {author} {\bibfnamefont {S.}~\bibnamefont
  {Shreve}},\ }\href@noop {} {\emph {\bibinfo {title} {Brownian Motion and
  Stochastic Calculus}}},\ Vol.\ \bibinfo {volume} {113}\ (\bibinfo
  {publisher} {Springer Science \& Business Media},\ \bibinfo {year}
  {2012})\BibitemShut {NoStop}%
\bibitem [{\citenamefont {Du}\ \emph {et~al.}(1998)\citenamefont {Du},
  \citenamefont {Pande}, \citenamefont {Grosberg}, \citenamefont {Tanaka},\
  and\ \citenamefont {Shakhnovich}}]{du1998transition}%
  \BibitemOpen
  \bibfield  {author} {\bibinfo {author} {\bibfnamefont {R.}~\bibnamefont
  {Du}}, \bibinfo {author} {\bibfnamefont {V.~S.}\ \bibnamefont {Pande}},
  \bibinfo {author} {\bibfnamefont {A.~Y.}\ \bibnamefont {Grosberg}}, \bibinfo
  {author} {\bibfnamefont {T.}~\bibnamefont {Tanaka}}, \ and\ \bibinfo {author}
  {\bibfnamefont {E.~S.}\ \bibnamefont {Shakhnovich}},\ }\bibfield  {title}
  {\enquote {\bibinfo {title} {On the transition coordinate for protein
  folding},}\ }\href@noop {} {\bibfield  {journal} {\bibinfo  {journal} {The
  Journal of Chemical Physics}\ }\textbf {\bibinfo {volume} {108}},\ \bibinfo
  {pages} {334--350} (\bibinfo {year} {1998})}\BibitemShut {NoStop}%
\bibitem [{\citenamefont {Bolhuis}, \citenamefont {Dellago},\ and\
  \citenamefont {Chandler}(2000)}]{bolhuis2000reaction}%
  \BibitemOpen
  \bibfield  {author} {\bibinfo {author} {\bibfnamefont {P.~G.}\ \bibnamefont
  {Bolhuis}}, \bibinfo {author} {\bibfnamefont {C.}~\bibnamefont {Dellago}}, \
  and\ \bibinfo {author} {\bibfnamefont {D.}~\bibnamefont {Chandler}},\
  }\bibfield  {title} {\enquote {\bibinfo {title} {Reaction coordinates of
  biomolecular isomerization},}\ }\href@noop {} {\bibfield  {journal} {\bibinfo
   {journal} {Proceedings of the National Academy of Sciences}\ }\textbf
  {\bibinfo {volume} {97}},\ \bibinfo {pages} {5877--5882} (\bibinfo {year}
  {2000})}\BibitemShut {NoStop}%
\bibitem [{\citenamefont {Metzner}, \citenamefont {Sch{\"u}tte},\ and\
  \citenamefont {Vanden-Eijnden}(2009)}]{metzner2009transition}%
  \BibitemOpen
  \bibfield  {author} {\bibinfo {author} {\bibfnamefont {P.}~\bibnamefont
  {Metzner}}, \bibinfo {author} {\bibfnamefont {C.}~\bibnamefont
  {Sch{\"u}tte}}, \ and\ \bibinfo {author} {\bibfnamefont {E.}~\bibnamefont
  {Vanden-Eijnden}},\ }\bibfield  {title} {\enquote {\bibinfo {title}
  {Transition path theory for {M}arkov jump processes},}\ }\href@noop {}
  {\bibfield  {journal} {\bibinfo  {journal} {Multiscale Modeling \&
  Simulation}\ }\textbf {\bibinfo {volume} {7}},\ \bibinfo {pages} {1192--1219}
  (\bibinfo {year} {2009})}\BibitemShut {NoStop}%
\bibitem [{\citenamefont {Yosida}(1971)}]{yosida1971functional}%
  \BibitemOpen
  \bibfield  {author} {\bibinfo {author} {\bibfnamefont {K.}~\bibnamefont
  {Yosida}},\ }\bibfield  {title} {\enquote {\bibinfo {title} {Functional
  analysis, 1980},}\ }\href@noop {} {\bibfield  {journal} {\bibinfo  {journal}
  {Spring-Verlag, New York/Berlin}\ } (\bibinfo {year} {1971})}\BibitemShut
  {NoStop}%
\bibitem [{\citenamefont {Lapelosa}\ and\ \citenamefont
  {Abrams}(2013)}]{lapelosa2013transition}%
  \BibitemOpen
  \bibfield  {author} {\bibinfo {author} {\bibfnamefont {M.}~\bibnamefont
  {Lapelosa}}\ and\ \bibinfo {author} {\bibfnamefont {C.~F.}\ \bibnamefont
  {Abrams}},\ }\bibfield  {title} {\enquote {\bibinfo {title} {Transition-path
  theory calculations on non-uniform meshes in two and three dimensions using
  finite elements},}\ }\href@noop {} {\bibfield  {journal} {\bibinfo  {journal}
  {Computer Physics Communications}\ }\textbf {\bibinfo {volume} {184}},\
  \bibinfo {pages} {2310--2315} (\bibinfo {year} {2013})}\BibitemShut {NoStop}%
\bibitem [{\citenamefont {Lai}\ and\ \citenamefont {Lu}(2018)}]{lai2018point}%
  \BibitemOpen
  \bibfield  {author} {\bibinfo {author} {\bibfnamefont {R.}~\bibnamefont
  {Lai}}\ and\ \bibinfo {author} {\bibfnamefont {J.}~\bibnamefont {Lu}},\
  }\bibfield  {title} {\enquote {\bibinfo {title} {Point cloud discretization
  of {F}okker--{P}lanck operators for committor functions},}\ }\href@noop {}
  {\bibfield  {journal} {\bibinfo  {journal} {Multiscale Modeling \&
  Simulation}\ }\textbf {\bibinfo {volume} {16}},\ \bibinfo {pages} {710--726}
  (\bibinfo {year} {2018})}\BibitemShut {NoStop}%
\bibitem [{\citenamefont {Khoo}, \citenamefont {Lu},\ and\ \citenamefont
  {Ying}(2018)}]{khoo2018solving}%
  \BibitemOpen
  \bibfield  {author} {\bibinfo {author} {\bibfnamefont {Y.}~\bibnamefont
  {Khoo}}, \bibinfo {author} {\bibfnamefont {J.}~\bibnamefont {Lu}}, \ and\
  \bibinfo {author} {\bibfnamefont {L.}~\bibnamefont {Ying}},\ }\bibfield
  {title} {\enquote {\bibinfo {title} {Solving for high dimensional committor
  functions using artificial neural networks},}\ }\href@noop {} {\bibfield
  {journal} {\bibinfo  {journal} {arXiv preprint arXiv:1802.10275}\ } (\bibinfo
  {year} {2018})}\BibitemShut {NoStop}%
\bibitem [{\citenamefont {Evans}(1998)}]{evans1998partial}%
  \BibitemOpen
  \bibfield  {author} {\bibinfo {author} {\bibfnamefont {L.}~\bibnamefont
  {Evans}},\ }\href@noop {} {\emph {\bibinfo {title} {Partial Differential
  Equations}}}\ (\bibinfo  {publisher} {Orient Longman},\ \bibinfo {year}
  {1998})\BibitemShut {NoStop}%
\bibitem [{\citenamefont {Thiede}(2018)}]{pyedgar}%
  \BibitemOpen
  \bibfield  {author} {\bibinfo {author} {\bibfnamefont {E.}~\bibnamefont
  {Thiede}},\ }\href@noop {} {\enquote {\bibinfo {title} {Py{EDGAR}},}\
  }\bibinfo {howpublished} {\url{https://github.com/ehthiede/PyEDGAR/}}
  (\bibinfo {year} {2018})\BibitemShut {NoStop}%
\bibitem [{\citenamefont {Wu}\ \emph {et~al.}(2017)\citenamefont {Wu},
  \citenamefont {N{\"u}ske}, \citenamefont {Paul}, \citenamefont {Klus},
  \citenamefont {Koltai},\ and\ \citenamefont
  {No{\'e}}}]{wu2017variationalkoopman}%
  \BibitemOpen
  \bibfield  {author} {\bibinfo {author} {\bibfnamefont {H.}~\bibnamefont
  {Wu}}, \bibinfo {author} {\bibfnamefont {F.}~\bibnamefont {N{\"u}ske}},
  \bibinfo {author} {\bibfnamefont {F.}~\bibnamefont {Paul}}, \bibinfo {author}
  {\bibfnamefont {S.}~\bibnamefont {Klus}}, \bibinfo {author} {\bibfnamefont
  {P.}~\bibnamefont {Koltai}}, \ and\ \bibinfo {author} {\bibfnamefont
  {F.}~\bibnamefont {No{\'e}}},\ }\bibfield  {title} {\enquote {\bibinfo
  {title} {Variational {K}oopman models: slow collective variables and
  molecular kinetics from short off-equilibrium simulations},}\ }\href@noop {}
  {\bibfield  {journal} {\bibinfo  {journal} {The Journal of Chemical Physics}\
  }\textbf {\bibinfo {volume} {146}},\ \bibinfo {pages} {154104} (\bibinfo
  {year} {2017})}\BibitemShut {NoStop}%
\bibitem [{\citenamefont {Chen}, \citenamefont {Tu},\ and\ \citenamefont
  {Rowley}(2012)}]{chen2012variants}%
  \BibitemOpen
  \bibfield  {author} {\bibinfo {author} {\bibfnamefont {K.~K.}\ \bibnamefont
  {Chen}}, \bibinfo {author} {\bibfnamefont {J.~H.}\ \bibnamefont {Tu}}, \ and\
  \bibinfo {author} {\bibfnamefont {C.~W.}\ \bibnamefont {Rowley}},\ }\bibfield
   {title} {\enquote {\bibinfo {title} {Variants of dynamic mode decomposition:
  boundary condition, {K}oopman, and {F}ourier analyses},}\ }\href@noop {}
  {\bibfield  {journal} {\bibinfo  {journal} {Journal of nonlinear science}\
  }\textbf {\bibinfo {volume} {22}},\ \bibinfo {pages} {887--915} (\bibinfo
  {year} {2012})}\BibitemShut {NoStop}%
\bibitem [{\citenamefont {Coifman}\ and\ \citenamefont
  {Lafon}(2006)}]{coifman2006diffusion}%
  \BibitemOpen
  \bibfield  {author} {\bibinfo {author} {\bibfnamefont {R.~R.}\ \bibnamefont
  {Coifman}}\ and\ \bibinfo {author} {\bibfnamefont {S.}~\bibnamefont
  {Lafon}},\ }\bibfield  {title} {\enquote {\bibinfo {title} {Diffusion
  maps},}\ }\href@noop {} {\bibfield  {journal} {\bibinfo  {journal} {Applied
  and Computational Harmonic Analysis}\ }\textbf {\bibinfo {volume} {21}},\
  \bibinfo {pages} {5--30} (\bibinfo {year} {2006})}\BibitemShut {NoStop}%
\bibitem [{\citenamefont {Berry}\ and\ \citenamefont
  {Harlim}(2016)}]{berry2016variable}%
  \BibitemOpen
  \bibfield  {author} {\bibinfo {author} {\bibfnamefont {T.}~\bibnamefont
  {Berry}}\ and\ \bibinfo {author} {\bibfnamefont {J.}~\bibnamefont {Harlim}},\
  }\bibfield  {title} {\enquote {\bibinfo {title} {Variable bandwidth diffusion
  kernels},}\ }\href@noop {} {\bibfield  {journal} {\bibinfo  {journal}
  {Applied and Computational Harmonic Analysis}\ }\textbf {\bibinfo {volume}
  {40}},\ \bibinfo {pages} {68--96} (\bibinfo {year} {2016})}\BibitemShut
  {NoStop}%
\bibitem [{\citenamefont {Ferguson}\ \emph {et~al.}(2010)\citenamefont
  {Ferguson}, \citenamefont {Panagiotopoulos}, \citenamefont {Debenedetti},\
  and\ \citenamefont {Kevrekidis}}]{ferguson2010systematic}%
  \BibitemOpen
  \bibfield  {author} {\bibinfo {author} {\bibfnamefont {A.~L.}\ \bibnamefont
  {Ferguson}}, \bibinfo {author} {\bibfnamefont {A.~Z.}\ \bibnamefont
  {Panagiotopoulos}}, \bibinfo {author} {\bibfnamefont {P.~G.}\ \bibnamefont
  {Debenedetti}}, \ and\ \bibinfo {author} {\bibfnamefont {I.~G.}\ \bibnamefont
  {Kevrekidis}},\ }\bibfield  {title} {\enquote {\bibinfo {title} {Systematic
  determination of order parameters for chain dynamics using diffusion maps},}\
  }\href@noop {} {\bibfield  {journal} {\bibinfo  {journal} {Proceedings of the
  National Academy of Sciences}\ }\textbf {\bibinfo {volume} {107}},\ \bibinfo
  {pages} {13597--13602} (\bibinfo {year} {2010})}\BibitemShut {NoStop}%
\bibitem [{\citenamefont {Rohrdanz}\ \emph {et~al.}(2011)\citenamefont
  {Rohrdanz}, \citenamefont {Zheng}, \citenamefont {Maggioni},\ and\
  \citenamefont {Clementi}}]{rohrdanz2011determination}%
  \BibitemOpen
  \bibfield  {author} {\bibinfo {author} {\bibfnamefont {M.~A.}\ \bibnamefont
  {Rohrdanz}}, \bibinfo {author} {\bibfnamefont {W.}~\bibnamefont {Zheng}},
  \bibinfo {author} {\bibfnamefont {M.}~\bibnamefont {Maggioni}}, \ and\
  \bibinfo {author} {\bibfnamefont {C.}~\bibnamefont {Clementi}},\ }\bibfield
  {title} {\enquote {\bibinfo {title} {Determination of reaction coordinates
  via locally scaled diffusion map},}\ }\href@noop {} {\bibfield  {journal}
  {\bibinfo  {journal} {The Journal of Chemical Physics}\ }\textbf {\bibinfo
  {volume} {134}},\ \bibinfo {pages} {03B624} (\bibinfo {year}
  {2011})}\BibitemShut {NoStop}%
\bibitem [{\citenamefont {Zheng}\ \emph {et~al.}(2011)\citenamefont {Zheng},
  \citenamefont {Qi}, \citenamefont {Rohrdanz}, \citenamefont {Caflisch},
  \citenamefont {Dinner},\ and\ \citenamefont
  {Clementi}}]{zheng2011delineation}%
  \BibitemOpen
  \bibfield  {author} {\bibinfo {author} {\bibfnamefont {W.}~\bibnamefont
  {Zheng}}, \bibinfo {author} {\bibfnamefont {B.}~\bibnamefont {Qi}}, \bibinfo
  {author} {\bibfnamefont {M.~A.}\ \bibnamefont {Rohrdanz}}, \bibinfo {author}
  {\bibfnamefont {A.}~\bibnamefont {Caflisch}}, \bibinfo {author}
  {\bibfnamefont {A.~R.}\ \bibnamefont {Dinner}}, \ and\ \bibinfo {author}
  {\bibfnamefont {C.}~\bibnamefont {Clementi}},\ }\bibfield  {title} {\enquote
  {\bibinfo {title} {Delineation of folding pathways of a $\beta$-sheet
  miniprotein},}\ }\href@noop {} {\bibfield  {journal} {\bibinfo  {journal}
  {The Journal of Physical Chemistry B}\ }\textbf {\bibinfo {volume} {115}},\
  \bibinfo {pages} {13065--13074} (\bibinfo {year} {2011})}\BibitemShut
  {NoStop}%
\bibitem [{\citenamefont {Ferguson}\ \emph {et~al.}(2011)\citenamefont
  {Ferguson}, \citenamefont {Panagiotopoulos}, \citenamefont {Kevrekidis},\
  and\ \citenamefont {Debenedetti}}]{ferguson2011nonlinear}%
  \BibitemOpen
  \bibfield  {author} {\bibinfo {author} {\bibfnamefont {A.~L.}\ \bibnamefont
  {Ferguson}}, \bibinfo {author} {\bibfnamefont {A.~Z.}\ \bibnamefont
  {Panagiotopoulos}}, \bibinfo {author} {\bibfnamefont {I.~G.}\ \bibnamefont
  {Kevrekidis}}, \ and\ \bibinfo {author} {\bibfnamefont {P.~G.}\ \bibnamefont
  {Debenedetti}},\ }\bibfield  {title} {\enquote {\bibinfo {title} {Nonlinear
  dimensionality reduction in molecular simulation: The diffusion map
  approach},}\ }\href@noop {} {\bibfield  {journal} {\bibinfo  {journal}
  {Chemical Physics Letters}\ }\textbf {\bibinfo {volume} {509}},\ \bibinfo
  {pages} {1--11} (\bibinfo {year} {2011})}\BibitemShut {NoStop}%
\bibitem [{\citenamefont {Long}\ and\ \citenamefont
  {Ferguson}(2014)}]{long2014nonlinear}%
  \BibitemOpen
  \bibfield  {author} {\bibinfo {author} {\bibfnamefont {A.~W.}\ \bibnamefont
  {Long}}\ and\ \bibinfo {author} {\bibfnamefont {A.~L.}\ \bibnamefont
  {Ferguson}},\ }\bibfield  {title} {\enquote {\bibinfo {title} {Nonlinear
  machine learning of patchy colloid self-assembly pathways and mechanisms},}\
  }\href@noop {} {\bibfield  {journal} {\bibinfo  {journal} {The Journal of
  Physical Chemistry B}\ }\textbf {\bibinfo {volume} {118}},\ \bibinfo {pages}
  {4228--4244} (\bibinfo {year} {2014})}\BibitemShut {NoStop}%
\bibitem [{\citenamefont {Kim}\ \emph {et~al.}(2015)\citenamefont {Kim},
  \citenamefont {Dsilva}, \citenamefont {Kevrekidis},\ and\ \citenamefont
  {Debenedetti}}]{kim2015systematic}%
  \BibitemOpen
  \bibfield  {author} {\bibinfo {author} {\bibfnamefont {S.~B.}\ \bibnamefont
  {Kim}}, \bibinfo {author} {\bibfnamefont {C.~J.}\ \bibnamefont {Dsilva}},
  \bibinfo {author} {\bibfnamefont {I.~G.}\ \bibnamefont {Kevrekidis}}, \ and\
  \bibinfo {author} {\bibfnamefont {P.~G.}\ \bibnamefont {Debenedetti}},\
  }\bibfield  {title} {\enquote {\bibinfo {title} {Systematic characterization
  of protein folding pathways using diffusion maps: Application to trp-cage
  miniprotein},}\ }\href@noop {} {\bibfield  {journal} {\bibinfo  {journal}
  {The Journal of Chemical Physics}\ }\textbf {\bibinfo {volume} {142}},\
  \bibinfo {pages} {02B613\_1} (\bibinfo {year} {2015})}\BibitemShut {NoStop}%
\bibitem [{\citenamefont {Berry}, \citenamefont {Giannakis},\ and\
  \citenamefont {Harlim}(2015)}]{berry2015nonparametric}%
  \BibitemOpen
  \bibfield  {author} {\bibinfo {author} {\bibfnamefont {T.}~\bibnamefont
  {Berry}}, \bibinfo {author} {\bibfnamefont {D.}~\bibnamefont {Giannakis}}, \
  and\ \bibinfo {author} {\bibfnamefont {J.}~\bibnamefont {Harlim}},\
  }\bibfield  {title} {\enquote {\bibinfo {title} {Nonparametric forecasting of
  low-dimensional dynamical systems},}\ }\href@noop {} {\bibfield  {journal}
  {\bibinfo  {journal} {Physical Review E}\ }\textbf {\bibinfo {volume} {91}},\
  \bibinfo {pages} {032915} (\bibinfo {year} {2015})}\BibitemShut {NoStop}%
\bibitem [{\citenamefont {M{\"u}ller}\ and\ \citenamefont
  {Brown}(1979)}]{muller1979location}%
  \BibitemOpen
  \bibfield  {author} {\bibinfo {author} {\bibfnamefont {K.}~\bibnamefont
  {M{\"u}ller}}\ and\ \bibinfo {author} {\bibfnamefont {L.~D.}\ \bibnamefont
  {Brown}},\ }\bibfield  {title} {\enquote {\bibinfo {title} {Location of
  saddle points and minimum energy paths by a constrained simplex optimization
  procedure},}\ }\href@noop {} {\bibfield  {journal} {\bibinfo  {journal}
  {Theoretica Chimica Acta}\ }\textbf {\bibinfo {volume} {53}},\ \bibinfo
  {pages} {75--93} (\bibinfo {year} {1979})}\BibitemShut {NoStop}%
\bibitem [{\citenamefont {Leimkuhler}\ and\ \citenamefont
  {Matthews}(2012)}]{leimkuhler2012rational}%
  \BibitemOpen
  \bibfield  {author} {\bibinfo {author} {\bibfnamefont {B.}~\bibnamefont
  {Leimkuhler}}\ and\ \bibinfo {author} {\bibfnamefont {C.}~\bibnamefont
  {Matthews}},\ }\bibfield  {title} {\enquote {\bibinfo {title} {Rational
  construction of stochastic numerical methods for molecular sampling},}\
  }\href@noop {} {\bibfield  {journal} {\bibinfo  {journal} {Applied
  Mathematics Research eXpress}\ }\textbf {\bibinfo {volume} {2013}},\ \bibinfo
  {pages} {34--56} (\bibinfo {year} {2012})}\BibitemShut {NoStop}%
\bibitem [{\citenamefont {Beauchamp}\ \emph {et~al.}(2011)\citenamefont
  {Beauchamp}, \citenamefont {Bowman}, \citenamefont {Lane}, \citenamefont
  {Maibaum}, \citenamefont {Haque},\ and\ \citenamefont
  {Pande}}]{beauchamp2011msmbuilder2}%
  \BibitemOpen
  \bibfield  {author} {\bibinfo {author} {\bibfnamefont {K.~A.}\ \bibnamefont
  {Beauchamp}}, \bibinfo {author} {\bibfnamefont {G.~R.}\ \bibnamefont
  {Bowman}}, \bibinfo {author} {\bibfnamefont {T.~J.}\ \bibnamefont {Lane}},
  \bibinfo {author} {\bibfnamefont {L.}~\bibnamefont {Maibaum}}, \bibinfo
  {author} {\bibfnamefont {I.~S.}\ \bibnamefont {Haque}}, \ and\ \bibinfo
  {author} {\bibfnamefont {V.~S.}\ \bibnamefont {Pande}},\ }\bibfield  {title}
  {\enquote {\bibinfo {title} {{MSMB}uilder2: Modeling conformational dynamics
  on the picosecond to millisecond scale},}\ }\href@noop {} {\bibfield
  {journal} {\bibinfo  {journal} {Journal of Chemical Theory and Computation}\
  }\textbf {\bibinfo {volume} {7}},\ \bibinfo {pages} {3412--3419} (\bibinfo
  {year} {2011})}\BibitemShut {NoStop}%
\bibitem [{\citenamefont {Su{\'a}rez}, \citenamefont {Adelman},\ and\
  \citenamefont {Zuckerman}(2016)}]{suarez2016accurate}%
  \BibitemOpen
  \bibfield  {author} {\bibinfo {author} {\bibfnamefont {E.}~\bibnamefont
  {Su{\'a}rez}}, \bibinfo {author} {\bibfnamefont {J.~L.}\ \bibnamefont
  {Adelman}}, \ and\ \bibinfo {author} {\bibfnamefont {D.~M.}\ \bibnamefont
  {Zuckerman}},\ }\bibfield  {title} {\enquote {\bibinfo {title} {Accurate
  estimation of protein folding and unfolding times: beyond {M}arkov state
  models},}\ }\href@noop {} {\bibfield  {journal} {\bibinfo  {journal} {Journal
  of Chemical Theory and Computation}\ }\textbf {\bibinfo {volume} {12}},\
  \bibinfo {pages} {3473--3481} (\bibinfo {year} {2016})}\BibitemShut {NoStop}%
\bibitem [{\citenamefont {Djurdjevac}, \citenamefont {Sarich},\ and\
  \citenamefont {Sch{\"u}tte}(2012)}]{djurdjevac2012estimating}%
  \BibitemOpen
  \bibfield  {author} {\bibinfo {author} {\bibfnamefont {N.}~\bibnamefont
  {Djurdjevac}}, \bibinfo {author} {\bibfnamefont {M.}~\bibnamefont {Sarich}},
  \ and\ \bibinfo {author} {\bibfnamefont {C.}~\bibnamefont {Sch{\"u}tte}},\
  }\bibfield  {title} {\enquote {\bibinfo {title} {Estimating the eigenvalue
  error of {M}arkov state models},}\ }\href@noop {} {\bibfield  {journal}
  {\bibinfo  {journal} {Multiscale Modeling \& Simulation}\ }\textbf {\bibinfo
  {volume} {10}},\ \bibinfo {pages} {61--81} (\bibinfo {year}
  {2012})}\BibitemShut {NoStop}%
\bibitem [{\citenamefont {Zwanzig}(2001)}]{zwanzig2001nonequilibrium}%
  \BibitemOpen
  \bibfield  {author} {\bibinfo {author} {\bibfnamefont {R.}~\bibnamefont
  {Zwanzig}},\ }\href@noop {} {\emph {\bibinfo {title} {Nonequilibrium
  {S}tatistical {M}echanics}}}\ (\bibinfo  {publisher} {Oxford {U}niversity
  Press},\ \bibinfo {year} {2001})\BibitemShut {NoStop}%
\bibitem [{\citenamefont {Takens}(1981)}]{takens1981detecting}%
  \BibitemOpen
  \bibfield  {author} {\bibinfo {author} {\bibfnamefont {F.}~\bibnamefont
  {Takens}},\ }\bibfield  {title} {\enquote {\bibinfo {title} {Detecting
  strange attractors in turbulence},}\ }\href@noop {} {\bibfield  {journal}
  {\bibinfo  {journal} {Lecture Notes in Mathematics}\ }\textbf {\bibinfo
  {volume} {898}},\ \bibinfo {pages} {366--381} (\bibinfo {year}
  {1981})}\BibitemShut {NoStop}%
\bibitem [{\citenamefont {Aeyels}(1981)}]{aeyels1981generic}%
  \BibitemOpen
  \bibfield  {author} {\bibinfo {author} {\bibfnamefont {D.}~\bibnamefont
  {Aeyels}},\ }\bibfield  {title} {\enquote {\bibinfo {title} {Generic
  observability of differentiable systems},}\ }\href@noop {} {\bibfield
  {journal} {\bibinfo  {journal} {SIAM Journal on Control and Optimization}\
  }\textbf {\bibinfo {volume} {19}},\ \bibinfo {pages} {595--603} (\bibinfo
  {year} {1981})}\BibitemShut {NoStop}%
\bibitem [{\citenamefont {Muldoon}\ \emph {et~al.}(1998)\citenamefont
  {Muldoon}, \citenamefont {Broomhead}, \citenamefont {Huke},\ and\
  \citenamefont {Hegger}}]{muldoon1998delay}%
  \BibitemOpen
  \bibfield  {author} {\bibinfo {author} {\bibfnamefont {M.~R.}\ \bibnamefont
  {Muldoon}}, \bibinfo {author} {\bibfnamefont {D.~S.}\ \bibnamefont
  {Broomhead}}, \bibinfo {author} {\bibfnamefont {J.~P.}\ \bibnamefont {Huke}},
  \ and\ \bibinfo {author} {\bibfnamefont {R.}~\bibnamefont {Hegger}},\
  }\bibfield  {title} {\enquote {\bibinfo {title} {Delay embedding in the
  presence of dynamical noise},}\ }\href@noop {} {\bibfield  {journal}
  {\bibinfo  {journal} {Dynamics and Stability of Systems}\ }\textbf {\bibinfo
  {volume} {13}},\ \bibinfo {pages} {175--186} (\bibinfo {year}
  {1998})}\BibitemShut {NoStop}%
\bibitem [{\citenamefont {Stark}\ \emph {et~al.}(1997)\citenamefont {Stark},
  \citenamefont {Broomhead}, \citenamefont {Davies},\ and\ \citenamefont
  {Huke}}]{stark1997takens}%
  \BibitemOpen
  \bibfield  {author} {\bibinfo {author} {\bibfnamefont {J.}~\bibnamefont
  {Stark}}, \bibinfo {author} {\bibfnamefont {D.}~\bibnamefont {Broomhead}},
  \bibinfo {author} {\bibfnamefont {M.}~\bibnamefont {Davies}}, \ and\ \bibinfo
  {author} {\bibfnamefont {J.}~\bibnamefont {Huke}},\ }\bibfield  {title}
  {\enquote {\bibinfo {title} {Takens embedding theorems for forced and
  stochastic systems},}\ }\href@noop {} {\bibfield  {journal} {\bibinfo
  {journal} {Nonlinear Analysis: Theory, Methods \& Applications}\ }\textbf
  {\bibinfo {volume} {30}},\ \bibinfo {pages} {5303--5314} (\bibinfo {year}
  {1997})}\BibitemShut {NoStop}%
\bibitem [{\citenamefont {Berry}\ \emph {et~al.}(2013)\citenamefont {Berry},
  \citenamefont {Cressman}, \citenamefont {Greguric-Ferencek},\ and\
  \citenamefont {Sauer}}]{berry2013time}%
  \BibitemOpen
  \bibfield  {author} {\bibinfo {author} {\bibfnamefont {T.}~\bibnamefont
  {Berry}}, \bibinfo {author} {\bibfnamefont {J.~R.}\ \bibnamefont {Cressman}},
  \bibinfo {author} {\bibfnamefont {Z.}~\bibnamefont {Greguric-Ferencek}}, \
  and\ \bibinfo {author} {\bibfnamefont {T.}~\bibnamefont {Sauer}},\ }\bibfield
   {title} {\enquote {\bibinfo {title} {Time-scale separation from
  diffusion-mapped delay coordinates},}\ }\href@noop {} {\bibfield  {journal}
  {\bibinfo  {journal} {SIAM Journal on Applied Dynamical Systems}\ }\textbf
  {\bibinfo {volume} {12}},\ \bibinfo {pages} {618--649} (\bibinfo {year}
  {2013})}\BibitemShut {NoStop}%
\bibitem [{\citenamefont {Wang}\ and\ \citenamefont
  {Ferguson}(2016)}]{wang2016nonlinear}%
  \BibitemOpen
  \bibfield  {author} {\bibinfo {author} {\bibfnamefont {J.}~\bibnamefont
  {Wang}}\ and\ \bibinfo {author} {\bibfnamefont {A.~L.}\ \bibnamefont
  {Ferguson}},\ }\bibfield  {title} {\enquote {\bibinfo {title} {Nonlinear
  reconstruction of single-molecule free-energy surfaces from univariate time
  series},}\ }\href@noop {} {\bibfield  {journal} {\bibinfo  {journal}
  {Physical Review E}\ }\textbf {\bibinfo {volume} {93}},\ \bibinfo {pages}
  {032412} (\bibinfo {year} {2016})}\BibitemShut {NoStop}%
\bibitem [{\citenamefont {Wang}\ and\ \citenamefont
  {Ferguson}(2018)}]{wang2018recovery}%
  \BibitemOpen
  \bibfield  {author} {\bibinfo {author} {\bibfnamefont {J.}~\bibnamefont
  {Wang}}\ and\ \bibinfo {author} {\bibfnamefont {A.~L.}\ \bibnamefont
  {Ferguson}},\ }\bibfield  {title} {\enquote {\bibinfo {title} {Recovery of
  protein folding funnels from single-molecule time series by delay embeddings
  and manifold learning},}\ }\href@noop {} {\bibfield  {journal} {\bibinfo
  {journal} {The Journal of Physical Chemistry B}\ }\textbf {\bibinfo {volume}
  {122}},\ \bibinfo {pages} {11931--11952} (\bibinfo {year}
  {2018})}\BibitemShut {NoStop}%
\bibitem [{\citenamefont {Fung}\ \emph {et~al.}(2016)\citenamefont {Fung},
  \citenamefont {Hanna}, \citenamefont {Vendrell}, \citenamefont {Ramakrishna},
  \citenamefont {Seideman}, \citenamefont {Santra},\ and\ \citenamefont
  {Ourmazd}}]{fung2016dynamics}%
  \BibitemOpen
  \bibfield  {author} {\bibinfo {author} {\bibfnamefont {R.}~\bibnamefont
  {Fung}}, \bibinfo {author} {\bibfnamefont {A.~M.}\ \bibnamefont {Hanna}},
  \bibinfo {author} {\bibfnamefont {O.}~\bibnamefont {Vendrell}}, \bibinfo
  {author} {\bibfnamefont {S.}~\bibnamefont {Ramakrishna}}, \bibinfo {author}
  {\bibfnamefont {T.}~\bibnamefont {Seideman}}, \bibinfo {author}
  {\bibfnamefont {R.}~\bibnamefont {Santra}}, \ and\ \bibinfo {author}
  {\bibfnamefont {A.}~\bibnamefont {Ourmazd}},\ }\bibfield  {title} {\enquote
  {\bibinfo {title} {Dynamics from noisy data with extreme timing
  uncertainty},}\ }\href@noop {} {\bibfield  {journal} {\bibinfo  {journal}
  {Nature}\ }\textbf {\bibinfo {volume} {532}},\ \bibinfo {pages} {471}
  (\bibinfo {year} {2016})}\BibitemShut {NoStop}%
\bibitem [{\citenamefont {Suarez}\ \emph {et~al.}(2014)\citenamefont {Suarez},
  \citenamefont {Lettieri}, \citenamefont {Zwier}, \citenamefont {Stringer},
  \citenamefont {Subramanian}, \citenamefont {Chong},\ and\ \citenamefont
  {Zuckerman}}]{suarez2014simultaneous}%
  \BibitemOpen
  \bibfield  {author} {\bibinfo {author} {\bibfnamefont {E.}~\bibnamefont
  {Suarez}}, \bibinfo {author} {\bibfnamefont {S.}~\bibnamefont {Lettieri}},
  \bibinfo {author} {\bibfnamefont {M.~C.}\ \bibnamefont {Zwier}}, \bibinfo
  {author} {\bibfnamefont {C.~A.}\ \bibnamefont {Stringer}}, \bibinfo {author}
  {\bibfnamefont {S.~R.}\ \bibnamefont {Subramanian}}, \bibinfo {author}
  {\bibfnamefont {L.~T.}\ \bibnamefont {Chong}}, \ and\ \bibinfo {author}
  {\bibfnamefont {D.~M.}\ \bibnamefont {Zuckerman}},\ }\bibfield  {title}
  {\enquote {\bibinfo {title} {Simultaneous computation of dynamical and
  equilibrium information using a weighted ensemble of trajectories},}\
  }\href@noop {} {\bibfield  {journal} {\bibinfo  {journal} {Journal of
  Chemical Theory and Computation}\ }\textbf {\bibinfo {volume} {10}},\
  \bibinfo {pages} {2658--2667} (\bibinfo {year} {2014})}\BibitemShut {NoStop}%
\bibitem [{\citenamefont {Durrett}(2010)}]{durrett2010probability}%
  \BibitemOpen
  \bibfield  {author} {\bibinfo {author} {\bibfnamefont {R.}~\bibnamefont
  {Durrett}},\ }\href@noop {} {\emph {\bibinfo {title} {Probability: Theory and
  Examples}}}\ (\bibinfo  {publisher} {Cambridge university press},\ \bibinfo
  {year} {2010})\BibitemShut {NoStop}%
\bibitem [{\citenamefont {Shaw}\ \emph {et~al.}(2010)\citenamefont {Shaw},
  \citenamefont {Maragakis}, \citenamefont {Lindorff-Larsen}, \citenamefont
  {Piana}, \citenamefont {Dror}, \citenamefont {Eastwood}, \citenamefont
  {Bank}, \citenamefont {Jumper}, \citenamefont {Salmon}, \citenamefont
  {Shan},\ and\ \citenamefont {Wriggers}}]{shaw2010atomic}%
  \BibitemOpen
  \bibfield  {author} {\bibinfo {author} {\bibfnamefont {D.~E.}\ \bibnamefont
  {Shaw}}, \bibinfo {author} {\bibfnamefont {P.}~\bibnamefont {Maragakis}},
  \bibinfo {author} {\bibfnamefont {K.}~\bibnamefont {Lindorff-Larsen}},
  \bibinfo {author} {\bibfnamefont {S.}~\bibnamefont {Piana}}, \bibinfo
  {author} {\bibfnamefont {R.~O.}\ \bibnamefont {Dror}}, \bibinfo {author}
  {\bibfnamefont {M.~P.}\ \bibnamefont {Eastwood}}, \bibinfo {author}
  {\bibfnamefont {J.~A.}\ \bibnamefont {Bank}}, \bibinfo {author}
  {\bibfnamefont {J.~M.}\ \bibnamefont {Jumper}}, \bibinfo {author}
  {\bibfnamefont {J.~K.}\ \bibnamefont {Salmon}}, \bibinfo {author}
  {\bibfnamefont {Y.}~\bibnamefont {Shan}}, \ and\ \bibinfo {author}
  {\bibfnamefont {W.}~\bibnamefont {Wriggers}},\ }\bibfield  {title} {\enquote
  {\bibinfo {title} {Atomic-level characterization of the structural dynamics
  of proteins},}\ }\href@noop {} {\bibfield  {journal} {\bibinfo  {journal}
  {Science}\ }\textbf {\bibinfo {volume} {330}},\ \bibinfo {pages} {341--346}
  (\bibinfo {year} {2010})}\BibitemShut {NoStop}%
\bibitem [{\citenamefont {Piana}\ \emph {et~al.}(2011)\citenamefont {Piana},
  \citenamefont {Sarkar}, \citenamefont {Lindorff-Larsen}, \citenamefont {Guo},
  \citenamefont {Gruebele},\ and\ \citenamefont
  {Shaw}}]{piana2011computational}%
  \BibitemOpen
  \bibfield  {author} {\bibinfo {author} {\bibfnamefont {S.}~\bibnamefont
  {Piana}}, \bibinfo {author} {\bibfnamefont {K.}~\bibnamefont {Sarkar}},
  \bibinfo {author} {\bibfnamefont {K.}~\bibnamefont {Lindorff-Larsen}},
  \bibinfo {author} {\bibfnamefont {M.}~\bibnamefont {Guo}}, \bibinfo {author}
  {\bibfnamefont {M.}~\bibnamefont {Gruebele}}, \ and\ \bibinfo {author}
  {\bibfnamefont {D.~E.}\ \bibnamefont {Shaw}},\ }\bibfield  {title} {\enquote
  {\bibinfo {title} {Computational design and experimental testing of the
  fastest-folding $\beta$-sheet protein},}\ }\href@noop {} {\bibfield
  {journal} {\bibinfo  {journal} {Journal of Molecular Biology}\ }\textbf
  {\bibinfo {volume} {405}},\ \bibinfo {pages} {43--48} (\bibinfo {year}
  {2011})}\BibitemShut {NoStop}%
\bibitem [{\citenamefont {Long}\ and\ \citenamefont
  {Ferguson}(2017)}]{long2017landmark}%
  \BibitemOpen
  \bibfield  {author} {\bibinfo {author} {\bibfnamefont {A.~W.}\ \bibnamefont
  {Long}}\ and\ \bibinfo {author} {\bibfnamefont {A.~L.}\ \bibnamefont
  {Ferguson}},\ }\bibfield  {title} {\enquote {\bibinfo {title} {Landmark
  diffusion maps ({L}-d{M}aps): Accelerated manifold learning out-of-sample
  extension},}\ }\href@noop {} {\bibfield  {journal} {\bibinfo  {journal}
  {Applied and Computational Harmonic Analysis}\ } (\bibinfo {year}
  {2017})}\BibitemShut {NoStop}%
\bibitem [{\citenamefont {Williams}\ and\ \citenamefont
  {Seeger}(2001)}]{williams2001using}%
  \BibitemOpen
  \bibfield  {author} {\bibinfo {author} {\bibfnamefont {C.~K.}\ \bibnamefont
  {Williams}}\ and\ \bibinfo {author} {\bibfnamefont {M.}~\bibnamefont
  {Seeger}},\ }\bibfield  {title} {\enquote {\bibinfo {title} {Using the
  {N}ystr{\"o}m method to speed up kernel machines},}\ }in\ \href@noop {}
  {\emph {\bibinfo {booktitle} {Advances in Neural Information Processing
  Systems}}}\ (\bibinfo {year} {2001})\ pp.\ \bibinfo {pages}
  {682--688}\BibitemShut {NoStop}%
\bibitem [{\citenamefont {Bronshtein}\ \emph {et~al.}(2007)\citenamefont
  {Bronshtein}, \citenamefont {Semendyayev}, \citenamefont {Musiol},\ and\
  \citenamefont {Muehlig}}]{bronshtein2007handbook}%
  \BibitemOpen
  \bibfield  {author} {\bibinfo {author} {\bibfnamefont {I.}~\bibnamefont
  {Bronshtein}}, \bibinfo {author} {\bibfnamefont {K.}~\bibnamefont
  {Semendyayev}}, \bibinfo {author} {\bibfnamefont {G.}~\bibnamefont {Musiol}},
  \ and\ \bibinfo {author} {\bibfnamefont {H.}~\bibnamefont {Muehlig}},\
  }\href@noop {} {\emph {\bibinfo {title} {Handbook of Mathematics}}},\
  \bibinfo {edition} {3rd}\ ed.\ (\bibinfo  {publisher} {Springer},\ \bibinfo
  {year} {2007})\BibitemShut {NoStop}%
\bibitem [{\citenamefont {E}\ and\ \citenamefont
  {Vanden-Eijnden}(2010)}]{vanden2010transition}%
  \BibitemOpen
  \bibfield  {author} {\bibinfo {author} {\bibfnamefont {W.}~\bibnamefont {E}}\
  and\ \bibinfo {author} {\bibfnamefont {E.}~\bibnamefont {Vanden-Eijnden}},\
  }\bibfield  {title} {\enquote {\bibinfo {title} {Transition-path theory and
  path-finding algorithms for the study of rare events},}\ }\href@noop {}
  {\bibfield  {journal} {\bibinfo  {journal} {Annual Review of Physical
  Chemistry}\ }\textbf {\bibinfo {volume} {61}},\ \bibinfo {pages} {391--420}
  (\bibinfo {year} {2010})}\BibitemShut {NoStop}%
\end{thebibliography}%

\section{Supplementary Information}

\subsection{Connection between DGA and Markov State Modeling}\label{ssec:dga_msm_connection}
Here, we describe in detail the connection between DGA and certain dynamical estimates calculated using a MSM.

To map the general dynamics onto the state space of the Markov Chain, we make three assumptions.
\begin{assumption}[]\label{asm:msm_obeys_bcs}
	Each Markov state $S_i$ is contained entirely in either $D$ or in $D^c$.
\end{assumption}
\begin{assumption}[]\label{asm:boundary_decomposable}
	The boundary conditions $b$ can be expressed as
	\begin{equation}\label{eq:msm_approx_boundary}
	b(x) = \sum_{l\in D^c}^{M'} b_l \1_{S_l} (x).
	\end{equation}
\end{assumption}
\begin{assumption}[]\label{asm:com_decomposable}
	For any $\mathcal{L}_{p}^\dagger$ considered, $p$ can be written as
	\begin{equation}\label{eq:msm_approx_com}
	p(x) = \sum_{j \in D}^M \frac{p_j}{\<\1_j\>} \1_j(x) +  \sum_{l \in D^c}^{M'} \frac{p_l}{\<\1_l\>}s \1_l(x).
	\end{equation}
\end{assumption}
The first assumption is necessary for the basis set to obey the homogeneous boundary conditions, and can be enforced explicitly in the construction of the MSM.
The second two assumptions will be required to make the action of $\mathcal{L}$ representable as the action of matrices on vectors over the MSM states.
While these assumptions should not be expected to hold for general $b$ and $p$, 
in the correct limit of infinite sampling and sufficiently small Markov states, we expect \eqref{eq:msm_approx_boundary} and \eqref{eq:msm_approx_com} to be arbitrarily good approximations.
In fact, for most $b$ in Section~\ref{sec:feynman_kac}, assumption~\ref{asm:boundary_decomposable} can hold exactly.
We also note that the vector $p_i$ sums to one, as
\begin{align*}
1 =& \int p(x) \mu(dx) \\
=&  \int \sum_{j \in D}^M \frac{p_j}{\<\1_j\>} \1_j(x) +  \sum_{l \in D^c}^{M'} \frac{p_l}{\<\1_l\>} \1_l(x) \mu (dx) \\
=&  \sum_{j \in D}^M p_j + \sum_{l \in D^c}^{M'}  p_l.
\end{align*}
Consequently, $p_i$ is a probability distribution over the MSM state-space.

\subsubsection{Equations with the Transition Operator}
We first consider equations that take the form of \eqref{eq:generic_inhomogeneous_eqn}.  
As our guess, we will use \eqref{eq:naive_guess}.
Substituting into \eqref{eq:exact_gkn_scheme}, applying Assumption~\ref{asm:boundary_decomposable}, and dividing by $\<1_{S_i}\>$, we arrive at
\begin{equation}
\sum_{j\in D}^M\frac{1}{\Delta t}\left(P-I\right)_{ij} a_j = \eta_i - \sum_{l\in D^c}^{M'} \frac{1}{\Delta t} \left(P-I\right)_{il}b_l.
\end{equation}
Here $P_{ij}$ is the MSM transition matrix defined in \eqref{eq:msm_tmat_defn} with a time lag of $\Delta t$, and $\eta_i$ is defined as
\begin{equation}
\eta_i = \frac{\<\1_i,h\>}{\< \1_i \>} 
\end{equation}
This can be rewritten as
\begin{equation}\label{eq:msm_from_dga}
\begin{split}
\sum_{j}\frac{1}{\Delta t} \left(P - I\right)_{ij} a_j &= \eta_i \text{ for } i\in D \\
a_i &= b_i \text{ for } i \in D^c
\end{split}
\end{equation}
where the sum is over states on the entire domain.
This is equivalent to \eqref{eq:generic_inhomogeneous_eqn} for the dynamics given by the MSM.

\subsubsection{Equations with Transition Adjoints}
For equations that take the form of \eqref{eq:generic_inhomogeneous_adjoint_eqn} we again begin with \eqref{eq:exact_gkn_scheme}, this time with terms defined by equations \eqref{eq:L_inner_product_adjoint}, \eqref{eq:r_inner_product_adjoint}, and \eqref{eq:h_inner_product_adjoint}.
Substituting in our guess function and Assumptions~\ref{asm:com_decomposable} and~\ref{asm:msm_obeys_bcs}, we have
\begin{equation}
\begin{split}
\sum_{j \in D}^M \<\mathcal{L} \1_i, \1_j \> & \left(\frac{p_j}{\<\1_j\>}\right) a_j \\
=  \left(\frac{p_i}{\<\1_i\>}\right)& \<\1_i, h \>  - \sum_{l\in D^c}^{M'}\<\mathcal{L} \1_i, \1_l \>b_l \left(\frac{p_l}{\<\1_l\>}\right)
\end{split}
\end{equation}
We then divide both sides by $p_i$.  Applying the definition of $P_{ij}$, we arrive at
\begin{equation}
\sum_{j \in D}^M p_i^{-1} \left( P - I\right)_{ij}^T p_j a_j 
	 = \eta_i -  \sum_{l\in D^c}^{M'} p_i^{-1} \left( P - I\right)_{il} p_l b_l .
\end{equation}
which, as before, is equivalent to solving 
\begin{equation}\label{eq:msm_adjoint_from_dga}
\begin{split}
\sum_{j} p_i^{-1} \left(P - I\right)^T_{ij} p_j &= \eta_i \text{ for } i\in D \\
a_i &= b_i \text{ for } i \in D^c.
\end{split}
\end{equation}
Comparing with \eqref{eq:adjoint_expression}, we see that the matrix with elements $p_i^{-1} \left(P-I\right)_{ij}^T p_j$ is the weighted adjoint of the MSM generator against $p_i$.
Consequently, \eqref{eq:msm_adjoint_from_dga} is equivalent to \eqref{eq:generic_inhomogeneous_adjoint_eqn} for the MSM.

\subsection{Details of Diffusion Map Construction}\label{ssec:dmap_details}
Here, we give the specific kernel and parameter choice used in our calculations used to construct the diffusion map in our calculations.  Our procedure closely follows work in references~\onlinecite{berry2015nonparametric} and~\onlinecite{berry2016variable}.  Specifically, our algorithm corresponds to the parameter choice $\alpha=0$ and $\beta=-1/d$ in reference~\onlinecite{berry2016variable} and not performing the bandwidth normalization in equation~(5).

\subsubsection{Kernel Construction}

As in Section~\ref{sec:dmap_basis_definition}, let $x_m$ be a collection of $N$ datapoints.
%We will assume the data is drawn from that the data is drawn from probability measure with density $q$ against the Lebesgue measure.
%We first construct an estimate of $q$.  
We define the initial bandwidth function
\begin{equation*}
\varsigma_0(x_m) = \frac{1}{k_0} \sum_{l=1}^{k_0} ||x_m-x_{I(m,l)}||^2
\end{equation*}
where $I(m,l)$ is the index of the $l$'th nearest neighbor to point $x_m$ (not including $x_m$). Here $k_0$ is a neighborhood parameter giving the number of nearest neighbors considered, we follow reference~\onlinecite{berry2016variable} and set it to 7.  We then construct the kernel density estimate 
\begin{equation*}
\begin{split}
q(x_m) = \frac{(2 \pi \eps_0)^{-d/2}}{N \varsigma_0(x_m)^d } \sum_{n=1}^{N} K_0 (x_m, x_n; \eps_0)\text{, where} \\
K_0 (x_m, x_n; \eps_0)=\exp\left(\frac{-||x_m-x_n||^2}{2 \eps_0 \varsigma_0(x_m ) \varsigma_0(x_n )}\right)
\end{split}
\end{equation*}
where $d$ is the intrinsic dimensionality of the data manifold and $\eps_0$ is a bandwidth parameter.  To estimate $d$ and a good choice for $\eps_0$, we consider all possible choices of $eps_0$ of the form $2^k$ with $k=-40, -39, \dots, 39, 40$.  We then set
\begin{equation}
d = \frac{2}{\ln(2)} \max_k \left[\ln\left(\frac{\sum_{m,n} K_0(x_m, x_n, 2^{k+1})}{\sum_{m,n} K_0(x_m, x_n, 2^{k})}\right)\right]
\end{equation} 
Reference~\onlinecite{berry2015nonparametric} suggests setting $\eps_0$ by using the $k$ where the right-hand-side attains its maximum.  In practice we find this can be overly aggressive, so we subsequently multiply $\eps_0$ by 2.
We then construct the Diffusion map kernel matrix as
\begin{equation}
K(x_m,x_n) = \exp \left(\frac{||x-y||^2}{\eps q(x_m)^{-1/d} q(x_n)^{-1/d}}\right)
\end{equation}
where we select the $\eps$ using the same procedure as before.

\subsubsection{Out-of-sample Extension for the Diffusion-Map Basis}

To predict the values of the quantities in Section~\ref{sec:feynman_kac} at new datapoints, we will need to extend the diffusion-map basis and guess functions to new configurations.
Initially, one might attempt this by constructing a new diffusion map matrix that contains both the old and the new points and recomputing the guess and eigenvectors.
However, not only would this procedure be expensive, it would change the values of the basis and guess functions on the old points.
Consequently, the estimates of $a_j$ would be incorrect, and the entire DGA scheme would need to be repeated.
We therefore seek a method for extending the basis and guess functions to new points that leave their values on older points unchanged.

Let $x_\nu$ be a new point added to the dataset.
To extend the basis functions to $x_\nu$, we can use the established method of Nystr\"om extension.\cite{williams2001using, long2017landmark}
Let $\varphi_i$ be an eigenvector of the submatrix discussed in Section~\ref{sec:dmap_basis_definition}, and let $\kappa_i$ be the associated eigenvalue.
The estimate of the basis function on $x_\nu$ is given by
\begin{equation}
\varphi_i (x_\nu) = \frac{1}{\kappa_i} \frac{\sum_m K_\eps (x_m, x_\nu) \varphi_i(x_m)}{\sum_m K_\eps (x_m, x_\nu)}
\end{equation}
To extend the guess function to new configurations, we introduce a new method based on the Jacobi method.\cite{bronshtein2007handbook}
We first consider $\hat{P}$, a new diffusion map matrix built using both the old datapoints $x_{1..N}$  and the new datapoint $x_\nu$.
The guess function associated with $\hat{P}$ would then solve the problem
\begin{equation}\label{eq:augmented_dmap_linsolve}
\left(\hat{P} - I\right) g = h
\end{equation}
for all of the points in $D$.
We will construct our estimate of $g$ at the new point by considering a single iteration of the Jacobi method for solving \eqref{eq:augmented_dmap_linsolve}. Our initial vector takes values of $g_m$ on $x_m$ and $0$ on $x_\nu$.
This gives us the following out-of-sample extension formula
\begin{align}
g_\nu = \frac{1}{\hat{P}_{\nu \nu} -1} \left(h_\nu - \sum_{m=1}^N \hat{P}_{\nu m} g_m \right)
\end{align}
where the sum runs only over points in the original dataset.
This can be further simplified using the definition of $\hat{P}$ to 
\begin{align}
g_\nu = \frac{\sum_{m=1}^N K_\eps \left(x_\nu, x_m\right) g_m }{\sum_{m=1}^N K_\eps \left(x_\nu, x_m\right)} - h_\nu \left(1 + \frac{ K_\eps \left(x_\nu, x_\nu\right)}{\sum_{m=1}^N K_\eps \left(x_\nu, x_m\right)}\right).
\end{align}

\subsection{Derivation of Transition Path Theory Reactive Flux and Rate in Discrete Time}\label{ssec:tpt_derivation}
Transition path theory was originally formulated for diffusion processes\cite{vanden2006transition} and was extended to finite-state Markov jump processes.\cite{metzner2009transition}  Here, we derive analogous equations for discrete-time Markov chains on arbitrary state spaces.
The derivation closely follows reference~\onlinecite{vanden2006transition}.
Let $x(t)$ be a single trajectory ergodically sampling the stationary measure.  We will extend the trajectory both forwards and backwards in time so the time index $t$ takes values from $-\infty$ to $\infty$.
For all $t$, let 
\begin{align}
t_{AB}^+(t) &=\min \left\{ t' |  t'\geq t, x(t')\in A\cup B   \right\} \\
t_{AB}^-(t) &=\max \left\{ t' |  t'\leq t, x(t')\in A\cup B   \right\}
\end{align}
be the next time the system entered $A$ or $B$ and the most recent time the system left $A$ or $B$, respectively.
Now let $C$ be as in \eqref{eq:defn_total_rxn_current}.
The total reactive current is defined as
\begin{align}
I_{B\to A} =& \lim_{T \to \infty} \frac{1}{2 T}\sum_{t\in[-T,T]} \left[\1_C\left(x(t)\right) \1_{C^c} \left(x(t+\Delta t)\right)\right. \nonumber \\
& \qquad\qquad -\left.\1_{C^c}\left(x(t)\right) \1_{C} \left(x(t+\Delta t)\right) \right] \\
& \qquad\qquad \left[\1_A \left(x(t_{AB}^-(t)\right) \1_B \left(x(t_{AB}^+(t+\Delta t)\right)\right] \nonumber
\end{align}
Using ergodicity and the strong Markov property, we can rewrite this as an average against $\rho_{\Delta t}$.
\begin{align}\label{eq:total_rxn_current_traj_level}
I_{B\to A} =& \int \1_{C^c}(y) q_+ (y) q_-(x) \1_C(x) 
\pi(x) \rho_{\Delta t}(dx, dy)  \\
                  &- \int \1_{C}(y) q_+(y) q_-(x) \1_{C^c}(x)  \pi(x) \rho_{\Delta t} (dx, dy)\nonumber \\
\end{align}
This is the discrete-time equivalent of equation~(30) in reference~\onlinecite{vanden2006transition}.
Applying the definition of the generator and observing that that $\1_C(x) \1_{C^c}(x)=0$ everywhere gives \eqref{eq:defn_total_rxn_current}.
We then arrive at \eqref{eq:defn_tpt_rate} in our work by the same arguments as in reference~\onlinecite{vanden2010transition}.

\subsection{Grid-Based Reference Scheme}
Here we discuss the scheme used to calculate the reference values for our test system in Sections~\ref{sec:dmap_basis_definition} and~\ref{sec:delay_embedding}.
Instead of considering the discrete time process directly, we will attempt to approximate the dynamics of the continuous-time Brownian dynamics on the test potential.  
To this end, we define a Markov hopping process on a grid that converges to the continuous time dynamics as the grid becomes finer.
Specifically, we allow nearest neighbor hops on a square grid with spacing $\epsilon$.  The hopping probabilities are given by 
\begin{align}\label{eq:grid_hopping_probabilities}
P(x+\epsilon,y) =& \left(\frac{1}{4}\right)\left(\frac{1}{1+\exp\left[U(x+\epsilon,y) - U(x, y)\right]}\right) \nonumber \\
P(x-\epsilon,y) =& \left(\frac{1}{4}\right)\left(\frac{1}{1+\exp\left[U(x-\epsilon,y) - U(x, y)\right]}\right)\nonumber  \\
P(x,y+\epsilon) =& \left(\frac{1}{4}\right)\left(\frac{1}{1+\exp\left[U(x,y+\epsilon) - U(x, y)\right]}\right)  \\
P(x,y-\epsilon) =& \left(\frac{1}{4}\right)\left(\frac{1}{1+\exp\left[U(x,y-\epsilon) - U(x, y)\right]}\right)\nonumber  \\
P(x,y) =& 1-  P(x+\epsilon,y) - P(x-\epsilon,y) \nonumber\\
& - P(x,y+\epsilon) - P(x,y-\eps).\nonumber
 \end{align}
Here $P(x\pm\epsilon,y)$ is the probability of hopping one grid point to the right or left, $P(x,y\pm\epsilon)$ is the probability of hopping up or down the grid, and $P(x,y)$ is the probability of remaining in place.

We will not give a full proof of convergence.
Instead we merely demonstrate that as $\epsilon \to 0$, we approximate the infinitesimal generator $\mathcal{L}^{\rm brwn}$ for Brownian Dynamics.
Let $P$ be the transition matrix associated with the transition probabilities given by \eqref{eq:grid_hopping_probabilities}, $f$ be a three-times continuously differentiable function, and the vector $\vec{f}$ the values of $f$ evaluated at each grid point.  In the limit of $\epsilon\to 0$,
\begin{equation}\label{eq:converged_generator}
\frac{16 (P-I) \vec{f}}{\epsilon^2}(x,y) =\mathcal{L}^{\rm brwn} f(x,y) + \mathcal{O}(\epsilon)
\end{equation}
where $\mathcal{L}^{\rm brwn}$ is the infinitesimal generator for Brownian dynamics with isotropic diffusion constant,
\begin{align*}
\mathcal{L}^{\rm brwn} f(x,y) =& -\partial_x U(x,y) \partial_x f(x,y) - \partial_y U(x,y) \partial_y f(x,y)  \\ &+ \partial_x^2 f(x,y) + \partial_y^2 f(x,y).
\end{align*}  

To demonstrate this, we write $(P-I)\vec{f}$ explicitly as
\begin{align*}
(P-I)f(x,y) =&  P(x+\epsilon,y)f(x+\epsilon  ,y) \\
& + P(x-\epsilon,y)f(x-\epsilon ,y)\\
& + P(x,y+\epsilon)f(x ,y+\epsilon) \\
&+P(x,y-\epsilon)f(x ,y-\epsilon)\\
 &+ P(x,y)f(x,y) + \mathcal{O}(\epsilon^3)\\
\end{align*}
If we expand $f$ to second order around $(x,y)$, the zeroth order term cancels, leaving
\begin{align*}
(P-I)f(x,y) =&  P(x+\epsilon,y)\left(\epsilon  \partial_x f  + \frac{1}{2}\epsilon^2\partial^2_x f   \right) \\ 
	&- P(x-\epsilon,y)\left( \epsilon \partial_x f - \frac{1}{2}\epsilon^2 \partial^2_x f   \right) \\
	&+ P(x,y+\epsilon r)\left(\epsilon \partial_y f  + \frac{1}{2}(\epsilon r)^2\partial^2_y f   \right) \\
	&- P(x,y-\epsilon r)\left(\epsilon \partial_y f  - \frac{1}{2}(\epsilon r)^2\partial^2_y f   \right) + \mathcal{O}(\epsilon^3)
\end{align*}
We then expand the transition probabilities to first order, giving
\begin{align*}
P(x \pm \epsilon,y) &= \frac{1}{8}\left( 1\mp \frac{1}{2}\partial_x U(x,y)\epsilon \right) + \mathcal{O}(\epsilon^2)  \\
P(x,y \pm \epsilon r) &= \frac{1}{8}\left( 1\mp  \frac{1}{2} \partial_x U(x,y)\epsilon \right) + \mathcal{O}(\epsilon^2).
\end{align*}
Substituting, simplifying, and multiplying by $16 / \epsilon^2$ gives~\eqref{eq:converged_generator}.

To estimate the reference quantities for our test system, we constructed a square grid on the interval $-2.5 \leq x \leq 1.5$ and $-1.5 \leq y \leq 2.5$ with grid spacing of 0.005.  We then construct the transition rate matrix $16 (P-I) / \epsilon^2$, and estimate the dynamical quantities using the corresponding formulas in Section~\ref{sec:feynman_kac}.

\subsection{Basis Size Choice for M{\"u}ller-Brown model}\label{ssec:mb_basis_size}
\bigadd{
In Figure \ref{fig:msm_basis_dependence}, we show the dependence of the root-mean-square error in the committor on basis size for the M{\"u}ller-Brown model.  While using 1000 basis functions gives a slightly better result at higher dimensions, it is  not enough to appreciably change the trends depicted in Figure~\ref{fig:committor_error_w_dimension}.  However, choosing 1000 or more basis functions gives worse results for the two-dimensional system.  We therefore chose to use $500$ dimensions to avoid giving the impression that the diffusion-map basis outperforms the MSM basis at low dimensions.
\begin{figure}
	\includegraphics[width=\columnwidth]{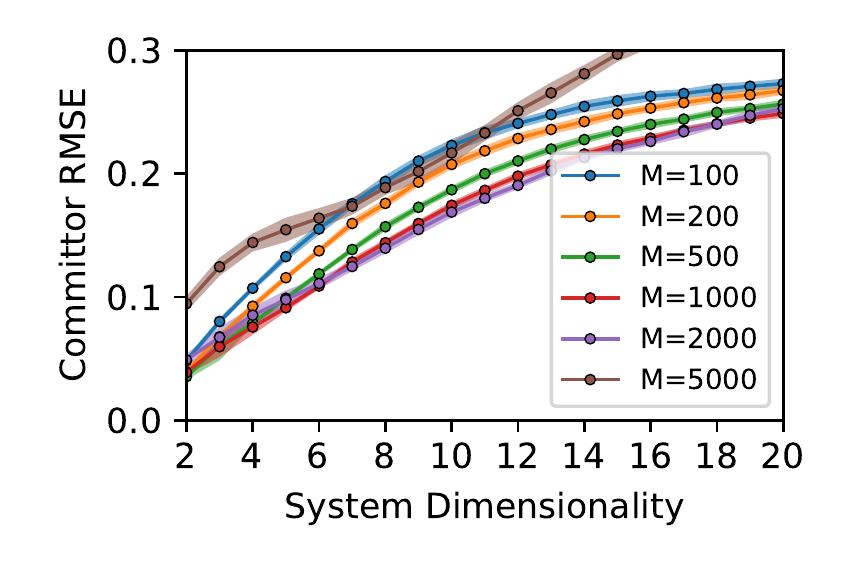}
	\caption{
		Dependence of the MSM committor root-mean-square error (RMSE) on the number of clusters.  Different curves correspond to different numbers of Markov states.
	}
	\label{fig:msm_basis_dependence}
\end{figure}
}

\subsection{Numerical Effect of Enforcing Detailed Balance}\label{ssec:detailed_balance_results}
\bigadd{
To test the effect of enforcing detailed balance in MSMs through a maximum likelihood procedure, we returned to our two-dimensional test potential without any additional nuisance degrees of freedom.
Using the clusterings described in Section~\ref{ssec:basis_set_d_comparison}, we constructed MSMs in PyEMMA both with and without the reversible option set to True.  We then estimated the mean first-passage time from state $B$ into state $A$, using the states depicted in Figure~\ref{fig:demo_basis_functions}A. As before, we repeated this procedure over thirty replicates.  Moreover, we also varied the number of short trajectories included in the dataset to observe trends in statistical convergence.
}

\begin{figure*}
	\includegraphics[width=\textwidth]{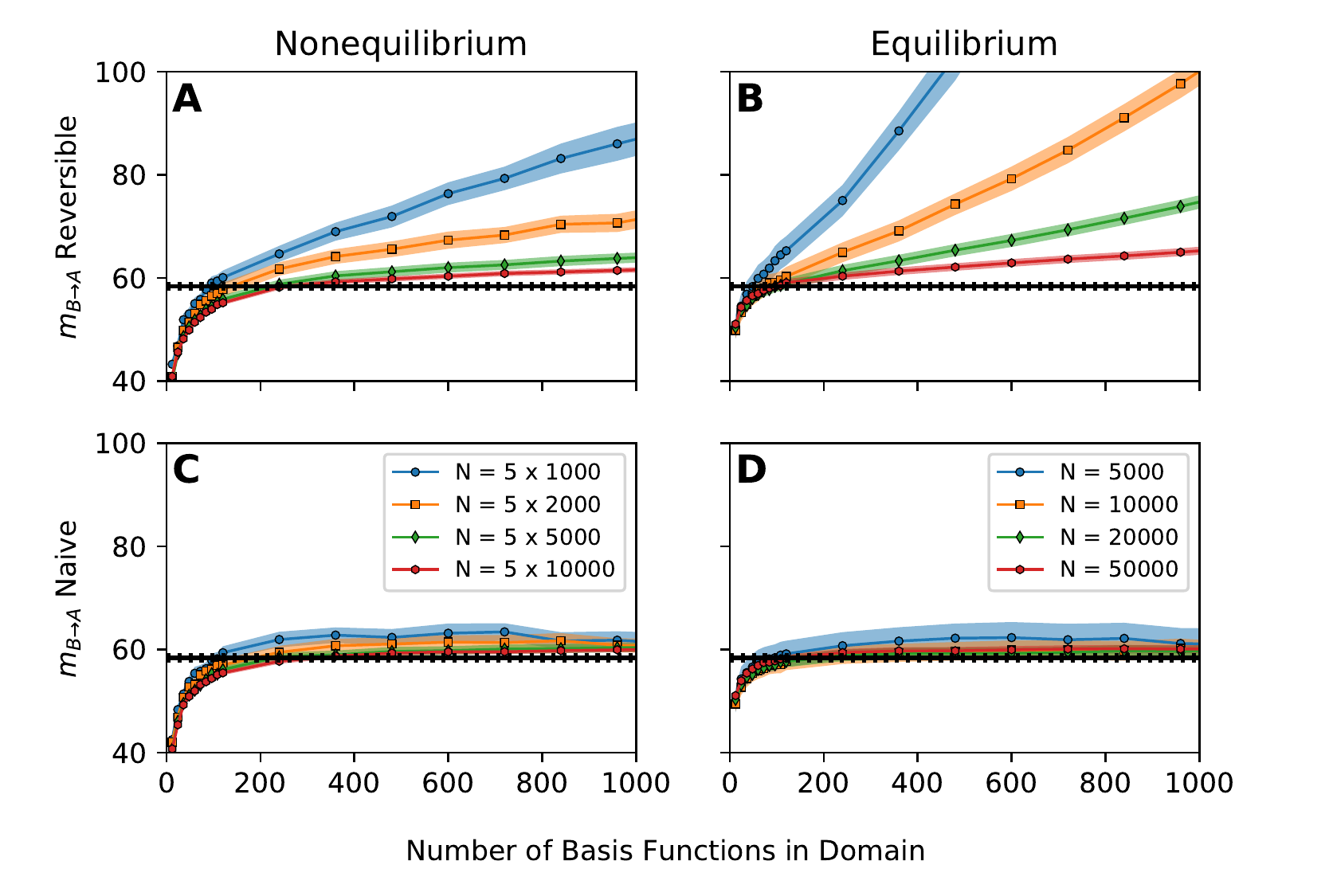}
	\caption{\bigadd{Effect of enforcing MSM reversibility on the estimated mean first-passage time from state $B$ to state $A$ on the scaled M{\"u}ller-Brown potential.  Estimates in the top row are constructed using the reversible MSM estimator and estimates in the bottom row are not.  The  columns correspond to two different datasets: the left column shows estimates constructed from the nonequilibrium dataset detailed in section~\ref{sec:dmap_basis_definition}, and the right column shows estimates constructed from a long equilibrium trajectory.  Different curves correspond to MSMs constructed from datasets of different sizes.}}
	\label{fig:reversibility_comparison}
\end{figure*}
\bigadd{
Our results are given in Figure~\ref{fig:reversibility_comparison}.
The mean first-passage time calculated using reversible MSMs, depicted in panel~A, grows unboundedly with increasing basis size.  To demonstrate that this not due to the nature of the data, we repeated the calculation on a long equilibrium trajectory of commensurate length.  Our results, shown in panel~B, exhibit the same phenomenon.  We also varied the error tolerance for convergence, as well as the minimum count required for connectivity.  Neither affected the results.  
Moreover, an in-house code for the iteration described in reference~\onlinecite{bowman2009progress} gave the same results as PyEMMA.
Rather, we see that the bias decays with increasing dataset sizes, suggesting that it is statistical in nature. 
}

\bigadd{
In panels~C and~D, we show estimates constructed without enforcing reversibility, which we term the naive estimator.  The naive estimator does not have the same bias.  This suggests that the maximum likelihood iteration introduces a large, slowly decaying statistical error.
}
\begin{figure*}
	\includegraphics[width=\textwidth]{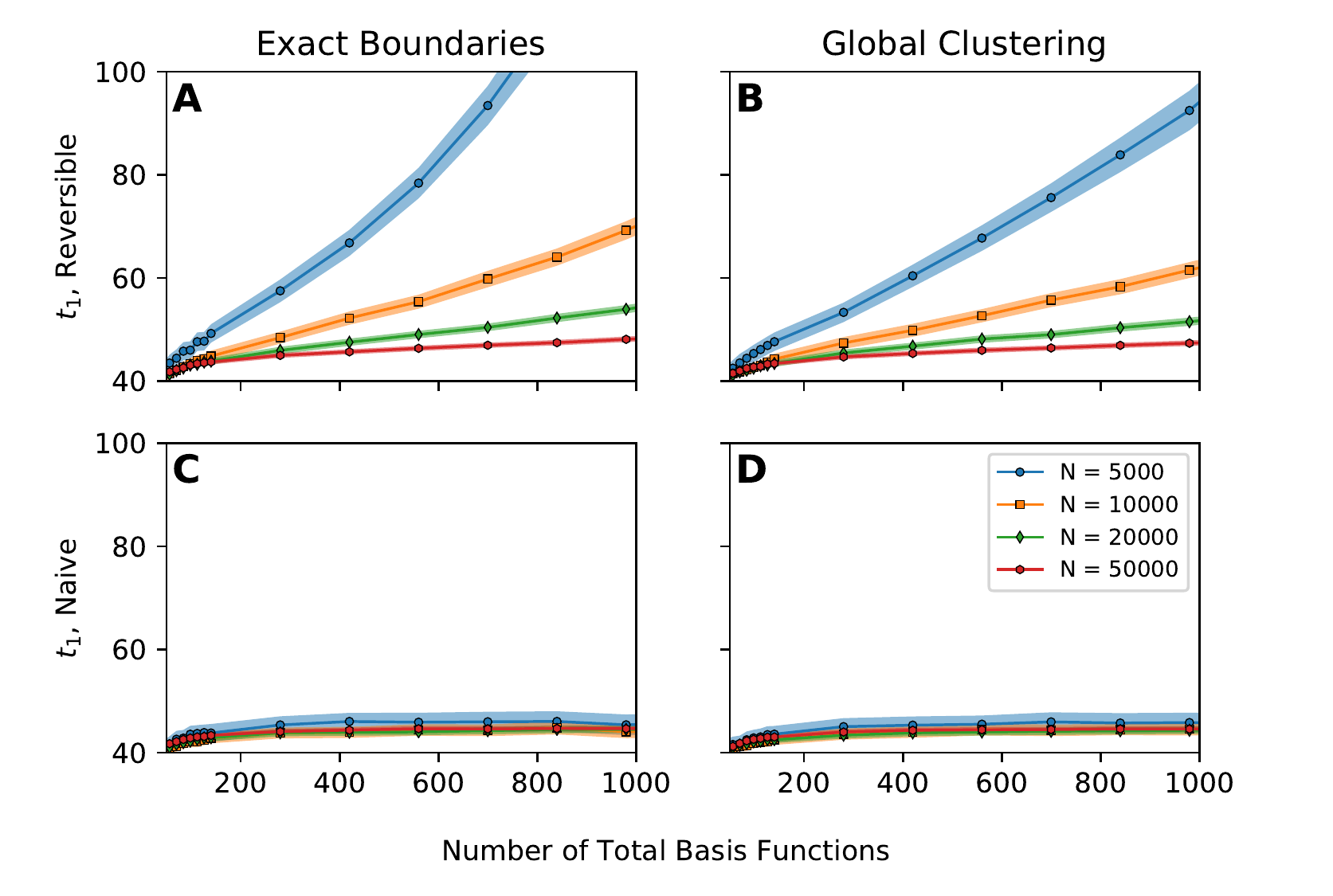}
	\caption{\bigadd{Dominant implied timescale for MSMs constructed on a long equilibrium trajectory on the scaled M\"uller-Brown potential. Estimates in the top row are constructed using the reversible MSM estimator and estimates in the bottom row use the naive estimator.  Columns correspond two different clustering schemes.  The left column gives estimates constructed using the clustering described in Section~\ref{ssec:basis_set_d_comparison}, and the right column gives estimates obtained by clustering the data without regard for the boundary conditions (i.e., globally).  Different curves correspond to MSMs constructed on different size datasets.}
	}
	\label{fig:reversibility_timescales}
\end{figure*}
\bigadd{
To ensure that this is not an artifact of the clustering procedure, we also constructed MSMs by applying $k$-means globally to the data, without regard to boundary conditions.  We then estimated the dominant implied timescale for both clustering schemes, which we plot in Figure~\ref{fig:reversibility_timescales}.  We see the same trends as in the mean first-passage time: for reversible MSMs, the implied timescale grows unboundedly with the number of basis functions for both clustering methods.  
%We do, however, see a slight difference in the rate of increase.  
In contrast, both clustering methods converge equally well when using the naive estimator.}

\subsection{Supplementary Plots for Delay Embedding the M{\"u}ller-Brown model}\label{ssec:deb_supp}
\begin{figure}
	\includegraphics[width=\columnwidth]{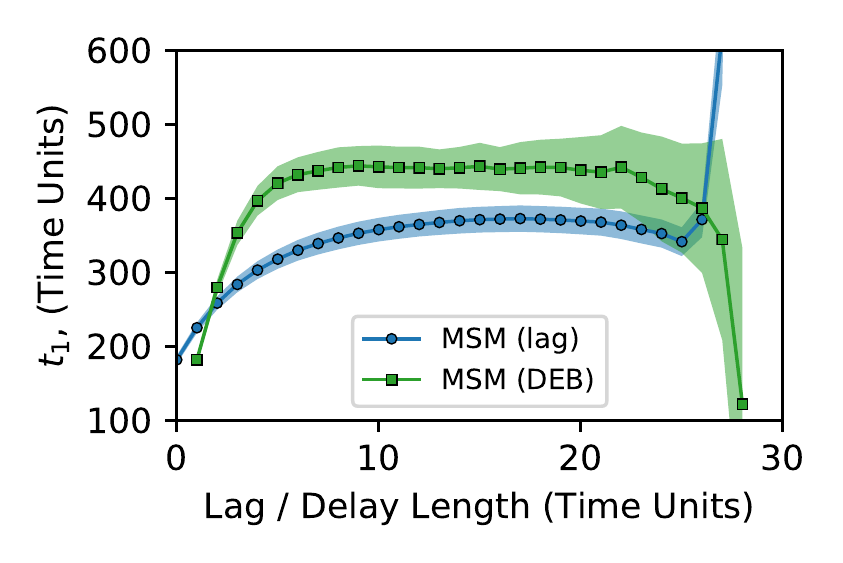}
	\caption{\bigadd{
    Implied timescales for the delay-embedded MSM and lagged MSMs in Section~\ref{sec:delay_embedding}.}
    }
	\label{fig:1d_msm_its}
\end{figure}

\begin{figure*}
  \includegraphics[width=\textwidth]{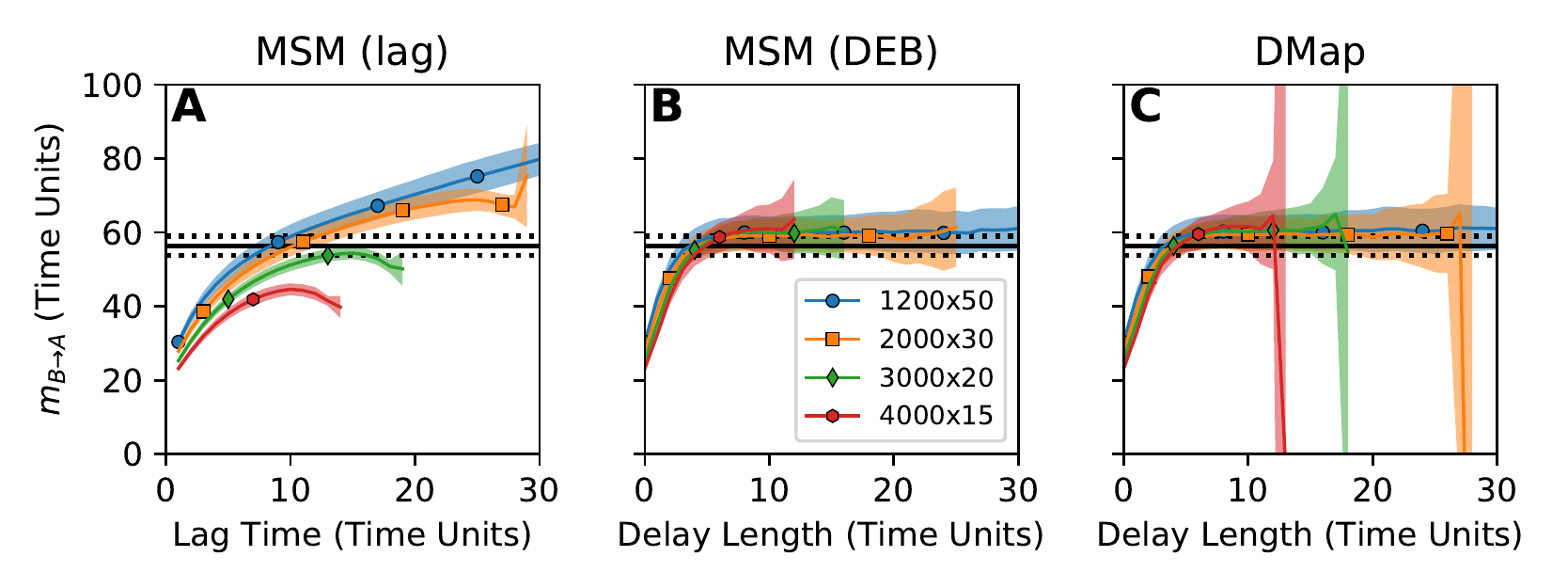}
  \caption{
		\bigadd{
Comparison of methods for controlling the projection error in an incomplete CV space.
		Plots are as in Figure~\ref{fig:delay_vs_lag_1d}, with the addition of three new datasets.  The curves correspond to datasets consisting of 1200 trajectories, each 50 time units long (blue circles), 200 points, each 30 time units long (orange squares, the same data as pictured in Figure~\ref{fig:delay_vs_lag_1d}), 3000 trajectories, each 20 units long (green diamonds), and 4000 trajectories, each 15 units long (red hexagons).}
		}
  \label{fig:delay_vs_lag_1d_Expanded}
\end{figure*}

\bigadd{
In Figure~\ref{fig:1d_msm_its}, we give implied timescales for the MSMs constructed in Section~\ref{sec:delay_embedding}.}
%We see that the delay-embedded MSM has a larger implied timescale for each lag time.  As the details obey detailed balance, the variational principle implies that the clustering on the delay-embedded space better captures the slow modes of the system.\cite{noe2013variational}
\bigadd{
To test the effect of trajectory length on the one-dimensional, delay-embedded data, we repeated the calculation for three additional datasets.  The total number of points in each dataset is fixed, but each nonequilibrium trajectory is of different length.  We plot the resulting curves in Figure~\ref{fig:delay_vs_lag_1d_Expanded}.  In all cases, we see an anomalous behavior when the delay length or lag time approaches the total length of the trajectory.}

\end{document}